\documentclass[aps,
 superscriptaddress,rmp,reprint,normalem, onecolumn,longbibliography,nofootinbib]{revtex4-2}
\usepackage{bm}
\usepackage{bm,dcolumn,amsmath,graphicx,url,textcase}
\usepackage{epsfig}
\usepackage{epstopdf}
\usepackage{euscript}
\usepackage{amsfonts}
\usepackage{float}
\usepackage{braket}
\DeclareMathAlphabet{\mathpzc}{OT1}{pzc}{m}{it}
\usepackage{threeparttable}
\usepackage{longtable}                  
\usepackage{changes} 

\usepackage{textpos}
\usepackage{amssymb,mathrsfs,amsthm}
\usepackage{booktabs}
\usepackage{array}
\usepackage{tabularx}
\usepackage{dcolumn}
\usepackage{rotating}
\usepackage{ulem}
\usepackage{soul}

\usepackage{upgreek}
\usepackage{textcomp}
\usepackage{xcolor}
\usepackage{datetime2}
\usepackage{feynmp-auto}  % for Feynman diagrams
\usepackage{tikz}  % for drawing
\usepackage[T1]{fontenc}
\usepackage{titlesec}
\usepackage{nicefrac}
\usepackage{color}
\usepackage{cancel}
\usepackage{makecell}
\newcommand{\cbox}[2]{\begin{minipage}[c]{#1}\centering #2\end{minipage}}
\newcommand{\lbox}[2]{\begin{minipage}[t]{#1}#2\end{minipage}}
\newcolumntype{Y}{>{\centering\arraybackslash}X}
\usepackage[colorlinks=true,linkcolor=black,citecolor=blue,urlcolor=blue,bookmarksopen]{hyperref}
\usepackage{orcidlink}

\usepackage{lineno}
\usepackage{siunitx}
\renewcommand{\theparagraph}{\alph{paragraph}.}

\titleformat{\paragraph}[runin]{\normalfont\normalsize\itshape}{\theparagraph}{0.4em}{}[ 
{\mbox{}\rule[0.5ex]{1.2em}{0.8pt}}]

\DeclareMathAlphabet{\mathpzc}{OT1}{pzc}{m}{it}

\newcommand{\DB}[1]{\textcolor{blue}{DB: #1}}
\newcommand{\YG}[1]{\textcolor{red}{YG: #1}}

\usepackage{comment}
\usepackage{graphicx} % Required for inserting images

\AtBeginDocument{%
  \setcitestyle{sort=false}%
}
\begin{document}

%\linenumbers  % Turn on line numbering

\title{Cavity, lumped-circuit, and spin-based detection of axion dark matter: differences and similarities}

\author{Deniz Aybas}
\affiliation{Department of Physics, Bilkent University, Ankara 06800, Turkey}

\author{Hendrik Bekker}
\affiliation{Johannes Gutenberg-Universit{\"a}t Mainz, 55122 Mainz, Germany}
\affiliation{Helmholtz Institute Mainz, 55099 Mainz, Germany}
\affiliation{GSI Helmholtzzentrum für Schwerionenforschung GmbH, 64291 Darmstadt, Germany}

\author{Dmitry Budker}
\affiliation{Johannes Gutenberg-Universit{\"a}t Mainz, 55122 Mainz, Germany}
\affiliation{Helmholtz Institute Mainz, 55099 Mainz, Germany}
\affiliation{GSI Helmholtzzentrum für Schwerionenforschung GmbH, 64291 Darmstadt, Germany}
\affiliation{Department of Physics, University of California, Berkeley, CA 94720-7300, United States of America}

\author{Wei Ji}
\affiliation{Johannes Gutenberg-Universit{\"a}t Mainz, 55122 Mainz, Germany}
\affiliation{School of Physics and State Key Laboratory of Nuclear Physics and Technology, Peking University, Beijing, 100871, China}

\author{On Kim}
\affiliation{Department of Physics and Astronomy, University of Mississippi, MS 38677, USA}
\thanks{Current address: University of Washington, Seattle, Washington 98195, USA}

\author{Younggeun Kim}
\email{kimyoung@uni-mainz.de}
\affiliation{Johannes Gutenberg-Universit{\"a}t Mainz, 55122 Mainz, Germany}
\affiliation{Helmholtz Institute Mainz, 55099 Mainz, Germany}
\affiliation{GSI Helmholtzzentrum für Schwerionenforschung GmbH, 64291 Darmstadt, Germany}

\author{Derek F. Jackson Kimball}
\affiliation{Department of Physics, California State University - East Bay, Hayward, California 94542-3084, USA}

\author{Jia Liu}
\affiliation{School of Physics and State Key Laboratory of Nuclear Physics and Technology, Peking University, Beijing 100871, China}
\affiliation{Center for High Energy Physics, Peking University, Beijing 100871, China}

\author{Xiaolin Ma}
\affiliation{School of Physics and State Key Laboratory of Nuclear Physics and Technology, Peking University, Beijing 100871, China}
\affiliation{International Center for Quantum-field Measurement Systems for Studies of the Universe and Particles (QUP, WPI),
High Energy Accelerator Research Organization (KEK), Oho 1-1, Tsukuba, Ibaraki 305-0801, Japan}

\author{Chiara P. Salemi}
\affiliation{Department of Physics, University of California, Berkeley, CA 94720-7300, United States of America}
\affiliation{Physics Division, Lawrence Berkeley National Laboratory, Berkeley, CA 94720-7300, United States of America}

\author{Yannis K. Semertzidis}
\affiliation{Department of Physics, Korea Advanced Institute of Science and Technology, Daejeon 34141, Republic of Korea}

\author{Alexander O. Sushkov}
	\affiliation{Department of Physics and Astronomy, Johns Hopkins University, Baltimore, Maryland 21218, USA}
	\affiliation{Department of Physics, Boston University, Boston, MA 02215, USA}

\author{Kai Wei}
\affiliation{School of Instrumentation Science and Opto-electronics Engineering, Beihang University, Beijing, 100191, China}
\affiliation{Hefei National Laboratory, Hefei, 230088, China}

\author{Arne Wickenbrock}
\affiliation{Johannes Gutenberg-Universit{\"a}t Mainz, 55122 Mainz, Germany}
\affiliation{Helmholtz Institute Mainz, 55099 Mainz, Germany}
\affiliation{GSI Helmholtzzentrum für Schwerionenforschung GmbH, 64291 Darmstadt, Germany}

\author{Yuzhe Zhang}
\affiliation{Johannes Gutenberg-Universit{\"a}t Mainz, 55122 Mainz, Germany}
\affiliation{Helmholtz Institute Mainz, 55099 Mainz, Germany}
\affiliation{GSI Helmholtzzentrum für Schwerionenforschung GmbH, 64291 Darmstadt, Germany}

\begin{abstract}
Axions and axion-like particles are compelling candidates for ultralight bosonic dark matter, forming coherent oscillating fields that can be probed by experiments known as haloscopes. A broad range of haloscope concepts has been developed, including resonant cavity haloscopes, lumped-element circuit detectors, and spin-based experiments, each sensitive to different axion couplings and mass ranges. Rather than attempting an exhaustive survey of all existing approaches, this comparative review provides a unified framework for the major haloscope classes, establishing a common language for the descriptions of signal generation, noise properties, data analysis, and scanning strategies.

Key properties of ultralight bosonic dark matter relevant for detection are summarized first, including coherence time, spectral linewidth, and stochasticity under the standard halo model. The discussion then compares cavity, Earth-scale, lumped-element, and spin haloscopes, focusing on expected signal shapes, dominant noise sources, and statistical frameworks for axion searches. Particular emphasis is placed on consistent definitions of signal-to-noise ratio and on how detector bandwidth, axion coherence, and noise characteristics determine optimal scan strategies.

By systematically comparing operating principles and performance metrics across these detector families, this framework clarifies shared concepts as well as the essential differences that govern sensitivity in different mass and coupling regimes. The resulting perspective synthesizes current search methodologies and offers guidance for optimizing future haloscope experiments.
\end{abstract}
%\date{November 2023---???}

\maketitle
\tableofcontents

\section{Introduction}\label{sec:Introduction}
Axions and axion-like particles (ALPs) are hypothetical spin-zero pseudoscalars that naturally arise in beyond-standard-model theories.
%Initially proposed as a solution to the strong CP problem, the axion preserves both charge and parity symmetries in the strong interaction on average through oscillating the nucleon electric dipole moment with the Peccei-Quinn mechanism. 
They are currently viewed as leading candidates for ultralight bosonic dark matter (UBDM), with particle masses  $\ll 10 \text{\,eV}$. UBDM is modeled as coherent oscillations of a pseudoscalar field at the Compton frequency\,\cite{Kimball2022Bibber}. The axion was first hypothesized as a solution to the Strong-CP problem\,\cite{Peccei_PhysRevLett.38.1440,PecceiQuinn:1977prd} and soon after was suggested as the constituent of the ``dark halo'' in a galaxy\,\cite{Sikivie:1983prl}. ALPs have similar properties but do not solve the strong-CP problem. In this article, we use the term ``axion'' to generically refer to both axions and ALPs.

Axions can have three non-gravitational interactions with the standard-model (SM) particles\,\cite{Graham:2013gfa,Sikivie_RevModPhys.93.015004} described by the Lagrangian terms: 
\begin{equation}
L \supset \frac{a}{f_a}F_{\mu\nu}\tilde{F}^{\mu\nu}, \,\,\,
L \supset  \frac{a}{f_a}G_{\mu\nu}\tilde{G}^{\mu\nu}, \,\,\,
L \supset  \frac{\partial_{\mu}a}{f_a}\bar{\Psi}_{f}\gamma^{\mu}\gamma_{5}\Psi_{f}\,,
\label{eq:Lint_axion}
\end{equation}
where the axion couples to the electromagnetic field described by the electromagnetic field strength tensor $F_{\mu\nu}$, to the gluon field $G_{\mu\nu}$, and to a SM fermion $\Psi_{f}$, respectively. These Lagrangians correspond to the Hamiltonians:
\begin{equation}
H_{\gamma} \supset g_{a\gamma\gamma}\,a\,\mathbf{E} \cdot \mathbf{B} , \,\,\,
H_{d} \supset g_{d}\,a\,\mathbf{\bm \sigma} \cdot \mathbf{E} , \,\,\,
H_{N} \supset g_\mathrm{aNN} \mathbf{\bm \nabla} a\cdot \mathbf{\bm \sigma}_N, \,\,\,
H_{e} \supset g_{aee} \mathbf{\bm \nabla} a\cdot \mathbf{\bm \sigma}_e\,,
\end{equation}
where $g_{a\gamma\gamma}$, $g_{d}$, $g_\mathrm{aNN}$, and $g_{aee}$ are effective couplings to the electric field $\mathbf{E}$ and magnetic field $\mathbf{B}$, the electric dipole moment (EDM, colinear with the spin of the particle $\mathbf{\bm \sigma}$), nuclear spin $\bm \sigma_{N}$, and electron spin $\mathbf{\bm \sigma}_{e}$, respectively.
%where axion effectively couples to a photon ($\vec{E},\ \vec{B}$) with coupling $g_{a\gamma\gamma}$, to an electric dipole moment (EDM), to a nuclear spin, and to an electron spin, respectively.
The predictions for the relative values of the coupling constants for the various interactions are model dependent; see, for example, a discussion by \citet{Kimball2022Bibber}.
Experiments searching for signatures of axions are designed to be sensitive to at least one of these interactions.
QCD axions that solve the strong-CP problem have prescribed relationships between their mass and particle couplings (see e.g. the KSVZ \cite{Kim:1979prl,Shifman:1980npb} and DFSZ \cite{Dine:1981plb} models), whereas coupling is not tied to mass for ALPs.

When designing a haloscope experiment, the overarching goal is detection of dark matter; however, 
one may choose different strategies depending on the circumstances. 
One reasonable objective could be to achieve a ``needle sensitivity'' by maximizing the sensitivity at a fixed axion mass (i.e., a narrow mass range). 
Although the probability of detecting an axion at one random frequency is small, a needle-sensitivity experiment is valuable for exploring the practical limits of sensitivity and for investigating systematic effects. 
A narrow-range search may also be motivated if a plausible candidate is found, necessitating targeted measurements to confirm or reject it at the observed mass. 
However, the most common scenario is a broad-range search aiming at a certain level of sensitivity. Although subjective, the ``value'' of such a search is reasonably considered to be proportional to the area of previously unexplored mass-coupling parameter space in log-log coordinates.

% \DB{We need to mention electron couplings somewhere but the current proposal is to focus on nuclear couplings not to blow up the scope infinitely. Younggeun will add something to the opening plot.} \DA{I mentioned the electron coupling above. I used $\sigma$ for spins, not $I$ as written in eqn. \ref{eq:NMRHaloscopeHamiltonian}, to make it more "universal" between nuclear and electron spins. Also, we should decide if we're going to use arrows or bold operators to signify vectors.}\\

\begin{figure}[h]
\includegraphics[width=1\linewidth]{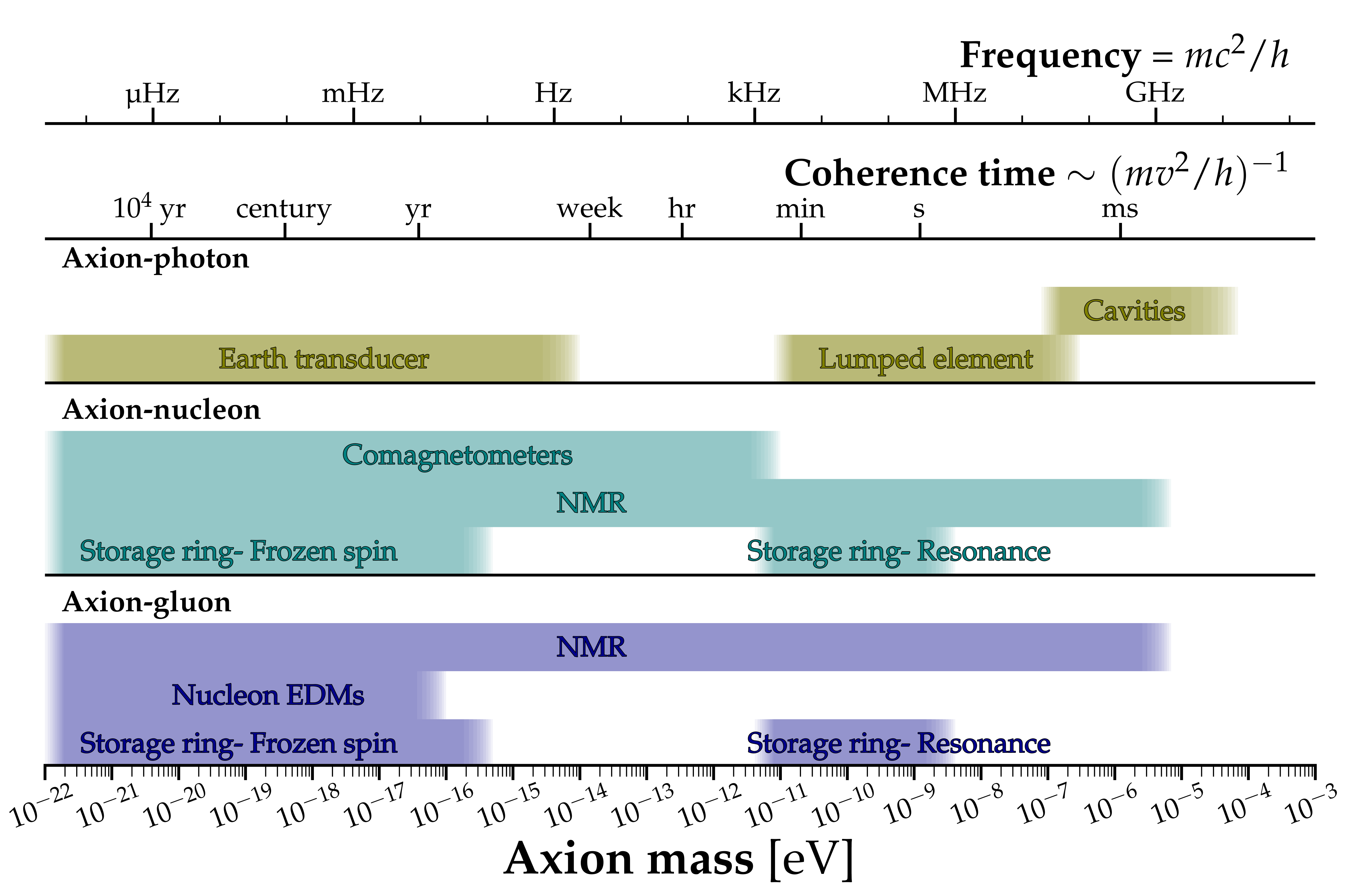}
\caption{We have classified various experimental methods for detecting dark matter axions based on their interactions with SM particles. While there are numerous experimental approaches, this manuscript focuses on several specific experiments of particular interest. As shown in the diagram, the preferred mass range varies depending on the type of axion interaction. We have compared the search strategies employed in each experimental method and discussed how, even for the same type of interaction, the tuning-strategy characteristics vary depending on the axion coherence time and the duration of the experiment. The upper bound which is of order 1\,meV is constrained by astronomical observations, and the lower bound, of order $10^{-22}\,\rm{eV}$, corresponding to the fuzzy dark matter limit~\cite{Chadha-Day:2021szb}.}
\label{fig:intro_compare}
\end{figure}

The first haloscope was designed to search for the axion-photon coupling\,\cite{Sikivie:1983prl}.
However, over time, the haloscope became an umbrella term for all axion dark matter searches that assume the standard halo model, where the axion particle density radially drops further away from the galactic center. Contemporary haloscope designs are highly sensitive to one or more of the possible couplings of the axion in different frequency ranges\,\cite{Aybas2021_SolidStateNMR_PhysRevLett,beadle2025_DM_NMR}, ideally limited by fundamental noise sources and detector bandwidth, as shown in Fig.\,\ref{fig:intro_compare}. In addition to the axion, haloscopes may also be sensitive to vector dark matter (with spin-1), such as the hidden or dark photon\,\cite{Ackerman:2009, Fabbrichesi:2021, Baryakhtar:2018}. Axions and dark photons require different data-analysis techniques as they have somewhat different characteristics and properties (for example, dark photons have a polarization that must be accounted for). 
%and in this work, we focus on searches for the axion. 

Haloscopes are not the only detectors sensitive to axions. There are also helioscopes, such as the CERN Axion Solar Telescope\,\cite{CAST:2017}. Helioscopes search for the photon couplings of solar axions produced in the Sun by converting them into detectable X-rays. Both haloscopes and helioscopes are Earth-bound searches for axion particles that reach the detectors. There are also various ``light shining through wall'' experiments reviewed, for example, in\,\cite{Semertzidis:2021rxs} that produce the axions they aim to detect, as well as various experiments searching for exotic forces produced by virtual axions as reviewed in\,\cite{Lei_RevModPhys2025}, for example, the ARIADNE experiment\,\cite{arvanitaki2014resonantly}. However, these experiments do not search for galactic axions, and so, by definition, are not haloscopes. An exception is the conversion of the CAST experiment\,\cite{CAST:2017} into the CAST-CAPP haloscope\,\cite{Adair:2022rtw} following its ``retirement'' as a helioscope.

In this work, our aim is to establish a common language between different types of haloscopes that search for the UBDM. We begin by summarizing the properties of the ultralight bosonic dark matter (UBDM) (Sec.\,\ref{sec: summary UBDM}). 
Then, we review various popular experimental approaches: cavity haloscopes, Earth transducer haloscopes, lumped circuit haloscopes, and spin haloscopes, including those based on storage rings. The approximate sensitive ranges are depicted in Fig.\,\ref{fig:intro_compare}. The mass range on the figure spans from the fuzzy-dark-matter regime ($10^{-22}\,\mathrm{eV}$) to the astrophysical limit ($10^{-3}\,\mathrm{eV}$).
In addition, there are also plasma haloscopes\,\cite{Lawson:2019plasma, Millar:2023plasma}, horn antenna haloscopes\,\cite{Jeong:2023horn}, and dielectric haloscopes\,\cite{Li:2020madmax, Majorovits:2020madmax}, which are not analyzed in this work. 
%\DA{Why not?}
The commonalities and differences between these haloscopes, such as the axion couplings and mass ranges they are sensitive to, are described (in Secs.\,\ref{sec:CavityHaloscopes}-\ref{sec:StorageRing}). These haloscopes rely on different search mechanisms for expected signals with different lineshapes, and have different main noise sources, and therefore, require specially optimized scanning strategies. After reviewing each type of haloscopes, we compare their key parameters and different scanning strategies, and establish  a common language describing these searches. Finally, we provide a general outlook on UBDM searches.\\ 

Specifically, this work addresses the following key questions:
\begin{itemize}
    \item What are the working and scanning mechanisms of haloscopes?
    \item What determines the axion mass range a haloscope is sensitive to? How much of this range can be covered in the resolution bandwidth (RBW) of a single scan and with what integration time?
    \item What are the definitions and values of scan rates in different haloscopes?
    \item How to find an optimized scan strategy? What is the figure of merit (FOM) for a DM search using a haloscope?
    \item How does the noise source influence the scan strategy? It can be inadvertently amplified in the detector, it can be suppressed (or not), it can be due to electromagnetic interference, it can be electronic Johnson-Nyquist noise, and its spectral profile, can range from white to monochromatic. 
\end{itemize}

Throughout this work, the signal-to-noise ratio is consistently defined as\,,
\begin{equation}
\mathrm{SNR} = \frac{P_\mathrm{total}-P_\mathrm{noise}}{\delta P_\mathrm{noise}} = \frac{P_\mathrm{signal}}{\delta P_\mathrm{noise}}\,,
\end{equation}
where $P_\mathrm{total}$ is the total power measured with a detector (for example, with a photo-diode in optical nuclear magnetic resonance (NMR) measurement or a microwave amplifier and a digitiser in a cavity haloscope), $P_\mathrm{signal}$ is the signal power, and $P_\mathrm{noise}$ is the average of noise power. $\delta P_\mathrm{noise}$ is the standard deviation of the noise power; it depends on the integration time and sets the sensitivity.\\

%\YG{$\bullet$ Let's unify the notation after first reading such as noise power, signal power, total power, fluctuation, etc...} %\DA{We should check if SNR defined in cavity (Eqn.   \ref{eq:cavity_SNR_frequent}) is the same for spin haloscopes)}.
%\YG{Eq 3 and 17 is the same.} 
%\DB{The new parts (like the Earh haloscope) should please adhere to the general notations in the m/s!}

% \Large
% Dear Collegues,
% \medskip

% This is the empty project I am sharing as discussed.

% \medskip
% Younggeun: Could you kindly set up the LaTex format and style the way you like?

% \medskip
% All: After Younggeun sets this up, we should fill out our names, affiliations, and funding acknowledgments

% \DK{We can add ``Earth transducer'' to axion-photon detection covering from about mHz to kHz. See\,\cite{fedderke2021earth,fedderke2021search,arza2022earth,sulai2023hunt,bloch2024curl,friel2024search}.} 
% \YG{Opnion of On or Yannis where the storage ring needs to be placed in this plot}
% \YG{Check the minimum mass of axion in the text. @ Derek, the upper limit of Earth transducer seems like $10^{-14}$eV in the text. I think we need to sync the numbers in the text and the plot.} 
\section{Summary of relevant UBDM properties}\label{sec: summary UBDM}

% {Simplified description of the axion field} 
The local average DM density in our part of the galaxy is on the order of \SI{0.45}{\GeV/\cm^{3}}\,\cite{localDM_Gaia,10.1093/mnras/stae034,Lim_2025}; if axions are its primary constituent, their number density is very high. For example, for an axion mass of $\SI{1}{\micro\eV}$ (de Broglie wavelength of $\lambda_{\rm{DB}} \approx 1.2\,\rm{km}$ for $v/c=10^{-3}$), the density is $\sim 4\times 10^{14}{\rm{cm}^{-3}}$. In this case, it is often convenient to think of DM as a field that can be written as
\begin{equation}
    a(r,t) = a_0 \cos(2\pi\nu_a t - \mathbf{k}\cdot \mathbf{r} + \phi)
    \,.\label{eq:axion_field}
\end{equation}
Here $a_0$ is the amplitude of the field, $\nu_a = m_a c^2/h$ is the axion Compton frequency, where $c$ is the speed of light and $h$ is the Planck constant, $\mathbf{k} = m_a \mathbf{v}_a/\hbar$ is the wave vector ($\mathbf{v}_a$ is the relative velocity of axion and the detector), $\mathbf{r}$ is the position vector, and $\phi$ is the phase in the interval $[0,\, 2\pi)$. 

\begin{figure}[h]
\includegraphics[width=.9\linewidth]{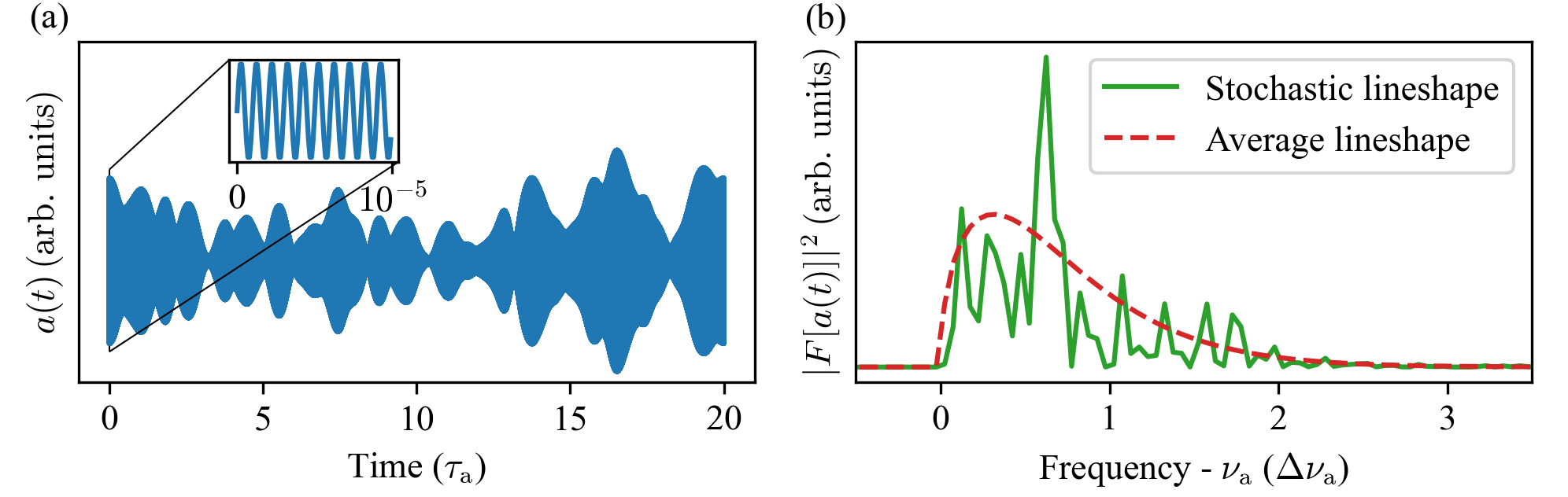}
\caption{Simulated axion fields in the time and frequency domains. (a) Time-domain axion field $a(t)$, with a magnified view showing the oscillation at axion Compton frequency in the inset. The envelope of the $a(t)$ changes over time, indicating the coherence time $\tau_a$ is finite. Here $\tau_a$ is chosen as $Q_a / \nu_a $. (b) The stochastic lineshape in the frequency domain is computed as $|F[a(t)]|^2$, where $F[a(t)]$ denotes the Fourier transform of the time-domain signal shown in (a). The average lineshape represents the expected amplitude spectrum after sufficient averaging. The linewidth $\Delta \nu_a$ is approximated as $\nu_a / Q_a $.}
\label{fig:axion_time_frequecny_domain}
\end{figure}

% {Stochastic nature of axion }
The axion field is stochastic in nature\,\cite{Foster2018Revealing_PhysRevD,Centers2021Stochastic,Lisanti:2021vij}. 
Due to the second-order Doppler effect, the frequencies of axions moving at speed $v$, are larger than $\nu_a$ by $v_{a}^2/2c^2$, which is on the order of $10^{-6}$ given that $v_{a}$ is comparable to the galactic virial velocity. Note that this is true in the so-called standard halo model\,\cite{Turner-1990,Evans2019SHM++_PhysRevD} and can be different if one considers models with axions forming a Bose-Einstein condensate\,\cite{Sikivie:2009qn} or in the case of the Sun and Earth centered local axion halos\,\cite{Budker_2023_DM_Formalism}. Therefore, due to the speed spread, the axion field, i.e. the superposition of all the individual axions with different speeds, is not monochromatic and has a finite coherence time on the order of $10^6$ Compton periods. The finite coherence time can be written as $\tau_a \approx Q_a / \nu_a$, where $Q_a \equiv (c/v)^2$ is the axion quality factor, $v$ is a typical value of on the order of Galactic virial velocity $10^{-3}c$. This corresponds to
a spectral linewidth of $\Delta\nu_a \approx (v/c)^2\nu_a \sim \nu_a/Q_a$\,\cite{Gramolin2022_SpectralPhysRevD}. 
Moreover, the summation of the signals of different phases and different frequencies indicates that the amplitude of the signal is not deterministic, but follows a probability distribution. %\sout{Considering the interference of axions of different frequencies, the power spectral density (PSD) of the axion field can be modeled by a two-dimensional random walk, and the PSD is found to follow the exponential probability distribution\,\cite{Centers2021Stochastic,Gramolin2022_SpectralPhysRevD} hence the amplitude and wave vector $\mathbf{k}$ acquire a slow time dependence.}
The summation of different axions in a certain frequency class can be modeled by a two-dimensional random walk in the phasor space with a fixed step size (axion amplitude) but a random direction (phase). The resulting PSD of the axion signal follows the exponential probability distribution\,\cite{Centers2021Stochastic,Gramolin2022_SpectralPhysRevD}.
An example of such stochastic lineshape is shown in Fig.\,\ref{fig:axion_time_frequecny_domain}.

\section{Cavity haloscopes}\label{sec:CavityHaloscopes}
%\YG{$\bullet$ \bf{This part answers the following: Can we model the scanning speed with given experimental parameters (such as Q, C, noise temperature)? From the modeled scanning speed equation, are there optimizable parameters to maximize the speed? The comment is retained for the convenience of the co-authors (as others below) but will be deleted before submission.} }

Resonant cavity haloscope experiments involve scanning the resonant frequency of a cavity to search for a match between the resonance of the cavity and the axion oscillation frequency. Since the concept was first proposed by Pierre Sikivie\,\cite{Sikivie:1983prl}, the initial experiments were conducted by scientists from Rochester University, Brookhaven National Laboratory (BNL), and Fermilab (RBF) at BNL, and separately University of Florida\,\cite{RBF-1987,UF-1990}. Currently, various experiments targeting a wide range of frequencies are being conducted worldwide, including the Axion Dark Matter Experiment [ADMX, \citet{ADMX_2025_PRL,ADMX-2001,ADMX:2011hrx,ADMX:2018gho,ADMX:2020ote,ADMX-2021,ADMX-2021-1}], HAYSTAC\,\cite{HAYSTAC-2017,HAYSTAC-2021}, CAPP\,\cite{CAPP-2020-1,CAPP-2020-2,CAPP-2021,CAPP-2023,CAPP-4,CAPP-2024-0,CAPP-2024-1}, QUAX\,\cite{Quax_2019,Quax_2021,Quax_2022,Quax_2025}, and ORGAN\,\cite{ORGAN_2017,ORGAN_2022,ORGAN_2024}, among others. A detailed survey of various haloscopes and other DM search experiments can be found at the \href{https://cajohare.github.io/AxionLimits/}{AxionLimits} repository\,\cite{AxionLimits}.

We focus on cavity haloscope experiments with double quadrature measurements, where both the in-phase and out-of-phase components of the electromagnetic field in the cavity are measured for axion dark matter detection. In Sec.\,\ref{sec:CavityHaloscopes:ExpectedSignal}, we describe the expected signal when the axion matches a resonant cavity mode. Section\,\ref{sec:CavityHaloscopes:Noise} outlines the dominant noise sources in cavity haloscopes. Subsequently, in Sec.\,\ref{sec:Cavity_frequentist}, we explain a frequentist statistical method, to analyze data for axion signal detection.
Following this in Sec.\,\ref{sec:CavityHaloscopes:ScanningStrategy}, we discuss scanning strategies in cavity haloscopes with a focus on frequency resolution. Specifically, we address single-bin searches, where the frequency resolution is comparable to the axion bandwidth, and multi-bin search schemes, where the resolution is finer than both the axion and cavity bandwidths. For each case, we explore how experimental parameters can be optimized to achieve the best sensitivity.
Finally in Sec.\,\ref{sec:CavityHaloscopes:casestudy}, we extend the discussion to identify the optimal scanning strategy under various noise conditions, providing a comprehensive framework for improving cavity haloscope experiments.

In this paper, we focus on haloscopes where the axion field interacts with a static magnetic field. In this case, the frequency and the corresponding wavelengths of the produced photon are determined by the axion mass. This puts a practical lower limit on the range of frequencies accessible to cavity haloscopes (see Fig.\,\ref{fig:intro_compare})---the magnet size determines the largest wavelength that can fit into it. Recently, an approach was proposed\,\cite{Berlin:2020aa}, where the static magnetic field is replaced with a radiofrequency field in a superconducting cavity (such cavities have been developed in conjunction with particle accelerators). The axion interacts with the oscillating field and produces photons at a sum or difference frequency between the frequency of the excited mode and the Compton frequency of the axion. Resonance detection is accomplished by tuning an unoccupied mode of the cavity to the resultant frequency and measuring the excitation. We do not discuss this type of experiment here. 

\subsection{Expected signal shape from a cavity haloscope}\label{sec:CavityHaloscopes:ExpectedSignal}
%\YG{$\bullet$ Axion photon interaction}

%Axion and photon fields interact as described by the Lagrangian\,\cite{Krauss1985prl,PhysRevD.64.092003,Sikivie_RevModPhys.93.015004}:
%\DA{Equation below is a Hamiltonian, not Lagrangian.}
%\begin{equation}
%\mathcal{L}_{int.} =  g_{a\gamma\gamma} a %\mathbf{E} \cdot \mathbf{B}\,,
%\label{eq:Lint_axion}
%\end{equation}
%where $g_{a\gamma\gamma}$ is the axion-photon coupling constant, $\mathbf{E},\ \mathbf{B}$ are electric field and magnetic fields, respectively\,\cite{Sikivie:1983prl}.
The axion and photon fields interact according to the first interaction Lagrangian of Eq.\,\eqref{eq:Lint_axion}.
\begin{comment}
For QCD axion, $g_{a\gamma\gamma}$ is defined with model-dependent parameter $g_{\gamma}$, Peccei-Quinn symmetry breaking level $f_{a}$, and fine structure constant $\alpha$\,\cite{Marsh:2015xka}:
\begin{equation}
    g_{a\gamma\gamma} = \frac{\alpha}{2\pi}\frac{g_{\gamma}}{f_{a}}.
\end{equation}
\end{comment}
This suggests that the electric and magnetic fields in the interaction must be parallel to each other. This interaction modifies Maxwell's equations, giving rise to an effective current described as:
\begin{equation}
\mathbf{J}_{\mathrm{eff}} = -\sqrt{\frac{\epsilon_{0}}{\mu_{0}}}g_{a\gamma\gamma}\left(\partial_{t}a\mathbf{B} +{\bm \nabla} a\times\mathbf{E}\right)\,.
\label{eq:effective_current}
\end{equation}
This oscillating effective current induces oscillating electromagnetic fields in the presence of boundaries\,\cite{Ouellet_Axion_EM_prd_2019,2019PDU....2600362K}. The spatial gradient is related to the momentum operator and is proportional to the velocity of axion.

%\YG{$\bullet$ Cavity haloscope} 
Cavity haloscopes employed in searches for dark matter axions operate on the following principle.
An external magnetic field $\mathbf{B}_{0}$ inside the cavity is typically applied using a superconducting magnet, giving rise to effective current $\mathbf{J}_{\mathrm{eff}}$ parallel to the external magnetic field in the leading order of $g_{a\gamma\gamma}$. The leading contribution to $\mathbf{J}_{\rm{eff}}$ arises from the time derivative term, since  the spatial gradient is suppressed by $|v/c| \ll 1 $ and there is no external applied electric field.
Then, the electromagnetic field induced by $\mathbf{J}_{\mathrm{eff}}$ in the cavity can be resonantly enhanced when the resonance frequency matches the frequency of the axion field. The induced electromagnetic field oscillates with a frequency given by the axion energy, close to the axion Compton frequency because one detects nonrelativistic galactic axions. The electric field of the converted photons, parallel to the external magnetic field, is captured with an antenna inserted in the cavity. The conversion power depends on the spatial overlap between the electric field profile $\mathbf{E}$ and the external magnetic field $\mathbf{B}_{0}$.
\begin{comment}
An external magnetic field is typically applied using a superconducting magnet
\end{comment}
This is referred to as the form factor $C$ defined as:
\begin{equation}
    C = \frac{\left|\int dV \mathbf{B}_{0}\cdot \mathbf{E}\right|^{2} }{\int dV |\mathbf{B}_{0}|^{2}\int dV \epsilon |\mathbf{E}|^{2}}
    \,,\label{eq:C}
\end{equation}
where $\epsilon$ denotes the relative permittivity of the material inside the cavity, including the so-called tuning rods used to adjust the resonance frequency.

%\YG{$\bullet$ Conversion power inside cavity and signal lineshape}
Then, the signal power measured with the antenna in a bandwidth $\Delta\nu$ centered at frequency $\nu$ near the cavity resonance $\nu_c$ is\,\cite{Sikivie:1983prl,2019PDU....2600362K,2020JCAP...03..066K}:
\begin{equation}
    P_\mathrm{signal} = \frac{b}{1+b}\frac{1}{\mu_{0}}\frac{g_{a\gamma\gamma}^{2}\rho_{a}}{m_{a}^{2}}2\pi\nu_{c}B_{0}^{2}VC\frac{Q_{l}Q_{a}}{Q_{l} + Q_{a}}\int_{\nu-\Delta\nu/2}^{\nu+\Delta\nu/2}D_{a}(u)L(\nu/\nu_{c},Q_{l}) \frac{d\nu'}{\nu_{a}}\,,
\label{eq:psd cavity}
\end{equation}
where $\rho_{a}$, $m_{a}$, $V$, and $b$ represent the dark matter density, axion mass, volume of the cavity, and antenna coupling respectively. $Q_{l}= Q_{c}/(1+b)$ is loaded quality factor of the cavity and $Q_{c}$ is the quality factor of cavity without the antenna present. The antenna coupling $b$ is defined by the ratio $P_\mathrm{sig}/P_{\rm{con}}$, where $P_{\rm{con}}$ is the axion-to-photon conversion power in the cavity without antenna.
$D_{a}(u)$ is the distribution of the kinetic energy of the cosmic axion dark matter with Compton frequency $\nu_{a}$\,\cite{PhysRevD.42.3572}, the term $u\equiv\nu / \nu_a - 1$ represents the axion kinetic energy normalized by its rest mass. 
% \WJ{shall we change it and the following to ``$D_a(u)$ is the ... while in natural units, u=... represents ...'', because it looks like this is not a function, but $D_a$ muliply the things in the bracket.} \YG{Done}
 Assuming the standard halo model, axion dark matter is virialized resulting in a Maxwell-Boltzmann velocity distribution. This can be translated into a kinetic energy distribution observed on Earth, $D_a(u)$, is given by:
\begin{equation}
    D_{a}(u) = \frac{2}{v_\mathrm{rms}^{2}}\sqrt{\frac{3}{2\pi}}\frac{1}{r}\exp\left[-\frac{3}{2}\left(r^{2} + \frac{2u}{v_\mathrm{rms}^{2}} \right)\right]\sinh\left[3r\sqrt{\frac{2u}{v_\mathrm{rms}^{2}}}\right]
    \label{eq:axion distribution},
\end{equation}
where $v_\mathrm{rms}$ is the root-mean-square velocity (in units of the speed of light, $c$) of the axion halo in the Galactic rest frame and $r$ is the boost factor $r = v_\mathrm{obs}/v_\mathrm{rms}\approx 0.85$ due to the relative velocity of the observer.
The Lorentzian shape of the cavity response is\,\cite{alesini2011power_coupling}:
\begin{equation}
    L(\nu/\nu_{c},\,Q_{l}) = \frac{1}{1+4Q_{l}^{2}(\nu/\nu_{c}-1)^{2}}\,.
    \label{eq:LorentzianFunction}
\end{equation}
\begin{comment}
The contribution of the antenna can be understood from the energy conservation. The power loss can be expressed as: $P_\mathrm{total} = P_{an} + P_{c}$, where \( P_\mathrm{total} \) is the total power dissipated in the cavity, and \( P_{c} \) is the power from axion-to-photon conversion. The antenna coupling, \( \beta \), is defined by the ratio \( P_{an}/P_{c} \). Using the definition of the Quality factor $Q \equiv \omega\frac{U}{P},$
U is total stored energy of electromagnetic field induced by axion inside of the cavity. It is sum over electric field and magnetic field, but near resonance, $U = 2U_{EM} = 2\int \vec{E}\cdot\vec{D}dV$.
We can relate the loaded Q factor (\( Q_{load} \)), which is \( \omega U/P_\mathrm{total} \), to the unloaded Q factor of the cavity (\( Q_{c} \)) and the Q factor of antenna in relation to the cavity (\( Q_{an} \)):
\begin{equation}
    \frac{1}{Q_{L}} = \frac{1}{Q_{c}} + \frac{1}{Q_{an}}.
\end{equation}
\end{comment}
The frequency-domain integral over the resolution bandwidth (RBW) gives the expected power to be measured with the antenna in the cavity.
Note that the signal power in Eq.\,\eqref{eq:psd cavity} depends on the coupling strength $b$, and it is maximized when $b =1$. However, the optimal coupling is $b=2$\,\cite{ALKENANY201711} to maximize scanning speed with the optimal scanning strategy, considering the contribution from the noise. Furthermore, the optimal coupling exceeds 2 when vacuum squeezing techniques are employed, owing to the larger bandwidth of the cavity at higher $b$\,\cite{HAYSTAC-2021, Colorado_Squeezing_1, Colorado_Squeezing_2}.

\subsection{Noise of a cavity haloscope}\label{sec:CavityHaloscopes:Noise}
\begin{figure}[h]
\includegraphics[width=0.5\linewidth]{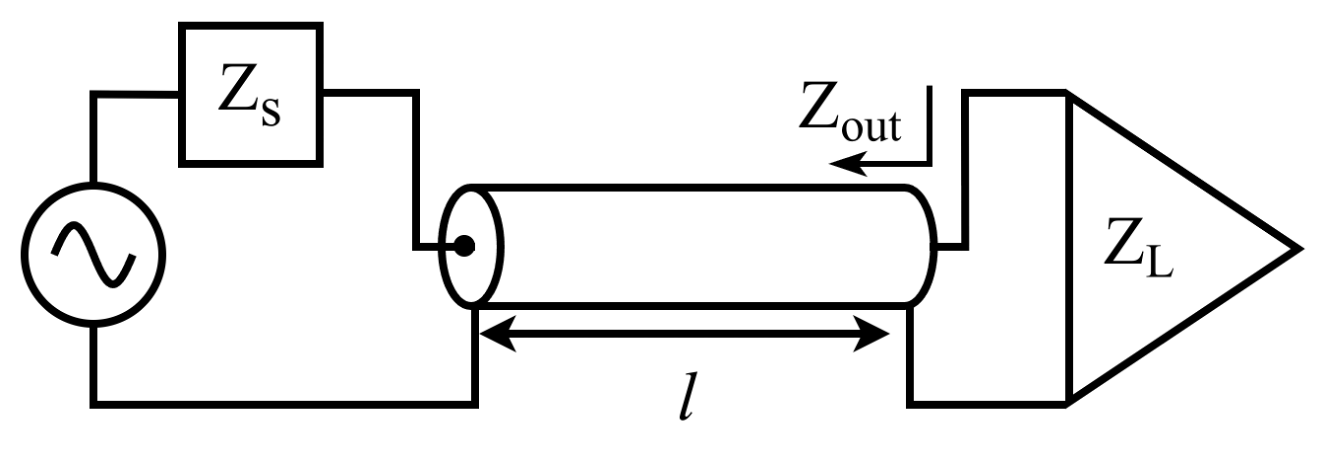}
\caption{A cavity haloscope RF chain features a source impedance $Z_{s}$ from the cavity, cavity thermal noise at temperature $T_{\mathrm{phy}}$ represented as AC voltage source $V_{n}$, a lossless transmission line of length $l$, connecting to a load with impedance $Z_{L}$, and the load-seen impedance $Z_\mathrm{out}(\nu,l)$. }
\label{fig:cavity_circuit}
\end{figure}

\begin{figure}[H]
    \centering
    \includegraphics[width=0.5\linewidth]{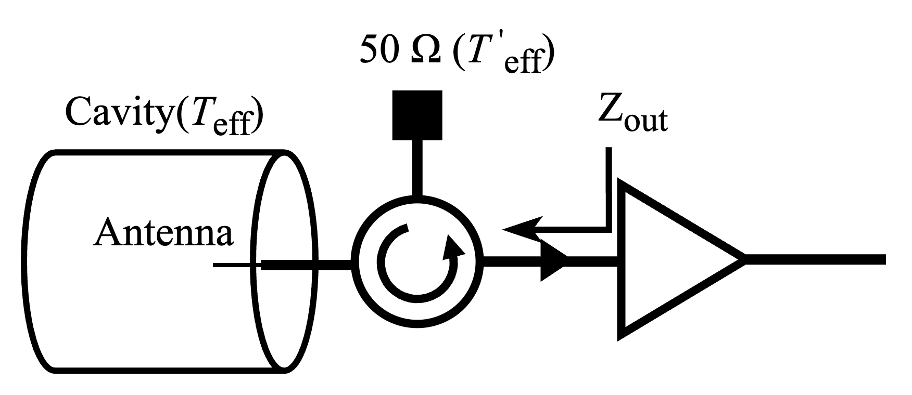}
    \caption{A circulator is placed between the cavity and the amplifier. A $50\,\Omega$ load is connected to one port of the circulator, and its thermal noise is reflected by the cavity and added to the thermal noise generated by the cavity. As a result, the impedance seen by the amplifier is $Z_{\rm{out}} =50\,\Omega$, and consequently the effective temperature seen by amplifier is  the cavity temperature ($T_{\rm{eff}}$) when cavity temperature ($T_{\rm{eff}}$) and the $50\,\Omega$ load temperature ($T’_{\rm{eff}}$) are equal.}
    \label{fig:cavity_circulator_diagram}
\end{figure}

The radio frequency (RF) chain of a cavity haloscope corresponds to the following circuit diagram, Fig.\,\ref{fig:cavity_circuit}. 
% \DA{This circuit diagram does not look correct. Characteristic impedance $Z_0$ here looks like a Voltage Divider, however $Z_0$ should be between the hot (positive) and cold (negative, ground) wires of the transmission line. A transmission line does not provide real resistance, it provides an "effective" impedance of 50 or 75 Ohms seen from both ends of the transmission line if nothing is connected on the other end. This Wikipedia figure is helpful: \url{https://en.m.wikipedia.org/wiki/File:Transmission_line_schematic.svg} }
% \YG{Yes you are right. The transmission line does not provide any loss. So, we set the propagation constant for the transmission line is set to purely imaginary. But, I omit this information in the text. Some description about it is in the appendix. In addition, for the circuit, yes you are right I draw like that just for simplicity :).}
% \DA{Perfect :)}
%\YG{$\bullet$ Lineshape of readout power impedance mismatched chain}
The cavity and antenna system can be regarded as a frequency dependent source impedance $Z_{s}(\nu)$. The transmission line connecting the cavity system to the RF amplifier chain has a length $l$, a characteristic impedance of $Z_{0}=50\,\Omega$ and propagation constant $\gamma$, and the linear amplifier is expressed as a load impedance $Z_{L}(\nu)$. The noise power $P_{\rm out}(\nu)$ seen by the amplifier can be written as a function of these impedances, the length of the cable, RBW $\Delta\nu$, and the effective temperature of the cavity $T_{\mathrm{eff}}$ as\,\cite{alesini2011power_coupling, Kim:2020_Revisiting_haloscope}:
\begin{equation}
    P_\mathrm{out}(\nu) = k_{B}T_{\mathrm{eff}}(T_{\mathrm{phy}},\nu)\Delta\nu\frac{4\mathrm{Re}[Z_{L}(\nu)]\mathrm{Re}[Z_{\rm out}(\nu,l)]}{|Z_{L}(\nu)+Z_{\rm out}(\nu,l)|^{2}}\,.
\end{equation}
The effective cavity temperature is expressed as\,\cite{PhysRevD.88.035020,Kim:2020_Revisiting_haloscope}:
\begin{equation}
T_{\mathrm{eff}}(T_{\mathrm{phy}},\,\nu) = \frac{h\nu}{k_{B}}\left(\frac{1}{e^{h\nu/k_{B}T_{\mathrm{phy}}} - 1 } + \frac{1}{2}\right),
\label{eq:cav_effective_temp}
\end{equation}
where $h$, $k_{B}$, and $T_{\mathrm{phy}}$ are Planck's constant, Boltzmann's constant, and the physical temperature of the cavity, respectively.
The $1/2$ term comes from vacuum fluctuations and is referred to as the quantum limit (QL). In the gigahertz range, if the cavity is at a sub-kelvin temperature, the effective temperature is dominated by the vacuum fluctuations and is at the QL. Under near-resonance conditions, the noise power, often referred to as the ``baseline" of the spectrum, represents the characteristic noise floor and is modeled with five parameters as described in \citet{PhysRevD.64.092003}, assuming the amplifier is designed to have a characteristic impedance of $Z_{0}$:
\begin{equation}
    P_{\rm out}(\nu) = k_{B}T_{\rm eff}(T_{\mathrm{phy}},\,\nu)\Delta \nu \frac{4b}{(1+b)^{2}}\left[\frac{a_{1} + a_{2}\delta + a_{3}\delta^{2}}{1+4\delta^2}\right]\,,
\label{eq:cav_baseline}
\end{equation}
where $\delta \equiv Q_{l}(\nu - \nu_{c})/\nu_{c}$, and $a_{1,2,3}$ are fitting parameters, which originate from the impedance of the circuit element and the noise figure of amplifier. 
Recent cavity haloscope experiments utilize a circulator between the RF components such as the cavity and amplifier to minimize impedance mismatch, as shown in Fig.\,\ref{fig:cavity_circulator_diagram}. A circulator is a non-reciprocal RF device that directs signals from one port to the next in a sequential manner, ensuring that power flows in a single direction through the connected components. One of the circulator ports is terminated with a 50$\,\Omega$, and the Johnson noise from this load is reflected by the cavity and propagates to the amplifier along with the Johnson noise transmitted through the cavity.

\begin{figure}[h!]
\includegraphics[width=0.8\linewidth]{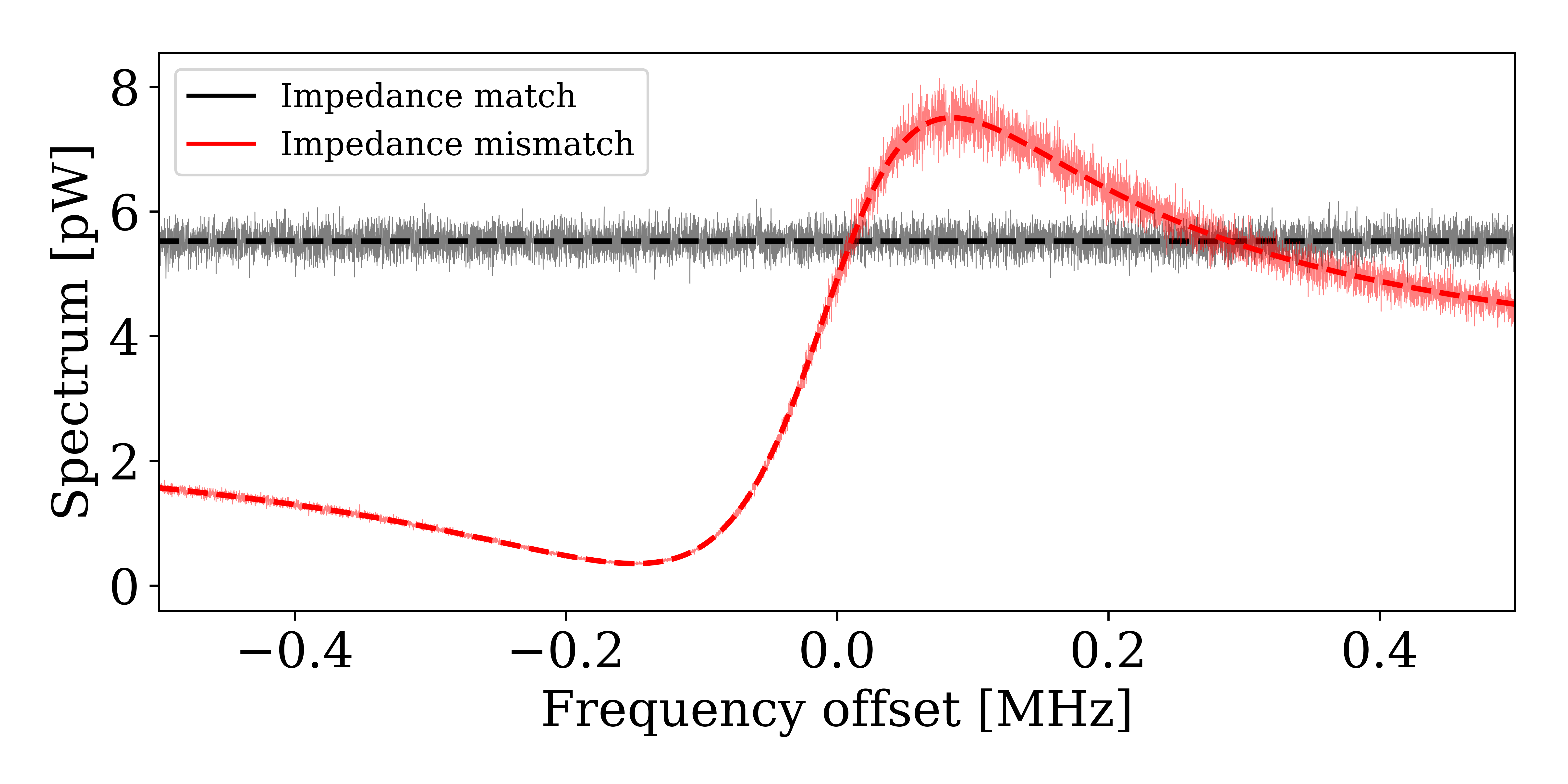}
\caption{ Example spectrum of the cavity haloscope RF chain with (in red) and without (in black) the impedance mismatch effect. The dashed line represents the baseline of each spectrum. To simulate the spectrum, the system noise temperature is set to 0.4\,K, the cavity Q factor to $4 \times 10^{4}$, the RBW to 100\,Hz, the integration time to 10 s, the antenna coupling to 2, and the total gain of the RF chain to 100 dB. For the impedance mismatch effect, $a_{2}/a_{1} = 0.3$ and $a_{3}/a_{1} = 0.7$ were used. }
\label{fig:noise_5_parameter}
\end{figure}

%\YG{$\bullet$ Lineshape of readout power impedance matched chain}
The ideal circulator makes the output impedance ($Z_{\rm{out}}$) to be $50\,\Omega$ for the working frequency range of the circulator. Near the resonance of the cavity, the noise power of circuit in Fig.\,\ref{fig:cavity_circulator_diagram} depends on the temperature of the cavity ($T_{\rm{eff}}$) and circulator ($T'_{\rm{eff}}$) and coupling of antenna ($b$) according to \citet{alesini2011power_coupling,J.Jeong:2022optimalhaloscope:design}:
\begin{equation}
    T_{\rm sys} = \frac{4b}{(1+b)^{2}}\frac{1}{1+4\delta^{2}}T_{\rm eff} + \frac{(1-b)^{2}+4(1+b)^{2}\delta^{2}}{(1+b)^{2}(1+4\delta^{2})}T_{\rm eff}' + T_{\rm add}\,,
\label{eq:system noise partition}
\end{equation}
where $T_{\rm add}$ is noise contribution from the amplifier. This implies that, if $T_{\rm eff}' = T_{\rm eff}$, the antenna coupling effect can be neglected. We obtain 
\begin{equation}
    T_{\rm sys} = T_{\rm eff} + T_{\rm add} \,.
\label{eq:system noise partition-simplified}
\end{equation}
Furthermore, if the noise contribution from the amplifier is the same across frequencies, the frequency dependence of the noise spectrum vanishes. Thus, the noise spectrum can be described as follows:
\begin{equation}
P_{\rm{noise}} = k_{B}T_{\mathrm{sys}}\Delta\nu\,,
\end{equation}
which is the Johnson-Nyquist noise\,\cite{Johnson:PR:1928, Nyquist:PR:1928}, and $T_{s}$ is called systematic noise temperature which is linear sum of effective temperature and added noise contributed by the RF chain, mainly from amplifier and series of circulators between the amplifier and the cavity. 
Figure\,\ref{fig:noise_5_parameter} shows an example of the spectrum with and without impedance mismatch effect. Note that the shape of the impedance mismatch spectrum resembles that of a Fano resonance, as the distortion arises from the interference between the forward and reflected signals at the cavity-amplifier interface (see \citet{Jiang:2023nan, Xu:2023vfn, Qin:2024wke} for spin haloscopes). 
 However, the spectrum is typically not flat, even when multiple circulators are inserted between the cavity and the amplifier, as circulators do not function perfectly in high magnetic fields.\footnote{This is because circulators contain materials that can get magnetized.} Consequently, the spectrum is modeled using a 5-parameter fit function, as shown in Eq.\,\eqref{eq:cav_baseline}.

Additionally, if the amplifier response exhibits nonlinear behavior or changes rapidly with frequency as in the case of Josephson Parametric Amplifier (JPA), the baseline cannot be described using only Eq.\,\eqref{eq:cav_baseline}, necessitating introducing additional parameters or using more sophisticated techniques like Savitzky-Golay (SG) filtering.
Note that fitting the spectrum in these ways can lead to a degradation of the axion signal, which should be taken into account in the analysis. This can be done, for example, with Monte Carlo simulations.

\subsection{How do we search for the axion in data?}\label{sec:Cavity_frequentist}
%\YG{$\bullet$Characteristics of signal and noise, distribution and statistics}
In contemporary cavity haloscopes, the power spectrum is acquired with RBW smaller than that of the standard-halo-model axion to obtain the axion spectral lineshape. The power spectrum is averaged $N$ times. A single spectrum of noise originating from Johnson-Nyquist noise from the thermal bath and the added noise from the RF-chain, follows a chi-square distribution with two degrees of freedom.\footnote{A power spectrum is a sum of squares of in-phase and out of phase components. For Gaussian noise, both components follow Gaussian distribution. The sum of squared $N$ Gaussian signals has a chi-square distribution with $N$ degrees of freedom. Therefore, the power spectrum of Gaussian noise follows chi-square distribution with two degrees of freedom. If one averages multiple ($N$) power spectra, the distribution of averaged spectrum follows a scaled-chi-square distribution with $2N$ degrees of freedom.} A large averaging number such as $N>100$ brings the spectrum to follow a  Gaussian distribution according to the central limit theorem if the noise is random\,\cite{book:stat_meta_analysi_Hartung}. Since the dark-matter axion has stochastic nature; the mean of in-phase and quadrature components of the axion signal are both zero. However, the axion signal produces an increase of the variance. Therefore, the axion signal is added to the background noise spectrum with its own spectral distribution. To quantitatively evaluate the detectability of the axion signal among the noise, SNR is defined as,
\begin{equation}
    \mathrm{SNR} = \frac{P_\mathrm{total}-\langle P_\mathrm{base}\rangle}{\delta P_\mathrm{base}}\,,
    \label{eq:cavity_SNR_frequent}
\end{equation}
where $P_\mathrm{total}, \langle P_\mathrm{base}\rangle$ and $\delta P_\mathrm{base}$ are the measured total power spectrum, the baseline of the spectrum that represents the noise power, and the deviation of the baseline noise power, respectively.

%\YG{$\bullet$Statistics test method Frequentest test and Bayesian method}
There are various methods for analyzing noisy data to search for axions, such as frequentist inference\,\cite{Brubaker:2017rna} and a Bayesian approach\,\cite{PhysRevD.101.123011}. In this section, we discuss the frequentist method, which involves structured hypothesis tests commonly used in axion cavity haloscope data analysis.

%\YG{$\bullet$ Hypothesis test}
Hypothesis testing is a method used to draw scientific conclusions from random events by evaluating two competing hypotheses: the null hypothesis ($H_{0}$) and the alternative hypothesis ($H_{1}$). 
In contrast to the terminology used outside the axion-search community, the null hypothesis here is that there is (!) an axion present in a given data set along with a noise, while the alternative hypothesis is that there is no such signal (noise only). 
If the null hypothesis is rejected with a certain level of confidence, we can set limits on the product of the local axion density and the coupling strength.

% Therefore, in hypothesis testing, the null hypothesis is set as a hypothesis that may not occur. Usually, it assumes that the data consists of noise only. By rejecting the null hypothesis, we can claim that the data contains non-noise phenomena.
%\YG{$\bullet$ Hypothesis test in axion search}
% In the search for dark matter axions, most of the events measured in experiments are expected to be noise. This is because the parameters of axions, such as mass, coupling constants, and the proportion of axions in dark matter, are not known. Therefore, the goal of hypothesis testing in this context is to evaluate whether the null hypothesis, which assumes the existence of axions with specific parameters, can be rejected with a certain level of confidence. This contrasts with conventional hypothesis testing, where the focus is typically on determining whether the data contains noise. 
A test statistic is utilized to perform the hypothesis test.
For a cavity haloscope, the test statistic used is the normalized power excess. It is typically calculated as the ratio of the normalized power to the deviation at each frequency. The normalized power spectrum is defined as the spectrum with the baseline removed, divided by the baseline itself. This normalized power excess is closely associated with SNR, see details in the
cavity haloscope references\,\cite{brubaker2018results,Brubaker_PhysRevD.96.123008,ADMX-2021, CAPP-2020-1,CAPP-2024-0}. 
% \YG{Most of axion haloscope search uses frequentiest test method. Do we need to cite all of them?}

The null hypothesis, which assumes the presence of an axion signal with a predetermined strength, usually assumes an excess of $5\sigma$, where $\sigma$ refers to the standard deviation under the alternative hypothesis. 
%Hence, $5\sigma$ signal refers to the axion signal having SNR of 5. 
Ideally, the distribution of the normalized power excess of null hypothesis follows a normal distribution centered around the targeted SNR (typically 5) when the number of spectra averages is large 
(in practice, this could be over 100), as predicted by the central limit theorem. The actual null distribution of the normalized power excess can be determined through Monte Carlo simulations by injecting an axion signal into the data. Let us assume that the standard deviation is normalized such that $\sigma = 1$.
% In this context, $\sigma = 1$ for the standard normal distribution, which is the distribution of the alternative hypothesis for the cavity haloscope.
\begin{figure}[h]
\includegraphics[width=0.8\linewidth]{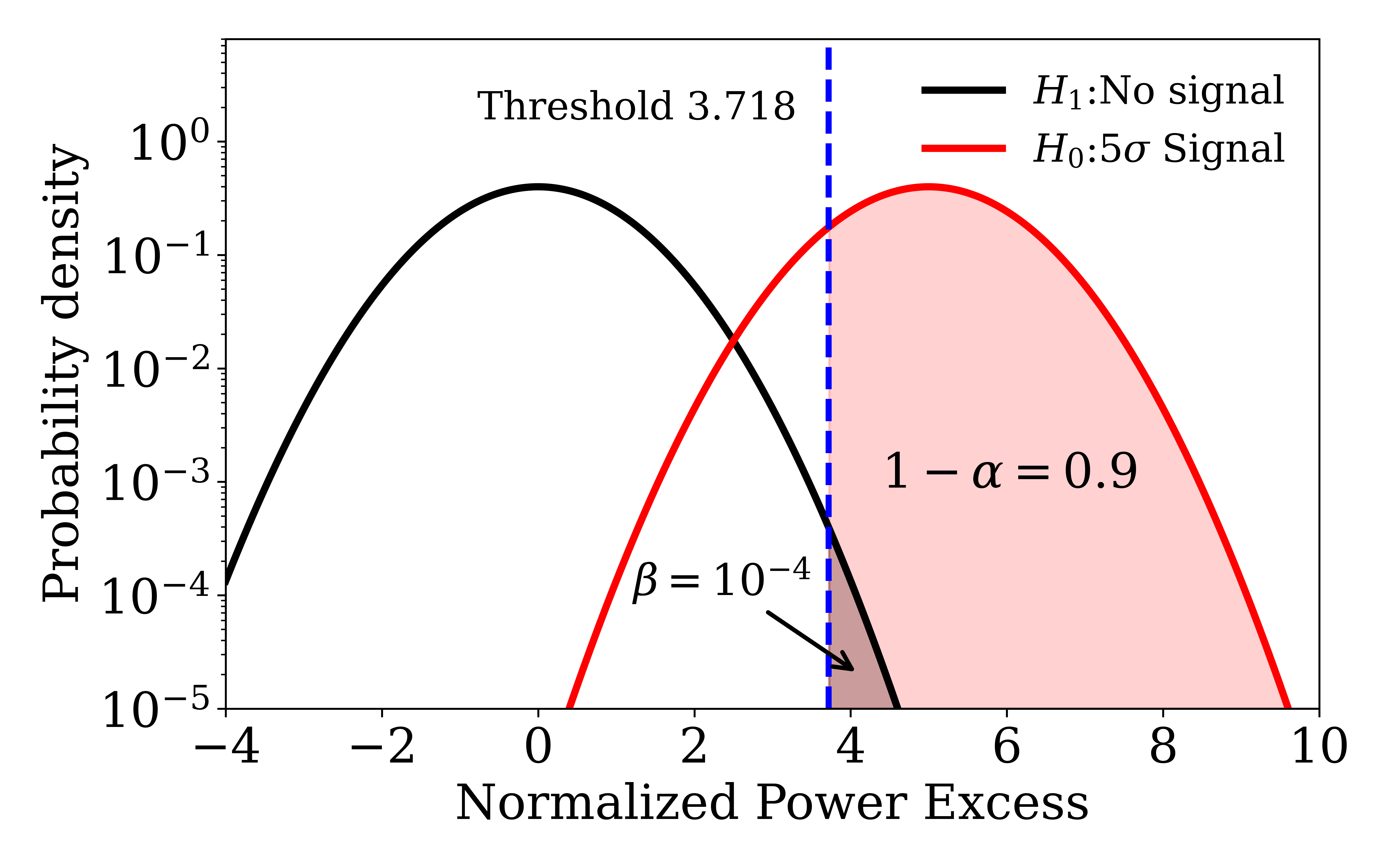}
\caption{Distribution of normalized power excess for the null hypothesis (red) and the alternative hypothesis (black). In cavity-haloscope experiments, one typically averages the power spectrum more than 100 times, and the normalized power excess is expected to follow a Gaussian distribution. The blue line is the threshold $\Lambda_{c}$ for a confidence level of 90\%; the corresponding Type II error is 0.01\% (see text). Figure from \citet{younggeunkim2022thesis}. }
\label{fig:cavity_frequntist}
\end{figure}
Figure\,\ref{fig:cavity_frequntist} shows the distribution of the normalized power excess corresponding to $H_{0}$ and $H_{1}$. There are two types of error: Type I error ($\alpha$) and Type II error ($\beta$)\footnote{The terminology ``Type I error'' (false positive) and ``Type II error'' (false negative) may be confusing in the case of cavity-haloscope experiments. This is because the definition of the null hypothesis in these experiments is the opposite of that in conventional hypothesis testing.
In conventional hypothesis testing, the null hypothesis is typically defined as ``the data contain only noise'' or ``the medicine is not effective.'' A false positive in this context means rejecting the null hypothesis when it is actually true, such as concluding that ``the data contain a signal when they actually contain only noise” or “the medicine is effective when it is actually not.''
However, in cavity haloscope experiments, the null hypothesis is set as ``axions exist in the data.'' Consequently, a false positive in this context means incorrectly concluding that there are no axions, even though they are actually present in the data.}. The Type I error is the probability of rejecting the null hypothesis when it is actually true, and it is defined as
\begin{equation}
    \alpha = \int_{-\infty}^{\Lambda_{c}}\mathcal{P}(x;H_{0})dx\,.
    \label{eq:typ1error}
\end{equation}
Typically, $\alpha$ is chosen to be 0.1 or 0.05, where $\Lambda_{c}$ is the threshold value. $\mathcal{P}(x;H)$ represents the probability distribution function of the variable $x$ under the hypothesis $H$. For example, in Fig.\,\ref{fig:cavity_frequntist},  $\mathcal{P}(x; H_{0})$ is a Gaussian distribution centered at 5 with a deviation of 1. The Type I error is also known as the significance level.
The Type II error is the probability of not rejecting the null hypothesis when it is actually false, and it is defined as
\begin{equation}
    \beta = \int_{\Lambda_{c}}^{\infty}\mathcal{P}(x;H_{1})dx
    \label{eq:typ2error}
\end{equation}
The threshold and target sensitivity in an experiment are determined by the required significance level and the Type II error rate. For a cavity haloscope that follows a normal distribution, if the target sensitivity is set to $5\sigma$, even with significance levels of 0.05 or 0.1, the $\beta$ value remains below $\sim 10^{-4}$. Therefore, the null hypothesis $H_{0}$ is set as a $5\sigma$ axion signal, and $\Lambda_{c}$ is determined according to $\alpha$.

\begin{comment}
The observed excess is then compared to the critical value: if it is less than the threshold, $H_{0}$ is rejected, indicating no significant evidence for the axion signal; if the observed excess exceeds the critical value, $H_{0}$ is favored and $H_{1}$ is disfavored, suggesting potential evidence for the axion signal. If $H_{0}$ is favored, the corresponding bin is tested again—this process is called a rescan—through additional experiments to determine whether the observed event is due to statistical fluctuations. After the rescan, the new data are analyzed together with the original data, to ensure that the initial findings are not ignored. If no excess exceeds the threshold in the combined dataset, we conclude the absence of the axion signal with the predetermined strength, with a confidence level of 90\% or 95\%, depending on the significance level.
\end{comment}
The observed excess signal is compared to the threshold $\Lambda_{c}$. If the excess signal is below the threshold, $H_0$ is rejected, indicating no significant evidence for the axion signal. Conversely, if the observed excess is larger than $\Lambda_{c}$, $H_0$ is favored, and $H_1$ is disfavored, suggesting potential evidence for the axion signal. However, when the threshold is set at the 90\% confidence level for $H_0$, see Fig.\,\ref{fig:cavity_frequntist}, exceeding the threshold corresponds to a
probability on the order of $10^{-4}$ for $H_1$, which means that even in the absence of an axion, some channels can exceed the threshold purely due to statistical fluctuations. In particular, when scanning over multiple independent frequency bins (e.g., more than $10^5$ bins), there is a significant probability of observing around ten or more statistical outliers. This is known as the ``look-elsewhere effect'', implying that an excess in a single frequency bin does not necessarily indicate a true physical signal.
To address this issue, a rescan procedure is implemented. Even if $H_0$ is initially favored, additional measurements are performed in the corresponding channel to determine whether the observed excess is due to statistical fluctuations or a genuine signal. In this process, new data are collected and analyzed together with the original data to test whether the initial excess was merely a statistical fluctuation. The probability of an excess appearing twice in the same channel is given by the product of two small probabilities and is generally negligible ($< 10^{-8}$ at 90\% confidence level). However, if the excess persists in the rescan, it becomes a strong candidate for an axion signal, necessitating further scrutiny to rule out systematic effects.
Through this analysis, the experiment can reliably assess whether the observed result is due to statistical fluctuations or represents a genuine physical signal.
% \DB{Proposal:}
% \DB{\begin{itemize}
%     \item Explain that in this nomenclature, the 90\%\,C.L. corresponds to a probability of $10^{-4}$ for a signal in a single channel to exceed the set threshold in the absence of an axion
%     \item If we have a scan over many (say, 10$^{4}$ independent channels, there is now a significant probability to see an outlier (or a few outliers) in the scan range; this is our expectation. This is called ``look-elsewhere effect''. 
%     \item To deal with the look-elsewhere effect, the standard procedure is to rescan over the outliers. The probability of an outlier occurring twice in the same channel is the product of two small numbers, and is usually negligible. If the outlier persists, this becomes a strong candidate for the axion and requires further scrutiny to check that this is not a systematic effect.
% \end{itemize}}

\subsection{Scanning Strategy of Cavity Haloscopes}\label{sec:CavityHaloscopes:ScanningStrategy}
%\YG{$\bullet$ Scanning strategy of cavity haloscope experiments depending on bin size}
 Since the mass of the axion is unknown, it is necessary to scan the resonance frequency. The scanning strategy (see Sec.\,\ref{sec:Introduction}) typically maximize the scanning speed (rate) depending on various factors, including cavity parameters (such as quality factor, form factor, and volume), antenna coupling, RBW, and system noise temperature.
 A figure of merit (FOM) that needs to be maximized to provide the maximum scanning speed for a given axion coupling\,\cite{Krauss1985prl,Ahn:2021JHEP} is
\begin{equation}
    \mathrm{FOM} = B_{0}^4V^{2}C^{2}Q_{c}\,, 
\end{equation}
and the cavity is designed to maximize this parameter. Note that FOM has a different dependence on cavity parameters compared to the signal power, Eq.\,\eqref{eq:psd cavity}, given by
\begin{equation}
P_\mathrm{\rm{signal}} \propto B_{0}^{2}VCQ_{l}Q_{a}/(Q_{l}+Q_{a})\,.
\end{equation}
This difference arises because the proper bandwidth of the measurement and the analysis are optimized to enhance the scanning speed rather than to maximize signal power. In the following section, the scanning-speed equation is derived, and the optimal experimental parameters for maximizing this scanning speed are described.

To derive the scanning-speed equation, SNR needs to be described with RBW $\Delta\nu$. The noise temperature of the detector chain is $T_{s}$ and the corresponding noise spectrum is assumed to be white. Then, the noise power within $\Delta\nu$ is $k_{B}T_{s}\Delta\nu$. On the other hand, the axion signal from the cavity has a frequency-dependent response due to the axion lineshape and the resonant cavity, Eq.\,\eqref{eq:psd cavity}. Therefore, the cavity response must be considered when evaluating the SNR during the calculation of the conversion power from the cavity. The expected power in the $\Delta\nu$ bandwidth that is sufficiently smaller than the axion bandwidth $\Delta\nu_{a}$, where axions are converted into photons inside the cavity and delivered to the antenna, can be derived from Eq.\,\eqref{eq:psd cavity}:
\begin{equation}
P_\mathrm{signal}(\nu_{a}) \approx \frac{b}{1+b}\frac{g_{a\gamma\gamma}^{2}\rho_{a}}{m_{a}^{2}}2\pi\nu_{c}B_{0}^{2}VC\frac{Q_{l}Q_{a}}{Q_{l} + Q_{a}}L(\frac{\nu_{a}}{\nu_{c}},Q_{l})D_{a}(\frac{\nu_{max}}{\nu_{a}} - 1)\Delta\nu/\nu_{a},
\label{eq:cavity_singlebin_response}
\end{equation}
where $\nu_{max}$ is the frequency satisfying $\partial_{\nu'}D_{a}(\nu'/\nu-1)=0$, and $\nu_{c}$ is the cavity resonance frequency. 
Equation\,\eqref{eq:cavity_singlebin_response} can be simplified using:
\begin{equation}
    \int_{\nu_{max} -\Delta\nu/2}^{{\nu_{max} +\Delta\nu/2}} L(\nu'/\nu_{c},Q_{l})D_{a}(\nu'/\nu_{a} -1)\frac{d\nu'}{\nu_{a}}\approx L(\nu_{a}/\nu_{c},Q_{l})D_{a}(\nu_{max}/\nu_{a} - 1)\Delta\nu/\nu_{a}.
\end{equation}
Here, $D_{a}(\nu_{max}/\nu_{a} - 1)$ is the height of the axion distribution, which depends on the motion of the Earth. The average value is $\langle D_{a}(\nu_{max}/\nu_{a} - 1)\rangle \approx 10^6$. (This can be understood from the fact that the relative width $\Delta \nu_a/\nu_a$ is assumed to be $\sim 10^{-6}$ and that the integral over the lineshape needs to give unity.)
\begin{comment}
\begin{figure}[h!]
\centering
\includegraphics[width=0.8\linewidth]{Figures/cavity_expected_power.png}
\caption{ The expected signal shape from the cavity ($Q_{c} = 5 \times 10^{4}$) at different RBW's is shown. For comparison, the expected cavity response at a low RBW and the axion bandwidth are also presented.}
\label{fig:caivty_response rbw}
\end{figure}
Figure\,\ref{fig:caivty_response rbw} shows the cavity response, which is the integral in Eq.\,\eqref{eq:cavity_singlebin_response}. When the axion search is conducted with $\mathrm{RBW} =\Delta\nu_{a}$, this is called single-bin search regime. On the other hand, when $\mathrm{RBW} \ll \Delta\nu_{a}$, this is called multi-bin search regime, since information about the dark matter axion is distributed across multiple frequency bins.
\end{comment}

For $Q_{c}<Q_{a}$ and $\Delta\nu\ll\Delta\nu_{a}$, the converted axion signal is spread into neighboring frequency bins. When searching for standard halo axions, the sensitivity can be increased by up to 6\% compared to the single-bin search mode, which uses an initial data acquisition frequency resolution of $\Delta\nu_{a}$, by taking advantage of the expected spectral lineshape of the signal, and this enhancement efficiency is denoted as $\eta_{h}$. This is done by evaluating the convolution of the measured spectrum with expected signal lineshape and evaluating the likelihood. This procedure is sometimes referred to as horizontal combination\,\cite{Brubaker_PhysRevD.96.123008}. Other terms used to describe this include optimal filtering\,\cite{Aybas2021_SolidStateNMR_PhysRevLett}. After this likelihood analysis, the acquired spectrum has bin-to-bin correlation along axion bandwidth $\Delta\nu_{a}$ and it effectively increases the RBW $\Delta\nu$ of the experiment to $\Delta\nu_{a}$. 

To derive the SNR at $\mathrm{RBW}= \Delta\nu_{a}$ using Dicke’s radiometer equation~\cite{Dicke:RSI:1946}, we need to account for the fact that the function $D_{a}(\nu/\nu_{a} - 1)$ does not have a rectangular shape. This is important because Dicke’s radiometer equation assumes a rectangular lineshape signal.
By numerically calculating the extent of this overestimation, we introduce the lineshape efficiency factor $\eta_1 \approx 0.81$ to more accurately represent SNR. 
%\YG{$\bullet$ Signal to noise ratio for single-bin search taken into account the lineshape efficiency $\eta_{1}$}
Then, the SNR can be expressed as:
\begin{equation}
    \mathrm{SNR} \approx \eta_{1}\eta_{h}\frac{g_{a\gamma\gamma}^{2}\rho_{a}}{m_{a}^{2}}\frac{b}{1+b}\omega_{c}\frac{B_{0}^{2}VC}{k_{B}T_{s}}\frac{Q_{l}Q_{a}}{Q_{l} + Q_{a}}L(\nu_{a}/\nu_{c},Q_{l})\sqrt{\frac{\Delta t}{\Delta\nu_{a}}}\,.
\label{eq:single bin snr}
\end{equation}

%\YG{$\bullet$ Strategy to have uniform sensitivity not only for the center but also for all frequency regions}
Now, SNR has a Lorentzian shape that is maximized at the cavity resonance frequency. We aim to achieve similar sensitivity not only at each resonance frequency of the resonator but also at frequencies around it. This goal can be achieved by selecting the optimal frequency tuning step $\delta\nu$, allowing off-resonance frequency bins to overlap appropriately, thereby increasing the statistics at each frequency to equalize sensitivity. Here, we consider the case of cavities with quality factors lower that that of the axion, though the same logic can be applied to those with a high $Q$ factor as well.

 %\YG{$\bullet$ Proof that uniform sensitivity can be achieved by combining spectra of different resonance frequencies}
 For a low-quality-factor cavity, the harmonic summation of two quality factors (cavity and axion) leads to overall low-quality-factor dominated by the cavity. 
If we tune the resonant frequency with tuning step $\delta \nu$, which is a fraction of the cavity bandwidth $\delta\nu = \Delta\nu_{c}/F$, we have for the scanning-rate:
\begin{equation}
    \frac{\delta\nu}{\Delta t} = \frac{\eta_{1}^{2}\eta_{h}^{2}}{\mathrm{SNR}^{2}}\left(\frac{g_{a\gamma\gamma}^{2}\rho_{a}}{m_{a}^{2}}\right)^2 \left(\nu_{c} \frac{B_{0}^{2}VC}{k_{B}T_{s}}\right)^2 \frac{b^{2}}{(1+b)^{3}}\frac{Q_{c}Q_{a}}{F}L(\nu_{a}/\nu_{c},Q_{l})^{2}\,.
    \label{eq:single_tune_scanning}
\end{equation}
Here, we can find that the antenna coupling maximizing the scanning rate is $b = 2$. Additionally, spectra from different tuning steps with overlapping frequencies can be added together for the overlapping frequency components, a process referred to as ``vertical combination''. This process can enhance the statistics for dark matter axions present in different tuning steps. Note that vertical combination does not change the measured lineshape of the axion. When the tuning range overlaps, vertical combination improves the statistics of frequency components not centered on the resonator, thereby flattening the overall sensitivity. In the scanning-rate equation, if $T_{s}$ is constant across different tuning steps, the Lorentzian function of the signal part can be expressed as a summation due to the vertical combination:
\begin{equation}
    L(\nu/\nu_{c},Q_{l})^{2}/F\rightarrow\frac{1}{F}\sum_{\nu = \nu_{c}-K\Delta\nu_{c}/F}^{\nu_{c}+K\Delta\nu_{c}/F}L(\nu/\nu_{c},Q_{l})^{2}\approx 0.8\,,
\end{equation}
where $K$ is an integer greater than $F$. For $K>2$, the summation converges to 0.8\,\cite{brubaker2018results}. This means that even if the cavity resonance frequency is tuned to less than one-third of the cavity bandwidth $\delta\nu = \Delta\nu_{c}/3$, the sensitivity is flat over the entire scanning range. This efficiency of 0.8 is labeled as $\eta^{2}_{v}$.

%\YG{$\bullet$ The final scanning rate equation or master equation takes into account the lineshape and combinational efficiency of the axion for single-bin search.}
Then, the total scanning rate of Eq.\,\eqref{eq:single_tune_scanning} can be evaluated as:
\begin{equation}
    \frac{d\nu}{dt}\equiv\frac{\delta\nu}{\Delta t} = \frac{\eta_{1}^{2}\eta^{2}_{v}\eta_{h}^{2}}{\mathrm{SNR}^{2}}\left(\frac{g_{a\gamma\gamma}^{2}\rho_{a}}{m_{a}}\right)^2 \left( \frac{B_{0}^{2}VC}{k_{B}T_{s}}\right)^2 \frac{b^{2}}{(1+b)^{3}}Q_{c}Q_{a}\,.
\label{eq: multi bin scanning rate}
\end{equation}
 As an example, in the CAPP-12T axion cavity-haloscope experiment\,\cite{CAPP-2024-0}, the scanning rate can be estimated to be 0.49\,MHz/day, based on key experimental parameters. The experiment operates under an external magnetic field of 9.828\,T and utilizes a cavity with a volume of 1.38\,L and a quality factor of \num{37000}. The system has a noise temperature of 380\,mK, corresponding to a noise power density of $5.30 \times 10^{-24}\,\mathrm{W/Hz}$ . The target SNR is set to 5, and the axion-photon coupling sensitivity is approximately 0.93 KSVZ. The form factor of the cavity is 0.68, and the antenna coupling is adjusted to $b = 2$ for optimal signal extraction\footnote{In practice, the effective scanning rate, 0.42\,MHz/day, was slightly lower than the estimated rate due to unavoidable dead time in data acquisition and slight signal degradation from baseline fitting. Although the measured spectrum has a peaked shape, the baseline noise is ideally flattened by fitting functions such as SG filters or a five-parameter model. However, these procedures inevitably distort the signal, leading to a degradation factor in sensitivity. For example, the Savitzky-Golay (SG) filter typically results in a degradation factor of 0.8–0.9,\cite{CAPP-2024-0,CAPP-2024-1}, while a 5-parameter fit function [Eq.\,\eqref{eq:cav_baseline}] shows slightly better performance, over 0.9,\cite{Bae:2024prl}. These effects are estimated and corrected via Monte Carlo simulations.}.

 In the case of $Q_{a} \ll Q_{c}$, the same approach can be applied. The scanning speed becomes the product of the quality factors of both the cavity and the axion, similar to the low quality factor cavity case, if the tuning step is a fraction of the axion bandwidth. 
 %However, there is lack of information of axion in the spectrum, therefore $\eta_{h}$ cannot be applied.
% \subsubsection{Recent study of increasing scanning speed using squeezing}

\subsection{Case study depending on physical temperature and amplifier noise}\label{sec:CavityHaloscopes:casestudy}
%\YG{$\bullet$ Generalization of scanning strategy study, dependence of ratio between resonator temperature and amplifier noise. What are the coupling strengths and tuning steps between the optimized resonator and readout antenna? In this case, I assume the baseline removing is not applied.}
The circulator introduces additional loss in the RF circuit while matching the impedance between the cavity and the amplifier. Additionally, the physical size of the circulator, see Sec.\,\ref{sec:CavityHaloscopes:Noise}, could become prohibitively large in the low frequency (megahertz) range, and it may be unusable in a cryogenic RF system\footnote{The physical size of the circulator is determined by the operating wavelength. Typically, the side length of the circulator is around 100\,mm in the megahertz range}. Therefore, the response of the noise power over the tuning range is not considered flat, as described in the previous section. The optimal tuning step and antenna coupling greatly depend on the shape of the noise power spectrum~\cite{Kim:2020_Revisiting_haloscope}. If the noise power spectrum is not flat, then the optimal tuning step changes depending on the physical temperature and the amplifier noise, since the contribution of physical temperature varies with frequency while the added noise from the linear amplifier is assumed to be independent of frequency. 

%\YG{$\bullet$ Power spectrum equation ignoring impedance mismatch}
For simplicity, we ignore the effect of impedance mismatch and consider the case of no circulator. Then, the noise power is expressed from Eq.\,\eqref{eq:cav_baseline}:
\begin{equation}
    P_{\rm{noise}}(\nu) =  k_{B}\Delta \nu_{c,a} 
    \left[\frac{4b}{(1+b)^{2}}\left(\frac{T_{\rm eff}}{1+4(\nu - \nu_{c})^{2}/\nu_{c}^{2}}\right) + T_{\rm add}\right]\,,
\end{equation}
where $\Delta\nu_{c,a}$ is the bandwidth of the cavity and the axion, respectively. Here, we only consider the limiting case of $Q_{c}\ll Q_{a}$.%\YZ{is it okay to write like this?}\YG{That’s a good point. The practical noise power spectrum doesn’t follow such a simple form due to impedance mismatch. However, without this assumption, it’s difficult to proceed with the discussion as outlined below.}.

%\YG{$\bullet$ Numerical calculation procedure to obatin optimal coupling and tuning step}
We can plug in the noise power and the axion signal power of Eq.\,\eqref{eq:psd cavity} to evaluate the SNR. In Sec.\,\ref{sec:CavityHaloscopes:ScanningStrategy}, we discussed how to use the SNR to evaluate the optimal scanning rate. Here, we consider the vertical combination and numerically evaluate the scanning rate with a varying scanning-step parameter $F$ defined as $\delta\nu = \Delta\nu_{c,a}/F$ and antenna coupling $b$ for different ratios of the physical temperature to the amplifier added noise.
\begin{figure}[h!]
\centering
\includegraphics[width=0.9\linewidth]{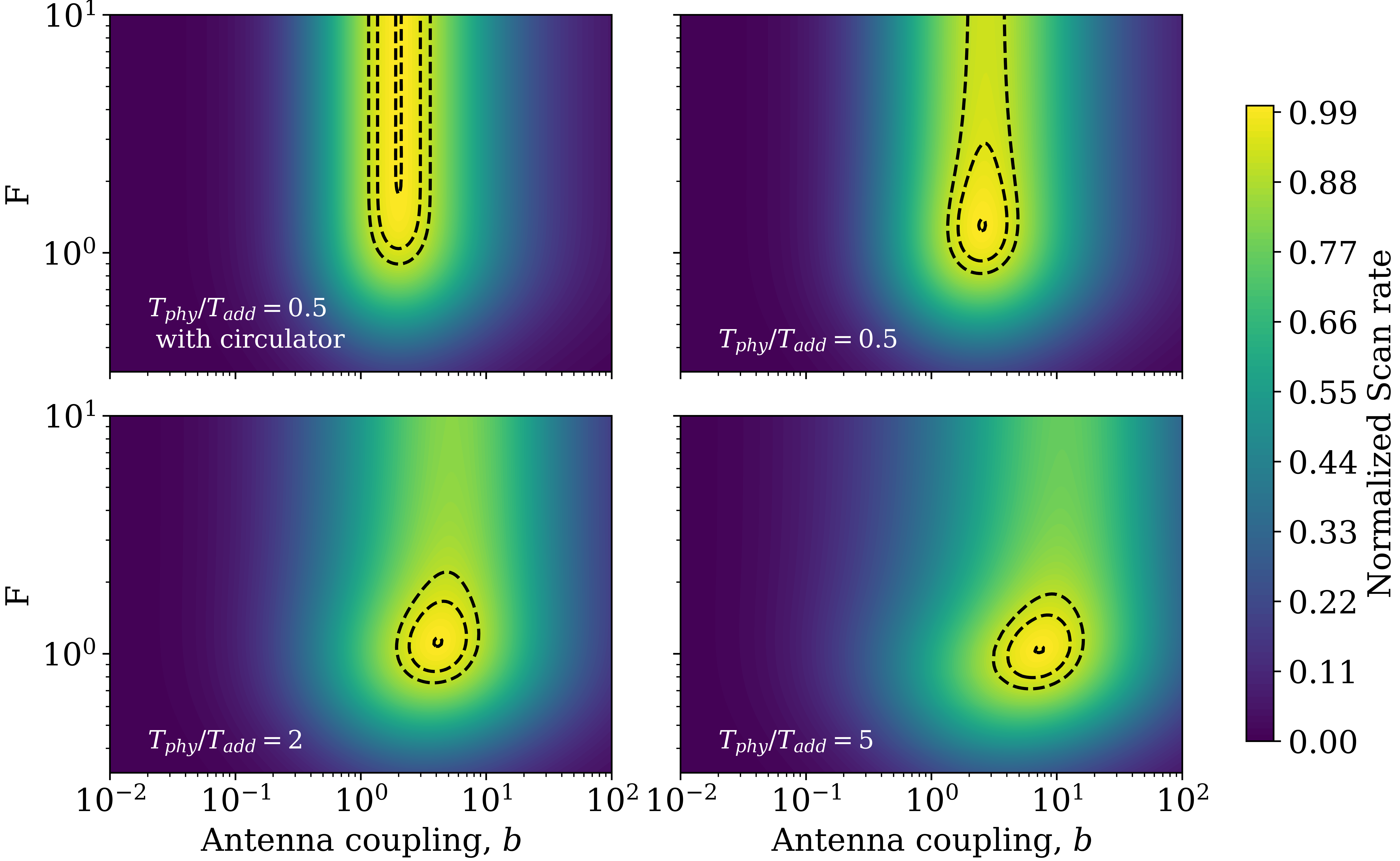}
\caption{Normalized scanning rate with different temperature ratio of $R\equiv T_{\rm phy} / T_{\rm add}$. The dashed lines indicate the scanning rates of 0.9,0.95, and 0.999. The top-left figure is for the case with a circulator between the cavity and the amplifier, and the optimum tuning step is $F\geq 2, b = 2$. As $R$ increases (noise from the resonator dominates), the tuning step converges to the bandwidth of the resonator, $F\rightarrow 1$.}
\label{fig:case study}
\end{figure}

%\YG{$\bullet$ Interpretation of numerical calculation results. Comparing with circulator (flat noise) case and without circulator. Dependence of Optimal parameters to physically temperature and added noise temperate ratio}
To facilitate comparison across different values of $R \equiv T_{\mathrm{phy}}/T_{\mathrm{add}}$, the scanning rate for each $R$ is normalized by its own maximum, such that the peak value is rescaled to unity. This normalization allows us to highlight how close a given $(F,\, b)$ configuration comes to the optimal condition within each noise setup, even though the absolute scanning rate differs between $R$ values.
Figure\,\ref{fig:case study} shows the normalized scanning rate for various $R$. In the figure, the dashed lines from the outermost to the innermost correspond to normalized scanning rates of 0.9, 0.95, and 0.999, respectively.
%The scanning rate was normalized to its maximum for each physical temperature to added noise ratio ($R\equiv T_{phy}/T_{add}$) cases, and Fig.\,\ref{fig:case study} shows the normalized scanning rate. In the figure, the dashed lines from the outermost to the innermost correspond to normalized scanning rates of 0.9, 0.95, and 0.999, respectively.
For comparison, realistic values of added noise and physical temperature in the presence circulator are shown. With the circulator inserted (upper left plot in Fig.\,\ref{fig:case study}), impedance is matched over a range of frequencies, which is a design feature of these devices. Correspondingly, the frequency dependence of the noise power and antenna coupling flatten out as shown in Fig.\,\ref{fig:noise_5_parameter} and Eq.\,\eqref{eq:system noise partition}, and the scanning rate converges to a maximum when $F\geq2$ and the antenna coupling is $b = 2$. On the other hand, there is a single solution $F\approx1.31$ and $b \approx 2.51$, maximizing the scanning speed if the circulator is not employed, as shown in the top-right of Fig.\,\ref{fig:case study}. This implies that even in the same noise-temperature configuration (i.e., the same value of $R$), the tuning step needs to be broader than in the case with the circulator to maximize the scanning speed. As the physical temperature is higher than the amplifier noise temperature, the optimal tuning step converges to the bandwidth of the resonator, $F\rightarrow1$. In practice, circulators are always used because the spectrum distortion due to impedance mismatch is more significant than the added noise from the circulator.

In summary, the cavity framework makes explicit how the signal lineshape and the resonator-shaped noise determine both the statistical test and the optimal scan parameters.

\section {Earth as a cavity haloscope (SNIPE hunt and SuperMAG searches) }\label{sec:Earth_Haloscope}

% \DB{Derek!!!!}

% \DB{Should we modify Fig.\,1 to add this type of haloscopes?} \DK{Yes please! See note in Fig. 1.}

% \DK{Here is a rough draft!}

There is another, rather different family of cavity experiments:
the Earth itself can act as a large-scale haloscope for ultralight axion-like dark matter, exploiting its geomagnetic field as a transducer to convert axion fields into detectable electromagnetic signals. 
The basic idea, proposed by \citet{arza2022earth}, is that an axion field $a(\mathbf{r},t)$ couples to photons via the interaction term described by Eq.\,\eqref{eq:Lint_axion}, namely $g_{a\gamma\gamma}a(\mathbf{r},t)\mathbf{E}\cdot\mathbf{B}$, where in this case $\mathbf{B} = \mathbf{B}_{\oplus}$ is the magnetic field of Earth. 
This coupling generates an effective current as described by Eq.\,\eqref{eq:effective_current}, $\mathbf{J}_\text{eff} \approx i g_{a\gamma\gamma} m_a a(\mathbf{r},t) \mathbf{B}_{\oplus}$, inducing an oscillating magnetic field pattern aligned with Earth's magnetic field. 
For axion masses roughly between $10^{-21}$ and $10^{-14}$\,eV (corresponding to frequencies ranging from $\sim \unit{\micro\Hz}$ up to several Hz), this signal is globally coherent and potentially detectable by sensitive magnetometer networks distributed across the Earth's surface\,\cite{arza2022earth}.

One can think of the Earth system as an electromagnetic cavity formed by the ionosphere and the conductive surface of the planet, with the lower atmosphere acting as a nonconductive gap in-between.
The axion-induced signal inherits its spatial configuration from the magnetic field of Earth since the effective current is proportional to $\mathbf{B}_{\oplus}$.
Under simplifying assumptions of a perfectly conducting Earth and ionosphere, the resulting global magnetic field pattern induced by $a(\mathbf{r},t)$ can be calculated to leading order\,\cite{arza2022earth}.
%\DB{What do you mean for the last part of this sentence}
%\DK{Sorry, that was quite unclear, I tried to revise accordingly with the above sentences, which also incorporate your suggestions.}
Note that the Earth-ionosphere cavity has a relatively low Q-factor of $\approx 3-8$ around the Schumann resonances\,\cite{sentman2017schumann} (global electromagnetic resonances of the Earth-ionosphere system) whose lowest frequency component is at about 7.8\,Hz, and so the searches discussed in this section are non-resonant, broadband searches that can access a wide range of axion masses simultaneously\footnote{While our discussion emphasizes non-resonant, broadband strategies, \citet{Taruya2025potep} develop a semi-analytical Earth–ionosphere cavity treatment that incorporates realistic, altitude-dependent atmospheric conductivity and proper boundary conditions at the Earth’s surface and lower ionosphere. This resolves resonant response of the cavity: the axion-induced magnetic field exhibits finite-$Q$ Schumann peaks for $m_a \gtrsim 10^{-14}\,\mathrm{eV}$ (notably enhanced near $m_a \sim 3\times10^{-14}\,\mathrm{eV}$), with quantitative predictions for the ground-level signal that are only weakly sensitive to conductivity-model choices and that show a characteristic geographic pattern.}.
The induced magnetic field is nearly monochromatic, oscillating near the axion Compton frequency $\nu_a$, and coherent over scales larger than Earth. 
Above frequencies of several hertz, the finite conductivity and inhomogeneity of the Earth ionosphere mean that the signal prediction is not robust, and the Schumann resonances (which exhibit variations over time due to seasonal changes and solar activity affecting the ionosphere) complicate the Earth's effective cavity behavior, setting a practical upper frequency limit for accurate theoretical modeling at about 5\,Hz\,\cite{fedderke2021earth}. 
Thus, so far, experimental searches are primarily conducted in the sub-Hertz range\,\cite{arza2022earth,fedderke2021search,sulai2023hunt,friel2024search,zhao2025scalar}, where predictions are most reliable.

The SuperMAG global magnetometer network\,\cite{gjerloev2009global}, originally designed for geophysical studies, provides decades of geomagnetic data recorded at roughly 500 stations around the world. 
\citet{arza2022earth} utilized this dataset to search for axion-induced signals, performing a vector spherical harmonic decomposition to isolate the characteristic axion signal pattern from background geomagnetic noise, similar to the approach of \citet{fedderke2021search} that was employed to search for hidden photon dark matter using the SuperMAG dataset. 
Despite several candidate spectral excesses, none survived detailed robustness checks, and the analysis resulted in constraints on $g_{a\gamma\gamma}$, achieving sensitivity comparable to the CAST helioscope\,\cite{CAST-2017} around axion masses of $3-4\times 10^{-17}$\,eV\,\cite{arza2022earth}.

Recently, the SuperMAG collaboration released high-fidelity, 1-s cadence geomagnetic data, allowing exploration of frequencies up to about 1\,Hz. 
\citet{friel2024search} utilized this dataset to improve constraints on axion-photon couplings for axion masses between $4\times10^{-18}$ and $4\times10^{-15}$\,eV. 
These constraints represent the strongest direct terrestrial bounds to date in this axion mass range\,\cite{friel2024search}.

Complementing the archival searches of SuperMAG data for axion dark matter, the Search for Non-Interacting Particles Experimental Hunt (SNIPE Hunt) is a dedicated experiment that uses an array of precision magnetometers at geographically separated sites –- remote, electromagnetically quiet ``wilderness'' locations chosen to minimize human-induced magnetic noise\,\cite{sulai2023hunt}.
By performing synchronized measurements of the local magnetic field variations at these distant sites, the SNIPE Hunt exploits the spatial correlations of the oscillating magnetic field induced by axion and hidden-photon dark matter: a true dark matter signal should appear with a coordinated pattern across the network, whereas local noise will be uncorrelated. 
The SNIPE Hunt instruments were unshielded to allow the Earth-sourced signal to reach them and had a sensitivity on the order of a few hundred pT/$\sqrt{\rm Hz}$ in the target frequency band.

A first-generation experiment used vector magnetoresistive sensors to search for signals at frequencies between 0.5-5\,Hz. 
No significant axion-induced signal was found, and the SNIPE Hunt established constraints on $g_{a\gamma\gamma}$ in the mass range $2\times10^{-15}$ to $2\times10^{-14}$\,eV. 
While these axion limits did not surpass the most stringent existing bounds (which come from astrophysical observations and helioscope experiments at somewhat higher masses), they represent direct terrestrial limits in this axion and hidden photon mass range\,\cite{sulai2023hunt}. 
Importantly, the SNIPE results were consistent with and complementary to the SuperMAG findings, covering the higher-frequency portion with an independent apparatus.

The SuperMAG and SNIPE Hunt searches represent the first terrestrial experiments directly probing the $10^{-17}-10^{-14}\,\rm{eV}$ mass decades for photon-coupled axion dark matter, exploring the parameter space previously accessible only via helioscope searches or astrophysical observations.
The next generations of the SNIPE Hunt experiment, currently underway, are using far more sensitive induction-coil magnetometers\,\cite{poliakov2017range} that can improve sensitivity to axion and hidden-photon dark matter by orders of magnitude. Note also the recent work of \citet{zhao2025scalar}, a dedicated single-site search for ultralight dark-photon dark matter using two scalar optically pumped magnetometers, improving on the first-generation SNIPE Hunt results by over two orders-of-magnitude.

Since the standard Earth-cavity haloscope approach is limited near and above the Schumann resonance regime ($\gtrsim 5$\,Hz), new strategies have been proposed to push into higher mass (frequency) territory. 
A particularly intriguing idea is to measure the spatial variation of the magnetic field (rather than the field itself) to cancel out environmental noise. 
\citet{bloch2024curl} propose a differential magnetometry scheme to measure the curl of the magnetic field ($\nabla \times \mathbf{B}$). 
The insight is that while the field at frequencies $\gtrsim 5$\,Hz may be difficult to interpret in terms of dark-matter-induced signals, by measuring $\nabla \times \mathbf{B}$ the experiment directly accesses the effective current $\mathbf{J}_\text{eff}$ and avoids the modeling uncertainties.
Practically, this scheme can be realized by placing two magnetometers separated by $\sim 100$\,m at ground level and another on a hillside nearby $\sim 100$\,m above the other two, forming a  triangle in the ``up-down/east-west'' plane.
The dark-matter-induced $\mathbf{J}_\text{eff}$, along $\mathbf{B}_{\oplus}$, would generate a nonzero curl which could be detected in the appropriate combination of signals from the magnetometer array. 
This method is predicted to enable robust searches for dark photons and axions at frequencies about a kilohertz\,\cite{bloch2024curl}.
Another approach is to improve the modeling of the electromagnetic signal by accounting more carefully for atmospheric conductivity and solving for the electromagnetic signal to higher order, and thus extending the frequency range for which reliable predictions of the dark-matter-induced signal can be obtained and data interpreted\,\cite{taruya2025hunting}.

%\DB{Ok, so this is a very nice review. Now we need to add a discussion of similarities and differences with other haloscope searches. Random thoughts: nonresonant like CASPEr-ZULF, MAinz-Krakow, ... No scanning. What else? Shall we also discuss somewhere the advantages of networks; these are briefly mentioned above; is the scaling with the number of nodeas out of scope?}
%\DK{Attempted a brief answer in the last paragraph.}

In comparison with other haloscope searches, the aforementioned non-resonant axion dark matter searches using the Earth as a transducer are most similar to the lumped-element circuit experiments, such as dark matter (DM) radio\,\cite{chaudhuri2015radio}, that are described in the next section, albeit at a much lower axion mass (frequency) range. Another important point is that the above searches employ a network of sensors, and in particular, they take advantage of the fact that the axion-induced magnetic field has a specific global pattern. This allows both suppression of uncorrelated noise by projecting the collective network signal onto the appropriate spherical vector harmonic components of the predicted signal and, perhaps more importantly, robust methods for rejecting spurious signals by testing subsamples of sensor data as discussed in detail by \citet{fedderke2021search}, \citet{arza2022earth}, and \citet{sulai2023hunt}.

Overall, Earth-scale searches are intrinsically broadband and exploit global coherence and distributed magnetometer networks, offering sensitivity in the $\mu$Hz–Hz band that is difficult to access with laboratory resonators.

\section {Lumped-element circuits}\label{sec:Lumped_circuits}

A coil, modeled as a lumped-element circuit, is typically used for electromagnetic detection of axion-photon conversion below gigahertz frequencies. Electrical resistance (or conductance), inductance, and capacitance are the defining parameters in the simplified circuit model of a coil, and together they determine the amplitude and frequency sensitivity. 
%\DB{We are not sure, how to understand ``detection frequency'' here. Is the the resonance frequency?} 
For the lumped-element model to be valid (as opposed to the distributed-element model, which treats the coil akin to a transmission line with standing waves), the coil should be much smaller in size than the wavelength of electromagnetic radiation at the detected frequency $\lambda$. Examples of such magnetoquasistatic systems where the magnetic field does not considerably change over length,
%\YG{May we define magnetoquasistatic system? In such a magnetic field the corresponding electric field is suppressed.} 
are lumped-circuit haloscopes that have coils with at most $\sim10\text{ cm}$ in diameter, which corresponds to roughly a gigahertz upper limit in the detection frequency through $\nu = c/\lambda $. Larger coils ($\sim100\text{ cm}$) can be used for detection at lower frequencies ($\sim100\text{ MHz}$). 
%\YG{The relation between diameter(system size) and frequency needs to be placed. In addition, if you refer resonance frequency corresponding diameter, then I am worrying about we can still apply magnetoquasistatic approximation. If it is resonance mode, we cannot ignore electric field and it seems not to apply magnetoquasistatic approximation} 
Coil-based detectors are designed to perform either a broadband or a narrowband frequency search.  Experimental parameters from major broadband and narrowband lumped-element circuit haloscopes are compared in Table \ref{tab:LC_comparison}. The experiments are further detailed in subsections \ref{sec:Lumped_circuits_broadband} and \ref{sec:Lumped_circuits_narrowband}.
%\YG{The distinction between broadband and narrowband seems to be defined by whether the quality factor is larger or smaller than unity. I have a few concerns: In narrowband lumped-element searches, the signal (or sensitivity) typically scales with some power of the quality factor. If the quality factor is also well defined in the broadband case, one might interpret Q < 1 as implying signal degradation. In addition, in the SHAFT schematic there does not appear to be an explicit physical capacitance. How, then, is the resonance of the circuit defined, and how is the quality factor of the circuit determined?}
\begin{table}[h]
\caption{Comparison of lumped-element circuit haloscopes. ``Limit'' status means that the experiment is already built, has performed science runs, and published axion coupling limits. ``Idle''--built but has not run for axion searches. ``Construction'' and ``design'' designate the corresponding stages of the experiments.}
\small
\label{tab:LC_comparison}
\renewcommand{\arraystretch}{1.3}
\setlength{\tabcolsep}{3pt}

\begin{ruledtabular}
\begin{tabular}{lccccc}
Experiment & Type & Bandwidth & Quality Factor & Sensitivity & Status \\
\hline
\lbox{0.20\linewidth}{SHAFT\\ \cite{Gramolin2021NaturePhysics}} &
\cbox{0.12\linewidth}{Broadband} &
\cbox{0.18\linewidth}{3\,kHz -- 3\,MHz} &
\cbox{0.15\linewidth}{--} &
\cbox{0.15\linewidth}{$10^{-10}$\,GeV$^{-1}$} &
\cbox{0.10\linewidth}{Limit} \\
\hline
\lbox{0.20\linewidth}{ABRA-10\,cm\\ \cite{Ouellet:2019prl}} &
\cbox{0.12\linewidth}{Broadband} &
\cbox{0.18\linewidth}{50\,kHz -- 2\,MHz} &
\cbox{0.15\linewidth}{--} &
\cbox{0.15\linewidth}{$10^{-10}$\,GeV$^{-1}$} &
\cbox{0.10\linewidth}{Limit} \\
\hline
\lbox{0.20\linewidth}{DMRadio-Pathfinder\\ \cite{Silva:2017}} &
\cbox{0.12\linewidth}{Narrowband} &
\cbox{0.18\linewidth}{100\,kHz -- 10\,MHz} &
\cbox{0.15\linewidth}{$10^6$} &
\cbox{0.15\linewidth}{N/A} &
\cbox{0.10\linewidth}{Idle} \\
\hline
\lbox{0.20\linewidth}{CAL-Pathfinder} &
\cbox{0.12\linewidth}{Narrowband} &
\cbox{0.18\linewidth}{$\sim$79\,MHz -- 80\,MHz} &
\cbox{0.15\linewidth}{$10^5$} &
\cbox{0.15\linewidth}{$10^{-15}$\,GeV$^{-1}$} &
\cbox{0.10\linewidth}{Design} \\
\hline
\lbox{0.20\linewidth}{DMRadio-m$^3$\\ \cite{Brouwer:2022m3}} &
\cbox{0.12\linewidth}{Narrowband} &
\cbox{0.18\linewidth}{30\,MHz -- 200\,MHz} &
\cbox{0.15\linewidth}{$10^5$} &
\cbox{0.15\linewidth}{$10^{-17}$\,GeV$^{-1}$} &
\cbox{0.10\linewidth}{Design} \\
\hline
\lbox{0.20\linewidth}{DMRadio-50L\\ \cite{Rapidis:2023}} &
\cbox{0.12\linewidth}{Narrowband} &
\cbox{0.18\linewidth}{5\,kHz -- 5\,MHz} &
\cbox{0.15\linewidth}{$10^6$} &
\cbox{0.15\linewidth}{$10^{-14}$\,GeV$^{-1}$} &
\cbox{0.10\linewidth}{Construction} \\
\hline
\lbox{0.20\linewidth}{DMRadio-GUT (10\,m$^3$)\\ \cite{Brouwer:2022gut}} &
\cbox{0.12\linewidth}{Narrowband} &
\cbox{0.18\linewidth}{100\,kHz -- 30\,MHz} &
\cbox{0.15\linewidth}{$10^7$} &
\cbox{0.15\linewidth}{$10^{-19}$\,GeV$^{-1}$} &
\cbox{0.10\linewidth}{Design} \\
\hline
\lbox{0.20\linewidth}{ADMX SLIC\\ \cite{ADMX:2020prl}} &
\cbox{0.12\linewidth}{Narrowband} &
\cbox{0.18\linewidth}{42\,MHz -- 44\,MHz} &
\cbox{0.15\linewidth}{$10^4$} &
\cbox{0.15\linewidth}{$10^{-12}$\,GeV$^{-1}$} &
\cbox{0.10\linewidth}{Limit} \\
\hline
\lbox{0.20\linewidth}{WISPLC\\ \cite{ZZhang:2022prd}} &
\cbox{0.12\linewidth}{Broadband\\Narrowband} &
\cbox{0.18\linewidth}{2\,kHz -- 2\,GHz\\2\,kHz -- 6\,MHz} &
\cbox{0.15\linewidth}{--\\$10^4$} &
\cbox{0.15\linewidth}{$10^{-12}$\,GeV$^{-1}$\\$10^{-16}$\,GeV$^{-1}$} &
\cbox{0.10\linewidth}{Construction} \\
\end{tabular}
\end{ruledtabular}
\end{table}
\subsection{Broadband searches with lumped circuits} \label{sec:Lumped_circuits_broadband}

%\DA{Even though narrowband searches result in greater sensitivities, one drawback is the necessity to change tuning of the circuit. 
%Broadband searches do not require any tuning, and in a single science run, provide reach in a wide frequency range, albeit with lower sensitivities.
Although narrowband searches achieve high sensitivity to $g_{a\gamma\gamma}$, they require tuning of the resonance frequency to cover a wide frequency range. In contrast, broadband searches dispense with tuning and, within a single run, can probe a wide frequency range, albeit with reduced sensitivity. An example of a broadband search is discussed in the Earth-haloscope Sec.\,\ref{sec:Earth_Haloscope}.

Although several geometries are possible, a common way to perform a broadband lumped-element search is with a superconducting toroidal coil with many turns that runs a direct current (DC), resulting in a static magnetic field inside the toroid. In the presence of this static magnetic field, axion-photon coupling results in an effective oscillating current parallel to the DC field, which in turn results in an induced magnetic field oscillating along the central axis of the toroid, see Eq.\,\eqref{eq:effective_current}, orthogonal to the static magnetic field. This oscillating magnetic field creates a time-varying magnetic flux, which induces an electromotive force (emf) $\mathcal{E}$ in a helical pickup coil placed inside the bore of the toroid. Both the toroidal coil and the helical pickup coil obey the lumped-element circuit model. 

%\YG{I think this description is very clear. I would suggest adding a brief introductory discussion on the broadband search. The original LC circuit concept proposed by Sikivie was not based on a toroidal system.}

One of the lumped-circuit haloscopes, A Broadband/Resonant Approach to Cosmic Axion Detection with an Amplifying B-field Ring Apparatus (ABRACADABRA, also known as ABRA), initially searched for the axion with a broadband readout circuit within the frequency range $11\text{ kHz} < \nu_a < 3\text{ MHz}$ using a 10\,cm diameter pickup loop (instead of a coil with many turns) in their Run 1\,\cite{Ouellet:2019prl}. An updated version of ABRA-10\,cm using a pickup cylinder searched for axions with a broadband readout circuit within the frequency range $50\text{ kHz} < \nu_a < 2\text{ MHz}$\,\cite{Salemi:2021} with an order of magnitude better sensitivity in their Runs 2 and 3.  Beyond the magnetoquasistatic limit, a coil or a cylinder is modeled as a distributed-element circuit, and self-resonances arise at higher frequencies. This causes a broadband lumped circuit haloscope to act as a multipole narrowband haloscope at higher frequencies, thus extending the sensitivity range from the original lumped circuit regime to the distributed circuit regime, albeit with gaps in frequencies\,\cite{Benabou:2023prd}. Another recent spin-off work, ABRA-GW, is sensitive to high-frequency gravitational waves alongside axions with a broadband detection scheme\,\cite{Pappas:2025}.

\begin{figure}[h!]
\centering
\includegraphics[width=0.9\linewidth]{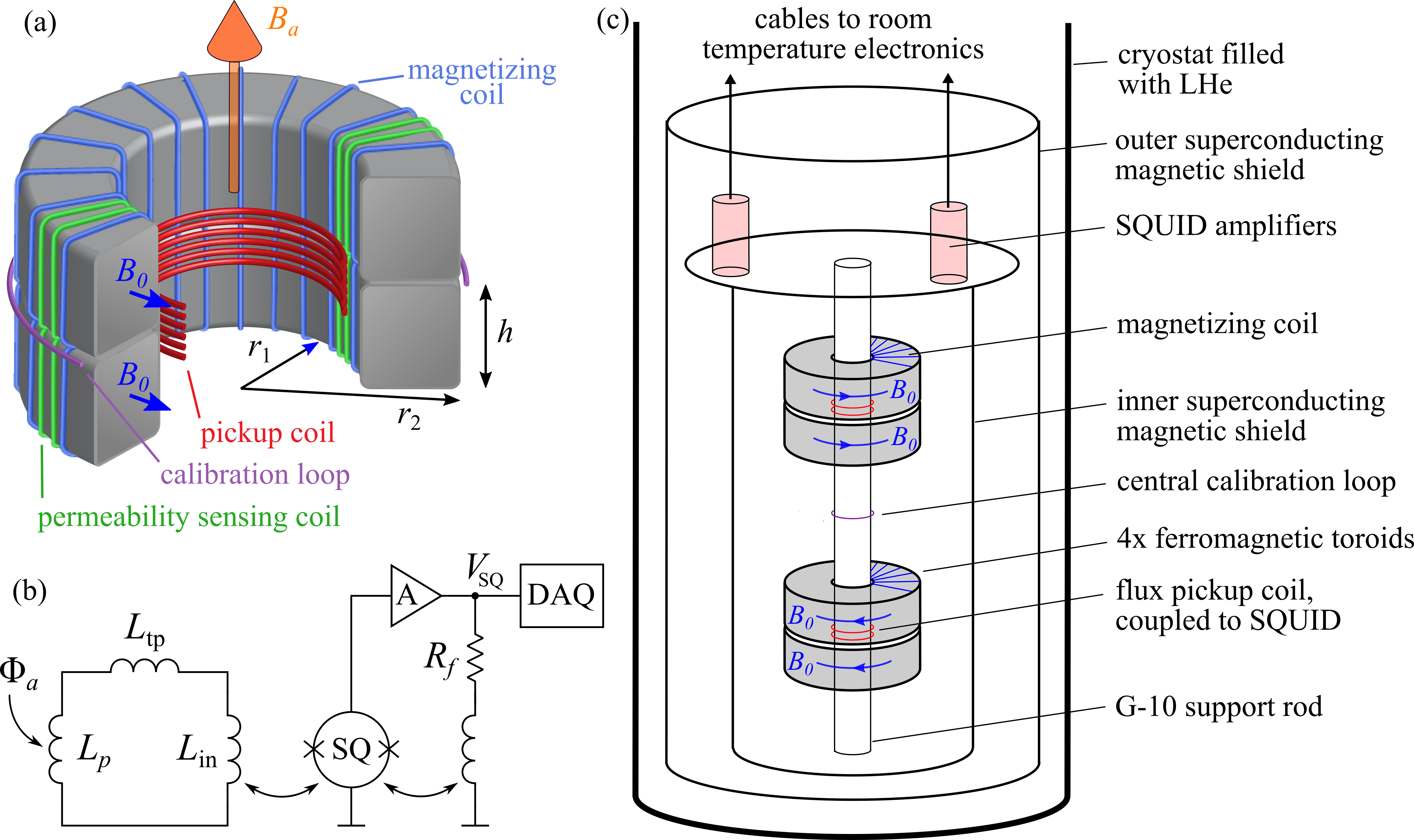}
\caption{An example of a lumped-circuit axion-detection experiment, SHAFT \cite{Gramolin2021NaturePhysics}:
(a) Schematic of the ferromagnetic core and pickup coil, (b) Detection chain diagram ($L_p$ is pickup coil, SQ is SQUID, A is amplifier, DAQ is data acquisition electronics, (c) full experimental schematic of the cryogenic insert at 4\,K.}
\label{fig:shaft scheme}
\end{figure}

A similar lumped circuit broadband haloscope, Search for Halo Axions with Ferromagnetic Toroids (SHAFT), achieved improved sensitivity by replacing the toroidal hollow core with a ferromagnet (see Fig.\,\ref{fig:shaft scheme})
%\DB{We suggest to import Fig. 1 from the SHAFT paper wholesale; however, changing the caption} 
to enhance the static magnetic field created by the toroidal coil by a factor of 24, and searched for axions within the frequency range of 3\,kHz\,$<$ $\nu_a\,<\,$3\,MHz\,\cite{Gramolin2021NaturePhysics}. In addition, SHAFT used two data acquisition channels and several counter-magnetized toroids to incorporate an experimental reversal, allowing real-time systematic rejection and signal enhancement using phase-sensitive data analysis.
Although their data acquisition and analysis methods differed, both SHAFT and ABRA-10\,cm had detection sensitivity limited %at $150\,\text{aT Hz}^{-1/2}$ 
by the input noise of the SQUID amplifier; they used commercially available Superconducting Quantum Interference Devices (SQUIDs).

In practice, searches are limited in frequency range.  On the upper end of frequencies, the practical limit is typically that of amplifier bandwidth, although with a sufficiently broadband amplifier one could theoretically search up to the point where geometric resonances dominate and the LC circuit begins to act more like a cavity.  At low frequencies, vibrational and 1/$f$ noise dominates, limiting sensitivity.

%\DA{Should we add drawings of experimental setups?} \DB{I would say, yes for SHAFT!}

%\DB{We have read this with YGK and Kai. We generally like it! When the narrowband part is discussed, it would be good to refer to a scanning strategy or lack thereof, as this is one of the main emphases of the paper, SASHA, \textbf{DENIZ, PLEASE HELP!!!}}

\subsection{Narrowband searches with lumped circuits} \label{sec:Lumped_circuits_narrowband}

%When an electromagnetic resonator is tuned to radio frequencies, it can be used for a narrowband search.
As an alternative to a broadband experiment with a purely inductive pickup, electromagnetic resonators can be used for a narrowband search.
Such a resonator is typically modeled as a lumped-element circuit with a capacitor $C$ placed in parallel with an inductor $L$ (the search coil). This results in an electromagnetic resonance at $\omega = 1/\sqrt{LC}$ with a quality factor of $Q = \omega L R^{-1}$, where $R$ is the total resistance of the resonator. These resonators are typically cooled from room temperature down to cryogenic temperatures to decrease $R$ and increase $Q$. 
The increase in Q factor enhances the signal strength with respect to non-resonator-sourced noise.
LC (inductor-capacitor) circuits as narrowband resonators for axion detection were first suggested by \citet{PhysRevLett.112.131301} with follow-up theoretical work by \citet{chaudhuri2015radio} and \citet{Kahn2016}. 
%\YG{Is this paragraph essential for the subsequent discussion? I think it would be more effective to emphasize why the resonance search approach is introduced, as well as its advantages and drawbacks compared to the broadband search. In addition, how the quality factor Q affects the sensitivity could be clarified.}

%There was some research in parallel on 3D lumped LC resonators that focused on novel cavity designs \cite{McAllister:2016prd}. \DB{What cavity designs? Are we not talking about lumped circuits?} \cs{That paper was for a kind of hybrid experiment in between the cavity and lumped-element limits with re-entrant cavities.  I would probably leave this reference out unless we want to talk about that regime.}
Dark Matter Radio (DMRadio) is a group of narrowband lumped-element-circuit haloscopes: Pathfinder, m$^3$, $50$L, CAL-Pathfinder, and GUT.
In DMRadio-Pathfinder, the capacitors are tunable with motor-controlled rods changing the conductive area around a dielectric \cite{Silva:2017}. It was built but did not have any science run (therefore did not provide any limit to axions), however, it is used as a testbed for the other, more ambitious experiments within the DMRadio program.  DMRadio-m$^3$ is a proposed lumped-element experiment inside a $\sim$4\,T peak field solenoidal magnet. Its frequency range extends beyond the magnetoquasistatic limit, so it uses a set of coaxial pickup coils with varying sizes to probe different frequencies, forming a complete scan range of $\sim$170\,MHz QCD axions with sensitivity to the pessimistic DFSZ model \cite{Brouwer:2022m3}. The CAL-Pathfinder experiment is a demonstrator of a similar geometry to reach these high frequencies.  DMRadio-$50$L scans lower frequencies than DMRadio-m$^3$, between 5\,kHz and 5\,MHz. The experiment, located at Stanford, has a toroidal magnet with a 1\,T average field and a high-Q LC-oscillator pickup \cite{Rapidis:2023}. DMRadio-GUT is an ambitious large-scale experimental proposal with plans to reach the sensitivity to the QCD-axion in over-GUT-scale axion masses (0.4 -- 120\,neV) with a $\sim$6.2\,y scan time. It is proposed to have a magnet $\sim$10\,m$^3$ in size with a $\sim$16\,T field; it will operate at $\sim$10\,mK temperatures with a Q$\sim 10^7$ pickup circuit with amplifier noise below the quantum-limit (QL) \cite{Brouwer:2022gut}.

The ADMX collaboration has put together an additional experiment called Superconducting LC circuit Investigating Cold axions (SLIC) that searched for axions in three narrow ranges, lower in frequency than the main ADMX experiments \cite{ADMX:2020prl}. Currently under construction, Weakly Interacting Slender Particle detection with an LC circuit (WISPLC) experiment has a broadband and a narrowband detection scheme within the same setup and can potentially search for axions up to 2\,GHz in frequency \,\cite{ZZhang:2022prd}, where the detection circuit would be beyond the magnetoquasistatic limit.

The scanning strategy analyzed for axion haloscopes in the gigahertz frequency range does not necessarily extend directly to the megahertz range in lumped-element narrowband resonators. 
At gigahertz frequencies, the resonators are in the ground state with a thermal occupancy of much less than one photon. 
However, at megahertz frequencies, even at millikelvin temperatures, there is a stochastic population of thermal photons in the resonator. %non-binary 
In this case, the on-resonance axion sensitivity of the tuned resonator is limited by its physical temperature. 
Experiments to-date amplify the current in the resonator using an electromagnetically coupled detector, such as a DC SQUID.
Development is underway for even lower-noise sensors such as RF Quantum Up-converters (RQUs), which are phase-sensitive amplifiers that would be able to reduce amplifier noise below the QL using backaction evasion, the low-frequency analog of squeezing \,\cite{Kuenstner2025}.

If a narrow-band-resonator measurement is dominated by thermal Johnson-Nyquist noise, the resonator frequency can be tuned in steps wider than the resonator bandwidth without significant loss of SNR\,\cite{Chaudhuri:2018rqn,Chaudhuri:2021}. Note that both the signal and the resonator-coupled noise share a similar Lorentzian response, while quantum noise and non-resonator coupled noise, such as amplifier output noise, have a profile that is flat in frequency space. Thus, the step size is determined as the frequency range where the thermal noise with a Lorentzian profile coincides with the amplifier output noise; this range is known as the `sensitivity' or `visibility' bandwidth, which can be significantly larger than the resonator bandwidth.  As a result, although SNR cannot be increased in this limit of large thermal noise, it can be made to remain approximately constant over a frequency range broader than the resonator bandwidth near resonance. Quantum optical techniques, such as squeezing and back-action evasion, can potentially be used to decrease the amplifier noise level below the QL and to extend the frequency range where the thermal noise prevails, thus increasing both the frequency step size and the search rate.

The accelerated search strategy used in spin haloscopes, as mentioned in Sec.\,\ref{sec:SpinHaloscopes:ScanningSpeed}, may be considered similar to the lumped-element experiments, since they both operate with constant SNR. However, the non-amplifiable white (flat) noise sources are different for each experiment type: shot noise in a spin haloscope, and amplifier output noise in a lumped-element circuit. The tuning strategy used in cavity haloscopes cannot be directly applied to lumped-element experiments because the lumped resonator thermal noise is typically not flat. Without a circulator, the equivalent impedance of the resonator as seen at the amplifier input is frequency dependent as also discussed in Sec.~\ref{sec:CavityHaloscopes:casestudy}. Therefore, the flat-noise assumption used in cavity-haloscope strategies does not hold.

Here we describe the scan rate for a lumped-element haloscope in analogy to the above equations for a cavity haloscope.  The first difference is that the amplifier input for a lumped-element haloscope is primarily reactive, rather than dissipative.  Thus rather than writing the SNR in terms of dissipated power, it is appropriately written in terms of currents, SNR$=|I_\mathrm{sig}|^2/|I_\mathrm{n}|^2$, where $|I_\mathrm{sig}|$ is the magnitude of the signal current induced in the resonance circuit and $|I_n|$ is the magnitude of the noise current also referred to the resonator.  Also, as discussed above, one scans a lumped-element resonator in steps of sensitivity bandwidth rather than resonator bandwidth.  Combined, these two effects give a general form for the lumped-element scan rate of
\begin{equation}
    \frac{d\nu_r}{dt}\sim\frac{\Delta\nu_\mathrm{sens}\Delta\nu_\mathrm{sig}}{\mathrm{SNR}^2}\frac{|I_\mathrm{sig}|^4}{|I_n|^4} \,,
\end{equation}
where $\nu_r$ is the resonator frequency, $\Delta\nu_\mathrm{sens}$ is the sensitivity bandwidth, $\Delta\nu_\mathrm{sig}=\nu_a/Q_a$ is the signal bandwidth (equivalently the axion linewidth, determined by the axion frequency and quality factor).   See the appendix of \citet{Brouwer:2022gut} for a derivation.  The noise term includes both thermal noise and amplifier input noise.  The signal term can be derived using the expected signal from axion interactions in Eq.\,\eqref{eq:effective_current} and applying a transfer function that includes both inductive coupling from the axion signal to the resonator and subsequent resonant enhancement.

Assuming $\Delta\nu_\mathrm{sens}>\Delta\nu_\mathrm{sig}$, the result is \cite{Brouwer:2022gut}
\begin{equation}
    \frac{d\nu_r}{dt}\sim\frac{1}{\mathrm{SNR}^2}\left(g_{a\gamma\gamma}^4\rho_a^2\nu_rQ_a\right)\left(\frac{c_\textrm{PU}^4Q_rB_0^4V^{10/3}}{k_BT\eta_A}\right) \,.
    \label{eq:lumped_scan_rate}
\end{equation}
Here $c_\mathrm{PU}^2\in[0,1]$ is a pickup-coupling efficiency factor, analogous to the cavity form factor $C$ describing the fraction of energy in the axion effective current that is coupled to the pickup circuit.  One can see that $c_\mathrm{PU}^2$ enters this scan rate in the same way that $C$ does in the analogous cavity scan rate equation, Eq.\,\eqref{eq:single_tune_scanning}.  The square magnitude of the axion effective current similarly depends on the magnetic field strength and volume $(B_0^2V)^2$, but here we have an additional factor of $(V^{2/3})^2$ compared to Eq.\,\eqref{eq:single_tune_scanning}.  This factor of squared area is a proxy for the square of the inductance of the LC pickup structure, $L\sim V^{2/3}$, which is required because the energy induced on the pickup depends on its inductance.

The lumped-element noise terms in Eq.\,\eqref{eq:lumped_scan_rate} are also different from the cavity case (Eq.\,\ref{eq:single_tune_scanning}) because this scan rate must take into account the different spectral shapes of noise that is either enhanced by the resonator or not.  $T$ here refers to the physical temperature of the resonator, which drives the Lorentzian-shaped background, while $\eta_A$ is the amplifier added noise in units of QL, which can be further split into flat-spectrum noise (noise that does not get shaped by the resonator) and resonator-shaped noise.  The different dependence on $T$, $\eta_A$, and frequency of Eq.\,\eqref{eq:lumped_scan_rate} compared to Eq.\,\eqref{eq:single_tune_scanning} is due to the inclusion of the scan rate speedup from larger sensitivity bandwidth, as calculated in appendix F.4 of \citet{Chaudhuri:2018rqn}.

Note that corrections in Eq.\,\eqref{eq:lumped_scan_rate} analogous to those in Eq.\,\eqref{eq:single_tune_scanning} for the halo-dependent signal shape are absorbed into the tilde, as are corrections for the difference between the inductor volume and the magnetic field volume that depend on the geometry of each detector.  For an example of a more precise treatment that takes these effects into account for the DMRadio-m$^3$ detector, see \citet{Brouwer:2022m3}.

In summary, lumped-element detectors extend axion–photon searches to frequencies where conventional cavities become impractical, with scan performance set by pickup coupling and by the interplay of resonator-shaped and broadband noise contributions.

\section{Spin haloscopes}
\label{sec:SpinHaloscopes}

\subsection{General remarks}\label{sec:SpinHaloscopes:General}
Axion-nucleon, axion-electron, and axion-gluon interactions can be searched for with spin haloscopes, which include haloscopes utilizing (co)magnetometers, NMR and electron-paramagnetic-resonance (EPR) spectrometers, and storage rings. Specifically, the axion-induced EDM or axion field gradient can induce spin precession measured in such haloscopes. In practice, spin haloscopes operate either in a broadband mode that continuously monitors spin-precession observables over a wide frequency range, or in a resonant mode where the spin-precession resonance is tuned (typically via the bias magnetic field) to scan across the axion frequency. Across both modes, sensitivity and scan performance are primarily set by the achievable spin polarization, coherence/relaxation times, and magnetometer/readout noise, together with the finite axion coherence time that dictates the optimal integration strategy.
%Recently, a variety of experiments using fermion spins have been carried out across a wide range of frequencies, targeting both broadband and narrowband signals\,~\cite{Garcon_2019_Constraints,Lee:2022vvb,CASPEr-LF-2025,Bloch2023NASDUCK_SERF, Gavilan-Martin2025,Smorra:2019aa,Wei:2023rzs,BlochItay2022Floquet_sciadv, Xu:2023vfn,Garcon:2017ixh, Aybas2021_SolidStateNMR_PhysRevLett, QUAX:2020adt, PhysRevX.7.041034,PhysRevLett.129.191801,PhysRevLett.126.171301}.
Recently, a variety of experiments using fermion spins have been carried out across a wide range of frequencies.
These efforts span
(i) \emph{broadband} magnetometry/comagnetometry searches\,\cite{Garcon_2019_Constraints,Lee:2022vvb,Bloch2023NASDUCK_SERF,Gavilan-Martin2025,Smorra:2019aa,Wei:2023rzs},
(ii) \emph{resonant} NMR-based searches (including alkali--noble-gas and condensed-matter NMR)\,\cite{BlochItay2022Floquet_sciadv,Xu:2023vfn,Garcon:2017ixh,Aybas2021_SolidStateNMR_PhysRevLett},
(iii) oscillating-EDM approaches\,\cite{PhysRevLett.129.191801,PhysRevLett.126.171301,PhysRevX.7.041034,Aybas2021_SolidStateNMR_PhysRevLett},
and (iv) electron-spin–based searches\,\cite{QUAX:2020adt}.

%%

%\jl{$\bullet$ Axion detection as pseudomagnetic field.}  
Axion dark matter couples to nucleons via the gradient and EDM interaction, described by the effective Hamiltonian:    
\begin{equation}
    H_{\rm axion} = g_{\rm aNN} \mathbf{\bm \nabla} a \cdot \boldsymbol{I}_N +\ g_{d}\,a\,\boldsymbol{I}_{N} \cdot \mathbf{E}\,, 
    \label{eq:NMRHaloscopeHamiltonian}
\end{equation}
where $g_{\rm aNN}$ is the coupling constant, and $\boldsymbol{I}_N$ is the nuclear spin for axion-nucleon interactions. Here $g_{d}$ is the axion to gluon coupling constant (also known as $g_{aN\gamma}$), and $\mathbf{E}$ is the effective electric field (possibly, an internal field in the case of composite systems). Axion dark matter behaves like a classical field, oscillating with a frequency close to its Compton frequency $\nu_a \approx m_a c^2/h $ as the dark matter is non-relativistic. 
Consequently, axion dark matter acts as a pseudomagnetic field affecting the nuclear spin.

The two terms in the Hamiltonian \eqref{eq:NMRHaloscopeHamiltonian} define two broad classes of spin haloscopes: the ``gradient'' haloscopes [Cosmic Axion Spin Precession Experiment (CASPEr)-gradient, (co)magnetometers~\cite{Graham:2013gfa}, ...] and ``electric'' (CASPEr-electric~\cite{Budker:2013hfa}, storage-ring~\cite{Chang2019}...), where the latter require application of electric fields to the spins.
In this section we mainly use the “electric” class to motivate storage-ring oscillating-EDM searches, while leaving a detailed discussion of low-frequency laboratory EDM experiments outside our scope.
We note that the effect of each of the two terms in Eq.\,\eqref{eq:NMRHaloscopeHamiltonian} can be cast as that of a corresponding pseudomagnetic field exerting torque on the spins.

Recent work by \citet{beadle2025_DM_NMR} pointed out that CASPEr and other ``magnetometer'' experiments are also sensitive to additional UBDM couplings, for instance, the axion-photon coupling and axion-electron coupling producing a real magnetic field on the probe spins. Further discussion of these effects is beyond our present scope.
\subsection{NMR Haloscopes}
In this subsection we focus on resonant NMR-based spin haloscopes, where the Larmor frequency is tuned (typically via the bias magnetic field) to search for the axion Compton frequency. We place particular emphasis on alkali–noble-gas (spin-amplifier) NMR experiments operated as resonant searches, and use them as a concrete reference architecture for the sensitivity and scan optimization discussed below. A broader overview of other spin-based approaches is provided in Sec.\,\ref{sec:SpinHaloscopes:General}.
In Sec.\,\ref{sec:sen-NMR-haloscopes} we derive the signal model and sensitivity for resonant NMR haloscopes (with emphasis on alkali–noble-gas spin-amplifier schemes). In Sec.\,\ref{sec:SpinHaloscopes:ScanningSpeed} we then discuss the corresponding scan strategy and optimal scanning speed, highlighting the trade-offs between sensitivity, bandwidth, and coherence times.

\subsubsection{Sensitivity of NMR haloscopes}
\label{sec:sen-NMR-haloscopes}

The technique of NMR is suitable for searching for weak oscillating (pseudo)magnetic fields such as the dark matter field. The NMR resonance frequency is given by the Larmor precession frequency $\nu_{c} = \gamma_N B_{\rm ext}$, where $\gamma_N$ is the gyromagnetic ratio of the nuclei, and $B_\mathrm{ext}$ is the external magnetic field. The exotic interaction of the gradient field of axion ${\bm \nabla} a$ may work as a pseudomagnetic field which drives the nuclear spin precession, just like standard magnetic fields. However, its coupling to spins is different from those of standard magnetic fields. Through this interaction, the gradient of the temporally oscillating axion field ${\bm \nabla} a$ may induce nuclear spin flips, with the probability of this process being maximal near the resonance. 

In spin-amplifier NMR experiments such as those by \citet{Jiang:2021dby}, one utilizes a comagnetometer containing alkali atoms and noble-gas atoms within a vapor cell, subjected to a static magnetic field. The alkali atoms are optically pumped with laser light, resulting in a high degree of electron-spin polarization. The noble-gas spins also attain high polarization through Fermi-contact interactions between the alkali-atom electrons and the noble-gas nuclei during atomic collisions between the two species. Once polarization stabilizes, the system is ready for measurement.
When exposed to an oscillating magnetic field or an axion pseudomagnetic field, if the oscillation frequency matches the Larmor frequency of the noble-gas spin system, the polarized noble-gas spins tilt away from the direction of the static field and precess around this direction. The tilt angle is proportional to the strength of the oscillating field and the precessing spins produce a rotating transverse magnetic field that can be detected \textit{in-situ} by the alkali spins, functioning as an atomic magnetometer. The detection is enhanced by local Fermi-contact interactions. 

In addition to atomic comagnetometers, one can detect the axion field gradient with NMR spectrometers.
%In this configuration, the prepared sample is enclosed in a sample holder.
The axion-induced spin precession is probed with pickup coils surrounding the sample. The bias magnetic field is applied to tune the Larmor frequency to search for the axion resonance. With superconducting magnets, the Larmor frequency can reach up to $\sim\SI{1}{\GHz}$. 
%\DB{Maybe we need to relate the Hamiltonian top the pseudomagnetic field, which will motivate the use of *magnetometers* or NMR devices; so the order of the second and third paragraph may need to be switched}  \jl{The switch of the paragraphs have been done.}

%\jl{$\bullet$ Axion detection as pseudomagnetic field in comagnetometer.} 

% which is denoted as the spin-based amplifier\,\cite{Jiang:2021dby}.

%felled by the \textit{in-situ} alkali spins, which an amplified magnetic field proportional to the number of polarized nuclei. 

%\DB{Here we want to explain what the measurement procedure is and why we choose in to be this way. \jl{Done.} Explain the choice of the acquisition time}

%\jl{$\bullet$ Measurement procedure.}  
The exact mass of the dark matter is unknown, so an experimental search should involve probing for pseudomagnetic fields in a range of frequencies, and compared with resonant-cavity experiments, spin haloscopes can cover a broad mass range of dark matter using broad-band methods \cite{Lee:2022vvb,Bloch2023NASDUCK_SERF,Wei:2023rzs} or by tuning the Larmor frequency \cite{jiang2021search,xu2023_ChangE} without changing the apparatus dimensions. In the scanning approach, the comagnetometer operates at a specific external magnetic field for a designated acquisition time to probe for axions at the NMR resonance frequency corresponding to the chosen value of the field. 
Once the desired sensitivity is achieved at this working point, the external magnetic field is adjusted to a different value to search for the next possible dark matter mass.
Typically, each single measurement can cover a dark matter mass range approximately equal to the NMR linewidth with similar sensitivity. By selecting a scan step size comparable to or narrower than the NMR linewidth, it is possible to continuously cover the entire range of dark matter masses within the target range without any gaps.

% \textcolor{red}{$>>>>>>>>\text{changes}>>>>>>>>>>>>>>>>>>>>>>>>>>>>>>>>>>>>>>>>>>>>>>$}

% \textcolor{red}{$>>>>>>>>>>>>>>>>>>>>>>>>>>>>>>>>>>>>>>>>>>>>>>>>>>>>>>>>>>>>>>>$}

%\jl{$\bullet$ Acquisition time comparing with coherence time and relaxation time.} 
The spin evolution in the presence the oscillating (pseudo-)magnetic field can be evaluated by solving the coupled Bloch equations. The relevant parameters are the acquisition time \(\Delta T\), axion coherence time \(\tau_{a}\), and spin transverse relaxation time \(T_{2}\). 
The axion coherence time is inversely proportional to the axion mass or Compton frequency. For experiments targeting low-frequency axions ($\nu_a\ll\SI{1}{\kHz}$ and $\tau_a\gg\SI{1}{\s}$), $\Delta T$ may not be much longer than $\tau_a$. In contrast, at higher frequencies ($\nu_a\gg\SI{1}{\kHz}$ and $\tau_a\ll\SI{1}{\s}$), $\Delta T$ can exceed $\tau_a$. The coherence time $\tau_a$, together with $T_2$, determines the linewidth of the axion signal, $\Delta\nu_\mathrm{sig}\approx \min(\tau_a^{-1}, T_2^{-1})$. 
Meanwhile, the RBW of the frequency spectrum is given by $\Delta T^{-1}$. 
Hence, we briefly discuss two extreme cases: (1) \(\Delta T\ll\tau_{a},\,T_2\), i.e., $\mathrm{RBW}\ll\Delta\nu_\mathrm{sig}$; (2) \(\Delta T\gg\tau_{a},\,T_2\), i.e., $\mathrm{RBW}\gg\Delta\nu_\mathrm{sig}$. 
These two cases represent whether the axion signal falls into a ``single bin'' or ``multiple bins'' in the frequency spectrum, respectively.

The axion signal size in the power spectrum scales with the square of the spin tipping angle. 
% \sout{We analyze the effects of axion-field decoherence, $T_2$ relaxation, and static (bias) magnetic field inhomogeneity. }
For $\Delta T\ll T_2$ or $\tau_a$, the orientation of spins is tilted from the initial orientation by an angle $\Delta \theta=\Omega_a \Delta T$, where $\Omega_a$ is the Rabi frequency corresponding to the axion pseudomagnetic field. If the measurement time is long ($\Delta T\gg\tau_{a},\, T_{2}$), we need to take decoherence or relaxation into account. The tipping angle reaches the steady state where $\Delta \theta \approx \Omega_a T_2$ when $T_2\ll \tau_a,\,\Delta T$. However, when $T_2\gg \tau_a $, the orientation of spins effectively undergoes a random walk due to the stochastic nature of the axion field\,\cite{Centers2021Stochastic}, yielding $\Delta \theta \approx \Omega_a \sqrt{\tau_a \,T_2}$. 
Hence, the root-mean-square of $\Delta \theta$ is
\begin{equation}
    \Delta \theta \approx  \Omega_a\mathrm{min}(\Delta T,\,T_2,\,\sqrt{\tau_a T_2})\,. 
    \label{eq:tipping_angle}
\end{equation}
Another potentially critical property is the static-field spatial inhomogeneity. 
When the inhomogeneity is so significant that the NMR signal in the spectrum is consequently broadened [linewidth $\gg 1/(\pi T_2)$], we should consider the fraction of spins on-resonance with the axion field gradient when calculating the axion signal power. 

For some NMR haloscopes, such as CASPEr, the scan ranges cover both low and high frequencies (sub-\unit{kHz} to $\sim$\SI{1}{\GHz}). The data analysis and hypothesis testing procedures for high-frequency measurements are similar to those used in cavity haloscopes, as discussed in Sec.\,\ref{sec:CavityHaloscopes}. On the other hand, at low frequencies, the axion-induced signal is contained in a few or a single frequency bin.
Conducting extended measurements to achieve frequency resolution finer than the signal width is not always practical. 
In the following text, we focus on the low-frequency (sub-\unit{kHz}) case, typically $\tau_a, T_2\gtrsim \Delta T$.
The intermediate-frequency case is analyzed by \citet{Yuzhe2023_FrequencyScanning}. 

%\jl{$\bullet$ Axion signal detection.} 
For an NMR search for an axion dark matter signal, we start by considering the local dark matter velocity, \(v \sim 10^{-3}c\), which leads to a fractional energy spread on the order of \((v/c)^2 \sim 10^{-6}\). Consequently, the signal appears nearly monochromatic in the frequency domain. Given that the noise is a continuous broadband background, focusing on a narrow frequency band to search for axions allows us to treat this background as random Gaussian white noise. %\sout{as constant with minor fluctuations}. 
For simplicity, we assume that the entire signal would fall into a single frequency bin.  %\jl{revised.}

If the axion dark matter signal is sufficiently strong, it will manifest as a distinct peak in the frequency bin corresponding to the axion mass, standing out against the relatively flat background noise. In the absence of a visible peak, we can employ statistical methods to set limits on the signal strength. In these statistical calculations, we use the measured power as the physical observable, which characterizes the mean signal and background, accounting for their fluctuations based on their probability distributions. This method ensures that, without a visible peak, the data can provide constraints on the potential presence of an axion dark matter signal.

In the analysis, we consider the background as random Gaussian white noise with zero mean\,\cite{Lee:2022vvb}. The noise in the experiment can be subdivided into two components. One is the magnetic noise that can be amplified with a spin amplifier. The other is non-amplifiable noise, e.g., photon shot noise. 
% Given the characteristics of random Gaussian white noise, the mean of the background is zero, but the variance is nonzero. 
Since the two contributions to the random background are independent of each other, their variances can be added together to obtain the total background variance, which is expressed as
\begin{align}
    \Sigma_{\rm bkg}(\nu) = \Sigma_{\rm nam} + \Sigma_{\rm am} \,\eta_{\rm F}^2(\nu)\,,
    \label{eq:total-bkg-whiteGaussian}
\end{align}
where \(\Sigma_{\rm nam}\) and \(\Sigma_{\rm am}\) denote the mean PSD (expressed in the units of $\mathrm{V}^2$/Hz) of the non-amplifiable and amplifiable noises, and $\eta_{\rm F}(\nu)$ is the amplification function. 
% {$\eta_{\rm F}^2$ denotes the amplification? }
% \new{For some NMR haloscopes, the background noise is mostly frequency-independent. Take NMR spectrometers utilizing DC superconducting quantum interference devices (DC-SQUIDs) as an example. The SQUID noise is approximately $\Sigma_{\rm bkg} = \SI{1e-12}{\Phi_0^2/\Hz^{1/2}}$ over the working frequencies from $\sim\SI{1}{\Hz}$ to \SI{4}{\MHz}.} 
In alkali-noble-gas NMR experiments (such as \citet{Jiang:2023nan, Xu:2023vfn}), due to the broadband nonresonant response amplitude from alkali spins and the resonant response amplitude from noble-gas spins, there is Fano interference between these two amplitudes. This interference leads to an asymmetric lineshape \(\eta_{\rm F}(\nu)\) in the spectral profile\,\cite{Jiang:2021fkw, Bloch:2021vnn}. However, when the amplification factor $G_\mathrm{NMR}$ at the resonant frequency is high, the lineshape simplifies to a Lorentzian function, resulting in \(\eta_{\rm F}(\nu) \to 1 + \eta(\nu) \)\,\cite{Xu:2023vfn}. The function \(\eta(\nu)\) is related to the Lorentzian in Eq.\,\eqref{eq:LorentzianFunction}:
% \begin{align}
% \eta(f) = \frac{G_\mathrm{NMR}}{\sqrt{1+ (2\pi)^2(f - f_{\rm res})^2 / \Gamma_{\rm NMR}^2}}\,,    
% \end{align}
\begin{align}
\eta^2(\nu) = G_\mathrm{NMR}^2L(\nu/\nu_c,\nu_c/\delta \nu_{\rm NMR})\,,   
\label{eq:eta-frequency}
\end{align}
where \(\delta \nu_{\rm NMR}\) is the NMR full width at half maximum (FWHM) in cyclic frequency units, and \(\nu_{c}\) is the nucleon Larmor resonant frequency. Here $G_\mathrm{NMR}=\tilde{b}_{\rm eff}\gamma_{\rm n}/(2\pi\delta \nu_{\rm NMR})$ is the amplification factor for the effective magnetic field due to the Fermi-contact interaction between alkali-atom electrons and the noble-gas nuclei and $\tilde{b}_{\rm eff}=\lambda M^{\rm n}P_0^{\rm n}$\,\cite{jiang2021search, jiang2021floquet, Xu:2023vfn}. 

If we measure the axion pseudomagnetic field for a time shorter than the coherence time and take the Fourier transform of the signal, we will see a signal near the Compton frequency with a certain amplitude. Now, repeating the measurement at a time many coherence times later, we should see a signal at the same frequency but with a different amplitude (and possibly, different sign).
Due to the stochastic nature of UBDM, the phase of the axion field is described by a flat distribution from 0 to $2\pi$.
The amplitudes measured this way follow a normal distribution with a zero mean. In this sense, the axion signal is, in a sense, also like-noise, stemming from the distribution of amplitudes to the PSD, one now finds that the power follows the $\chi^2$ distribution\,\cite{Lee:2022vvb}. This is because the power of a signal consisting of two independent components of equal average power (here, the two quadratures of the signal) follows the $\chi^2$ distribution\,\cite{ross2009introduction}.

The mean power spectral density (PSD) of the axion dark matter signal can be parameterized as  
\begin{align}
    \Sigma_{\rm sig}(\nu) = g_{\rm sig}^2(\nu) \times \eta^2(\nu) \,, 
\end{align}
where $g_{\rm sig}$ is the phenomenological coupling that represents the axion signal strength and encodes all the axion dark matter information like the axion couplings, DM distribution, etc. 
Since the axion gradient field acts as an oscillating external magnetic field affecting the nuclear spin, it also gets amplified by the NMR resonance and is thus multiplied by $\eta^2(\nu)$. 

%\jl{$\bullet$ Checking if there is a signal} 
The power in each bin of the observed power-spectral density $\Sigma_{\rm obs}$ follows a $\chi^2$ distribution with two degrees of freedom, which is equivalent to an exponential distribution. If the measurement time is much longer than the axion coherence time, as is typically the case for cavity haloscopes Sec.\,\ref{sec:Cavity_frequentist}, the spectrum is averaged over many independent realizations.  The resulting statistics are well approximated by a Gaussian distribution.

Similar to Sec.\,\ref{sec:Cavity_frequentist} discussing cavities, frequentist hypothesis testing is adopted. As described in Sec.\,\ref{sec:Cavity_frequentist}, test statistics are required to conduct a hypothesis test, and they can be modeled using a probability distribution function. For spin haloscope, since the observed power-spectral density $\Sigma_{\rm{obs}}$ follows the exponential distribution, the normalized power spectrum density with and without axion signal ($\Sigma_{\rm{sig}}$) under given noise ($\Sigma_{\rm{bkg}}$) also follows the exponential distribution. Their distribution functions are 
\begin{align}
&\,\,\,\mathcal{P}_0(x;H_{0})=\frac{1}{c_0+1}{\rm Exp}\left(-\frac{x}{c_0 +1}\right)\,,\\
&\,\,\,\mathcal{P}_1(x;H_{1})={\rm Exp}\left(-x\right),
\end{align}
where $x=\Sigma_{\rm{obs}}/\Sigma_{\rm{bkg}}$ is the normalized power-spectral density and $c_{0}\equiv\Sigma_{\rm{sig}}/\Sigma_{\rm{bkg}}$ is the target signal strength normalized by $\Sigma_{\rm{bkg}}$. Since the test statistic depends on the analysis procedure, we can see that the parameters of the probability distribution function are different from the cavity-haloscope case (normalized power excess). The Type I error [see Eq.\,\eqref{eq:typ1error}] is chosen as  0.1\,(0.05), see Sec.\,\ref{sec:Cavity_frequentist}, representing the probability of falsely rejecting the null hypothesis of $10\%\, (5\%)$.
If the measured $\Sigma_{\rm obs}$ is large enough that the probability [type II error \eqref{eq:typ2error} $\beta<0.01\%$] for noise alone to fluctuate to this value is small, the alternative hypothesis ($H_1$) is rejected.
%\YG{In statistics terminology, we say "the null hypothesis ($H_{0}$) is not rejected". In addition, "how large"?}
If no excess exceeds the 90\%\,(95\%) threshold of $H_0$, which means that no significant axion signal is observed, the null hypothesis is rejected. Then, we may exclude the axion parameter space with the confidence level that we set. 

If no excess is observed and $\Sigma_{\rm obs}\approx \Sigma_{\rm bkg}$, one may derive the upper limit on the signal power at the 90\% confidence level for a single frequency bin analogously to Eq.\,\eqref{eq:typ1error} and Eq.\,\eqref{eq:typ2error} in Sec.\,\ref{sec:Cavity_frequentist}, but the integration range starts from zero since the difference of probability distribution function:
\begin{align}
 \alpha = &\int^{\Lambda_{c}}_{0}\mathcal{P}_0(x;H_{0}) dx\,,\label{eq:c0_calc}
\end{align}
\begin{align} 
\beta=&\int_{\Lambda_{c}}^{\infty}\mathcal{P}_1(x;H_{1})dx \,,
\end{align}
with $\Lambda_c$ calculated from Eq.\,\eqref{eq:c0_calc} to be $\Lambda_c=(1+c_{0}){\rm log} (\frac{1}{1-\alpha})$. For $\beta=0.01\%$  and  $\alpha=0.1$ we get $\Lambda_c=9.21$ and $c_0\approx 86.41$.

\begin{figure}[h]
\includegraphics[width=0.6\linewidth]{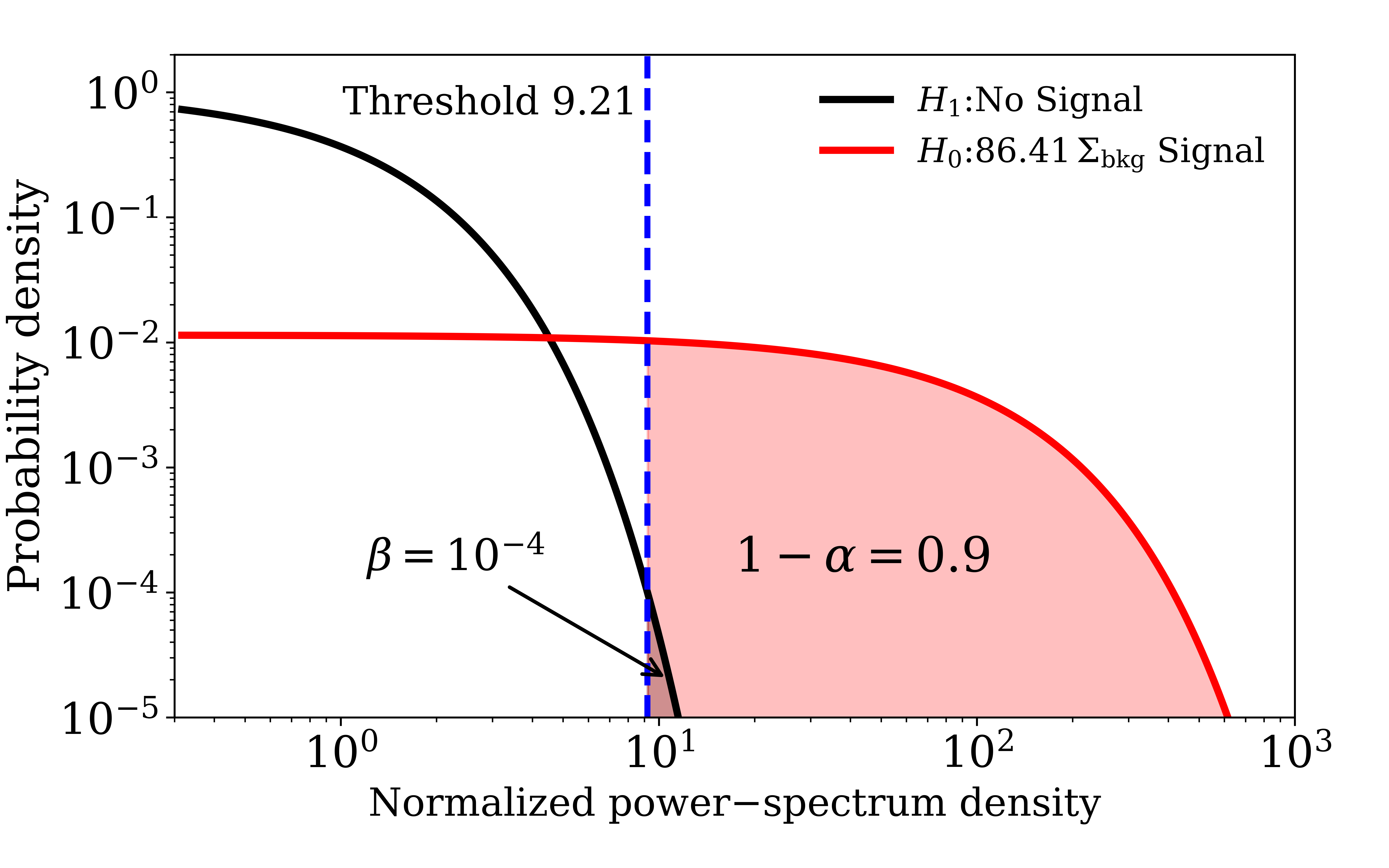}
\caption{Distribution of null hypothesis (red) and alternative hypothesis (black). The confidence level is set to 90\%. 
% \DB{I propose a very detailed explanation here in the caption of what is going on in this picture. Do not worry about the length. }
For $\Sigma_{\rm obs}$ larger than the threshold, $H_0$ is accepted. For the normalized power-spectrum density  smaller than the threshold, $H_0$ is rejected, and accept alternative hypothesis $H_{1}$. This plot is similar to Fig.~\ref{fig:cavity_frequntist}, but the test statistics differs since the number of average is limited to 1 due to the measurement time being smaller than the axion coherence time.} 
 \label{fig:spin_frequntist}
\end{figure}

The above is illustrated in Fig.\,\ref{fig:spin_frequntist} , with the black and red curves representing the alternative and the null hypothesis, respectively.  If $\Sigma_{\rm obs}/\Sigma_{\rm{bkg}}$ is equal or larger than the threshold $\Lambda_{c}$ (the vertical dashed line), we take additional data at that frequency and check whether it originates from a real axion signal or is just a statistical fluctuation. Similarly, for the null hypothesis with large signal strength like for the red curve, the probability of  $\Sigma_{\rm obs}/\Sigma_{\rm{bkg}}$ being smaller than the threshold is less than $10\%$; thus the null hypothesis may be falsely rejected with this (small) probability. 

\begin{figure}[h]
    \includegraphics[width=0.8\linewidth]{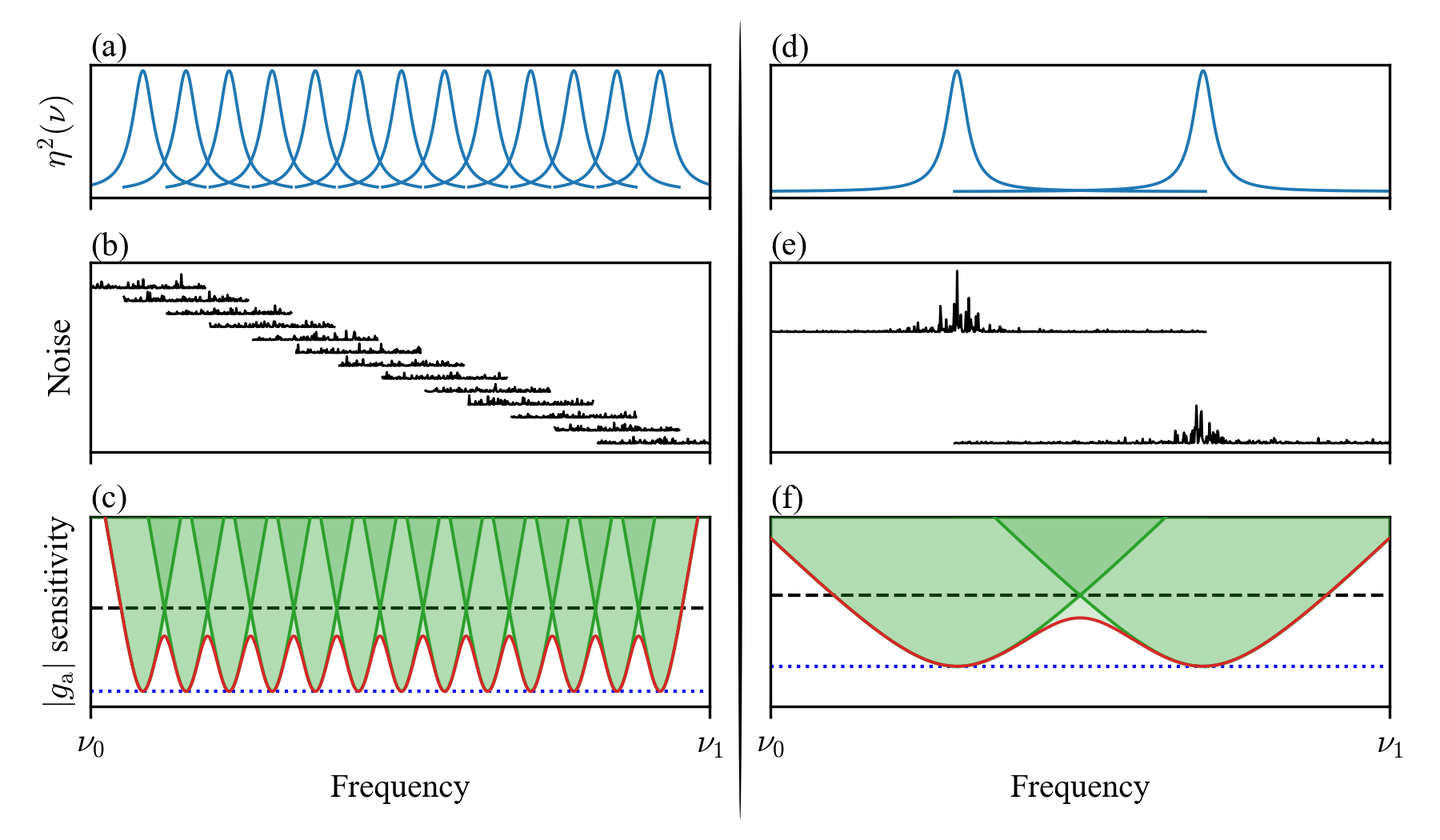}
    \caption{Examples of experiments scanning over the frequency range $[\nu_0,\,\nu_1]$. 
    % Left column: non-amplifiable noise dominating. Right column: amplifiable noise dominating. Rows from top to bottom: frequency spectra of amplification $\eta^2(\nu)$, PSD noise, and corresponding $g_\mathrm{a}$ sensitivity. 
    Left column: Dominance of non-amplifiable noise ($\kappa\ll1$).
    Right column: Dominance of amplifiable noise ($\kappa\gg1$).
    Rows (top to bottom): Frequency spectra of the square of amplification $\eta(\nu)$, PSD noise, and the corresponding $g_\mathrm{a}$ sensitivity. Offsets have been added to the PSD noise spectra for clarity.
    In panels (c) and (f), the blue dotted line represents the best sensitivity $\bar{g}_{\rm sig}^{\rm best}$, the black dashed line represents $2\bar{g}_{\rm sig}^{\rm best}$, and the red line shows the combined sensitivity of individual steps.
    The scan-step size is larger when amplifiable noise dominates due to the broader sensitivity bandwidth. } 
     \label{fig:flat_peak}
\end{figure}

Following Eq.\,\eqref{eq:c0_calc}, a limit $\bar{g}_{\rm sig}$ on the signal strength as a function of DM frequency $\nu$ can be derived as:
\begin{align}
    \bar{g}_{\rm sig}^2 (\nu) =  c_0 \left[ \Sigma_{\rm am}
    + \frac{\Sigma_{\rm nam}}{G_\mathrm{NMR}^2} \left(1+ \frac{4(\nu-\nu_c)^2}{\delta\nu_{\rm NMR}^2} \right) \right],
    \label{eq:coupling-gsig}
\end{align}
where we have utilized $G_\mathrm{NMR} \gg 1$.
% and  the fact $\Sigma_{\rm nam} \gg \Sigma_{\rm am}$.
It is reasonable that the best limit $\bar{g}_{\rm sig}^{\rm best}$ is achieved at the resonant frequency $\nu= \nu_c$, which leads to 
\begin{align}
    \bar{g}_{\rm sig}^{\rm best } {}^2 (\nu) =  c_0 \left[ \Sigma_{\rm am}
    + \frac{\Sigma_{\rm nam}}{G_\mathrm{NMR}^2}  \right] \equiv c_0 \Sigma_{\rm am} \left(1 + \kappa^{-1} \right),
    \label{eq:coupling-gsig-best}
\end{align}
where we define
\begin{align}
    \kappa \equiv G_\mathrm{NMR}^2 \Sigma_{\rm am}/\Sigma_{\rm nam}\,,
    \label{eq:kappa-definition}
\end{align}
which is a characteristic quantity for NMR experiments. Typical NMR experiments have $\kappa \lesssim 1 $, which means the background noise is dominated by non-amplifiable noise\,\cite{Jiang:2021dby, Bloch:2021vnn, budker2014proposal}, called the flat regime. For $\kappa \gg 1$, the background noise is dominated by the amplifiable noise\,\cite{Xu:2023vfn} and called the peaked regime.
Examples of these two scenarios are plotted in Fig.\,\ref{fig:flat_peak}. 

%\jl{$\bullet$ The sensitive bandwidth for the signal coupling in a single measurement.} 
In addition, we need to know what the bandwidth is within which the experiment has a 
similar sensitivity as that on resonance. We can define a sensitivity full width at half maximum, $\delta \nu_{\rm FWHM}^{\rm sen}$, where the limit on the coupling increases at most by a factor of 2, $\bar{g}_{\rm sig} (\nu_{\rm res} \pm  \delta\nu_{\rm FWHM}^{\rm sen} /2 ) = 2 \bar{g}_{\rm sig}^{\rm best}$. Using Eqs.\,\eqref{eq:coupling-gsig} and \eqref{eq:coupling-gsig-best}, this leads to

\begin{align}
   \displaystyle \delta \nu_{\rm FWHM}^{\rm sen}&=\delta \nu_{\rm NMR}\left[\frac{G_\mathrm{NMR}^2}{\Sigma_{\rm nam}}\left(3\Sigma_{\rm am}+4\Sigma_{\rm nam}/G_\mathrm{NMR}^2\right)-1\right]^{1/2}\nonumber\\
    &= \sqrt{3\left( 1+ \kappa \right)} \delta \nu_{\rm NMR}\,.
    \label{eq:NMR-acce-step}
\end{align}

%\jl{$\bullet$ How wide is the sensitive frequency bandwidth.}
We see that the parameter $\kappa$ affects the sensitivity bandwidth of NMR experiments. For $\kappa \ll 1$, the sensitivity bandwidth is roughly the same as the NMR bandwidth. For $\kappa \gg 1$, the sensitivity bandwidth a factor of $\sqrt{\kappa}$ larger compared to the NMR bandwidth. Such comparison is illustrated in Fig.\,\ref{fig:flat_peak}. 
%\jl{jl: The simplified calculation for NMR is presented above.}
%To conclude, in this section, we discussed the sensitivity and the sensitivity bandwidth for a single-bin search with a spin amplifier.\YG{Arne: Figure 10. includes a lot of bins}

\subsubsection{The scanning speed}\label{sec:SpinHaloscopes:ScanningSpeed}

%\jl{$\bullet$ The assumption for spin experiment scanning speed calculation.}
Similarly to the scanning speed in cavity experiments\,\cite{Krauss:1985ub, Kim:2020_Revisiting_haloscope} (see Sec.\,\ref{sec:CavityHaloscopes:ScanningStrategy}), we can derive a corresponding optimal scanning speed for the NMR experiment. For example, CASPEr-gradient experiment\,\cite{CASPEr-LF-2025} has adopted some of the approaches developed for the cavity based haloscopes\,\cite{Gramolin2022_SpectralPhysRevD}. Since in low-frequency NMR experiments, the measurement time $\Delta T$ is shorter or comparable to the axion coherence time $\Delta T \lesssim \tau_a$ and longer than the spin coherence time $T_{2}$, the axion signal mostly falls into a single bin in the frequency spectrum, simplifying the analysis.\footnote{Note that below axion frequencies of 11\,Hz, a sideband pattern becomes resolved in the spectrum due to the Earth rotation at 11.6\,$\mu$Hz.}

%\jl{$\bullet$ The pseudomagnetic field of signal.}
 Axion dark matter induces an external pseudomagnetic field component of the apparatus 
\begin{align}
    \mathscr{B}_{\rm axion} = \frac{g_{\rm aNN}}{\mu_{\rm N}} {\bm \nabla}a (t) \cdot \hat{\textbf{m}}(t)\,,
\end{align}
where $t$ is the time of the measurement and $\mu_{\rm N}$ is the magnetic dipole moment of the nucleus. For a simplified analysis, we disregard the sidereal motion of the Earth and consider a sensitive axis $\hat{\mathbf{m}}$, which is perpendicular to the pump and probe beam\,\cite{Lee:2022vvb}. 
Measurements are taken with a sampling time $\Delta t$, so that $t=n \Delta t$, where $n$ is an integer ranging from $0$ to $N-1$, with $ N = \Delta T/\Delta t$. 
% \sout{and $ \Delta T$ being the total duration of the measurement}\textcolor{orange}{$\Delta T$ is explained previously}.

%\jl{$\bullet$ Measuring pseudomagnetic field and its PSD.}
In the NMR experiments, we measure the magnetic field or the pseudomagnetic field $\mathcal{B}$ in units of [fT] in a time series $\mathcal{B}_n\,[\textrm{fT}]=\left\{\mathcal{B}_0,..\mathcal{B}_{N-1} \right\}$. The PSD we characterize here in the units of $[\textrm{fT}^2/\text{Hz}]$ for a given time series $\mathcal{B}_n\,[{\rm fT}]$ is
\\
\begin{align}
&P_k =\frac{1}{f_s N}\left|\tilde{\mathcal{B}}_k\right|^2=\frac{\Delta t^2}{\Delta T}\left|\tilde{\mathcal{B}}_k\right|^2, 
    \quad\nonumber\\
   &\tilde{\mathcal{B}}_k \equiv \sum_{n=0}^{N-1} \mathcal{B}_n \exp^{-i\omega_k n \Delta t} , \quad 
   \omega_k \equiv \frac{2\pi k}{N\Delta t}\,,
\end{align}
where $\Delta t=1/f_s$, with $f_s$ being the sampling rate, satisfying $f_s  \Delta T=N$. $\tilde{\mathcal{B}}_k$ is the Fourier component of the pseudomagnetic field in the frequency space, with $k$ being an integer from $0$ to $N-1$.
The complex Fourier transformed readout $\tilde{\beta}_k$ can be separated to two real quantities for real and imaginary parts\,\cite{Lee:2022vvb} 
\begin{align}
    R_k \equiv \frac{2}{N} \Re \left[ \tilde{\mathcal{B}}_k \right], \quad 
    I_k \equiv \frac{2}{N} \Im \left[ \tilde{\mathcal{B}}_k \right].
\end{align}

The fluctuations of pseudomagnetic field arising due to the stochastic nature of dark matter are described by the variances, from which one can form covariance coefficients $\sigma(R_k, R_r)$, $\sigma(R_k, I_r)$, $\sigma(I_k, R_r)$ and $\sigma(I_k, I_r)$, explicitly given in \citet{Lee:2022vvb}, with $k$ and $r$ indicating frequency-bin indices.

In the simplified treatment where the measurement duration is shorter than the coherence time of the axion, the signal is contained in a single, e.g. the $k$-th, frequency bin. In this case, the covariance between the different frequency bins vanishes, 
% (\textcolor{red}{$\sigma(R_k/I_k, R_r/I_r) = 0$}, which is more clear?) 
$\sigma(R_k/I_k, R_r/I_r) = 0$ when $r \neq k$.
Moreover, the covariance matrices between the real and imaginary parts of the axion signal, $\sigma(R_k, I_k)$ and $\sigma(I_k, R_k)$, also vanish\,\cite{Lee:2022vvb} because of the uniformly distributed random phases.
For a particular sensitive axis $\hat{\mathbf{m}}$, the variances for the real and imaginary components can be simplified to
\begin{align}
\hat{\sigma}_k \equiv {\sigma}(R_k,R_k)={\sigma}(I_k,I_k)= 
\pi \rho_\text{DM}\left(\frac{g_{\rm aNN}}{\mu_{\rm N}}\right)^2\int_0^\infty dv f_{\rm DM}(v) \epsilon_2^2(v) v^4 {\rm sinc}^2\left[\frac{\Delta T(\omega(v)-\omega_k)}{2}\right], 
\label{eq:signalvariance}
\end{align}
where the angular frequency is $\omega(v) \equiv m_a (1+v^2/2)$. The function $f_{\rm DM}$ describes a reduced velocity distribution and the function $\epsilon_{2, v}$ describes the effects of the sensitive axis, which can be found in Eq.\,B13 and B20, respectively, in the paper by \citet{Lee:2022vvb}.
%where $2$ means the sensitive direction $\hat{\mathbf{m}}=\left\{0,1,0\right\}$ and we use continuous velocity $v$ in the integration instead of discrete summation of the velocity $\beta_j$\,\cite{Lee:2022vvb}.
%\YG{What does $\epsilon_{2}$ describe?}

The simplified results above coincide with those of the NASDUCK collaboration \cite{Bloch:2021vnn}.
%readout of the  signal $\sigma(R_k, I_k)$ and $\sigma(I_k, R_k)$ vanishes. 
If we consider a much longer measurement time such that the signal spreads to several frequency bins and include the sidereal motion of the Earth, the covariance among different frequency bins terms, $\sigma(R/I_k, R/I_r)$ with $r \neq k$ are nonzero in general\,\cite{Wei:2023rzs, Xu:2023vfn}.

%\jl{$\bullet$ PSD of background noise.}
Next, we need to calculate the average PSD of the axion signal $P^{\rm sig}$ and compare it with the average (background) PSD of the measured raw data ($P^{\rm bkg}$), which is in units of [\({\rm V^2/Hz}\)]. Since the background is random Gaussian white noise, the background noise PSD can be described by its Gaussian variance as [see Eq.\,\eqref{eq:total-bkg-whiteGaussian}]
\begin{align}
    P^{\rm bkg}(\nu) = \Sigma_{\rm bkg }(\nu) g^2_\text{cali}(\nu) \approx \Sigma_{\rm am } \eta^2_F(\nu) g^2_\text{cali}(\nu)\,,
    \label{eq:Pbkg-approximate}
\end{align}
where \(g_\text{cali}(\nu)\) is the calibration function that transforms the magnetic field strength \([{\rm fT}]\) to the readout voltage response \([\mathrm{V}]\), and \(\Sigma_{\rm am}\) is the PSD variance of the white Gaussian magnetic noise in units of \([{\rm fT}^2/{\rm Hz}]\) in the frequency domain. In the second equality in \eqref{eq:Pbkg-approximate}, we have assumed the dominance of magnetic noise in Eq.\,\eqref{eq:total-bkg-whiteGaussian}.

%\jl{$\bullet$ Averaged PSD of signal.}
The average PSD of the axion signal, is given as\,\cite{Xu:2023vfn}  
\begin{align}
P^{\rm sig}_\nu&= 2\Delta T \eta^2(\nu) g^2_\text{cali}(\nu) \hat{\sigma}_{k}\,,
\end{align}
in the units of $[{\rm V^2/Hz}]$. The factor of 2 comes from the fact that the real and imaginary parts of the variance are the same, and the time $\Delta 
T$ originates from translating the signal power to PSD. Since the axion signal is stochastic, the expectation of the measured average PSD for the signal is also described by its covariance. Therefore, the PSD is proportional to the signal covariance $\hat{\sigma}_{k}$ defined in Eq.\,\eqref{eq:signalvariance}. Here, \(k\) denotes the frequency bin index for the frequency \(f\), \(\eta(\nu)\) represents the unitless Lorentzian amplification function for signal.

Hence, the SNR is calculated as the ratio of the signal PSD  to the noise PSD in the resonant frequency bin ($\nu$):
\begin{align}
% {\rm SNR}\equiv \frac{P^{\rm sig}_f}{P^{\rm bkg}_f}=\frac{1}{2}\frac{ T \eta(f)^2 g_\text{cali}(f)^2 {\sigma}(R_{[f]},R_{[f]}) }{P_{\rm bkg}  } \approx \frac{1}{2}\frac{ T  g_\text{cali}(f)^2 {\sigma}(R_{[f]},R_{[f]}) }{P_{\rm bkg}/G_\mathrm{NMR}^2  }\,.
{\rm SNR}\equiv \frac{P^{\rm sig}_{\nu}}{P^{\rm bkg}_{\nu}}=\frac{ 2 \Delta T \eta^2(\nu) g^2_\text{cali}(\nu) \hat{\sigma}_{k}}{P^{\rm bkg}(\nu)  } \approx \frac{ 2 \Delta T \hat{\sigma}_{k} }{\Sigma_{\rm am}}\,,
\label{eq:snr}
\end{align}
where in the last approximate equality, we have used the assumptions $G_\mathrm{NMR} \gg 1$ and $\kappa \gg 1$ [see Eq.\,\eqref{eq:kappa-definition}].

%\jl{$\bullet$ Derivation of scan speed.}
Since we have taken the scan step as $\Delta f=\delta\nu_{\rm FWHM}^{\rm sen}$ and the measurement time for a single scan $\Delta t = \Delta T$, we obtain the scanning speed by solving for $\Delta T$ in Eq.\,\eqref{eq:snr} with a desired ${\rm SNR}$,
\begin{align}
& \frac{d\nu}{dt}\approx \frac{\Delta \nu}{\Delta t}= \frac{\delta \nu_{\rm FWHM}^{\rm sen}}{\Delta T}= \sqrt{1+\kappa} \times \sqrt{3} \delta\nu_{\rm NMR}   \frac{2 \hat{\sigma}_{k}}{{\rm SNR} \,\Sigma_{\rm am}} \,,
\label{eq:spin Scanning speed}
\end{align}
where we made use of  Eq.\,\eqref{eq:NMR-acce-step} in the last equality. As a result, for the magnetic noise dominated NMR experiment, $\kappa \gg 1$, the scan speed is accelerated by a factor of $\sqrt{1 + \kappa}$ because of the increase in the sensitivity bandwidth. 
%\xlm{Hi Alex, this section of scan speed amplification is not made by enlarging the $T_2^{*}$ to increase the FWHM of the Lorenzian response of the magnetic field. We just simply point out that if the ratio of magnetic noise level and white noise level $\kappa = \Sigma_{a}/\Sigma_{\rm na}$ could be larger, we will end up with a larger bandwidth, and this relation holds true without tuning the resonant bandwidth. I think you also pointed out in later comments that it is similar to Kent and Graham's paper.}
% \YG{It is good place to remind that it is hard to average to obtain $\sqrt{N}\propto \sqrt{\Delta t}$ due to $T<\tau_{a}$. This arises a difference from cavity haloscope. Cavity case, $\Delta t\neq T$.}
This equation does not imply that increasing NMR linewidth is a way to boost the scanning speed because this also degrades the SNR. It rather prescribes the scanning speed once the NMR parameters have been fixed.

% \jl{There is some inconsistency in the calculation above, because $P_{\rm bkg} = T \Sigma_{\rm bkg}$. The measurement time $T$ gets canceled. The logic broke down here.}

%\jl{$\bullet$ Other methods to enhancing scanning speed.}
There are alternative methods for enhancing scanning and sensitivity discussed by \citet{ Dror:2022xpi, Yuzhe2023_FrequencyScanning} for spin-precession experiments, as well as by \citet{Malnou:2018dxn, Chaudhuri:2018rqn, Chaudhuri:2021, Chen:2021bgy} for cavity experiments. For example, \citet{Yuzhe2023_FrequencyScanning} discusses how sensitivities to axion couplings depend on the dwell time of the measurement at each step under various experimental parameters. 
The conclusion is that, when using the area of the accessible parameter space in log-log coordinates as the FOM, the optimal strategy is to operate at the maximum nuclear spin relaxation times and to choose the largest step size, whether determined by the NMR or axion linewidth.
We emphasized that the approach used by \citet{Xu:2023vfn} is compatible with the aforementioned optimal strategy but employs a significantly larger step size beyond the physical widths mentioned, due to the enhanced bandwidth of the coupling sensitivity. 
%\DB{We need to read and think about this paragraph on the next path}
%%%

\subsection{Storage-ring based searches}
% \section{Storage-ring based searches}
\label{sec:StorageRing}
Storage rings, which confine charged particles through a pure magnetic field or a combination of electric and magnetic fields, are also being used for axion searches. A variety of the confined particles—such as electrons, muons, protons, deuterons, or heavier ions—can be used to search for axion-induced effects on spins. Such effects could arise from gradient (axion-fermion) coupling and the effective electric dipole moment (EDM) coupling arising from the axion coupling to gluons. 
%\sout{Axion-like particles (ALPs)  can couple to Standard Model (SM) fermions in two primary ways: via a gradient (axion-fermion) coupling and via an effective electric dipole moment (EDM) coupling. }
The relevant Hamiltonian interaction terms are expressed as Eq.\,\eqref{eq:NMRHaloscopeHamiltonian}. The \(g_{aNN}\,{\bm \nabla} a \cdot \mathbf{I}_{N}\) term is the axion-fermion or axion-nucleon coupling, and the \(g_{d}\,a\,\mathbf{I}_{N}\cdot \mathbf{E}\) term arises from the axion EDM coupling. It manifests as an oscillatory EDM.

% \sout{A variety of charged particles in a storage ring—such as electrons, muons, protons, or heavier ions—can be used to search for these ALP-induced effects on spin. }
In the laboratory frame, the effect of the two couplings of the axion can be represented as induced spin-precession angular frequencies:
\begin{align}
    \label{eq:omega_aNN}
    \bm{\omega}_{aNN} &\propto g_{aNN}\,\sqrt{2 \rho_\mathrm{DM}} \,\bm{v} \cos(m_a t)\,, \\
    \label{eq:omega_EDM}
    \bm{\omega}_{aN\gamma} &\propto d_{aN\gamma}(t)\,\bigl(\mathbf{E} + \bm{v}\times \mathbf{B}\bigr)
    \;=\;
    g_{d}\,\frac{\sqrt{2 \rho_\mathrm{DM}}}{m_a}\,\cos(m_a t)\,\bigl(\mathbf{E} + \bm{v}\times \mathbf{B}\bigr)\,,
\end{align}
where $d_{aN\gamma}$ is the oscillating electric dipole moment of the particle induced by the axion field, and \(\rho_\mathrm{DM}\) is the local dark matter density,
%\approx 0.45\,\mathrm{GeV/cm^3}\) 
%\YS{Every field uses their own traditional value for the local dark matter density. It turns out, that our value is the average from the most recent measurement of GAIA. We could say, ...and rho is the local dark matter density. Also, I agree, all parameters should be defined locally.} \YG{The numerical value is not used in this text. So, shall we move the detailed numerical value to the introduction part?}
%  is the local dark matter density, 
%\(\beta \equiv v/c\) {I can see three different $\beta$'s: antenna coupling, type II error, and here. } \YG{It is true that $\beta$ represents three different quantities, but it seems to be a standard symbol in each field. Therefore, it should be clearly specified when used in equations for each case.}
$\bm{v}$ is the velocity of the stored particle (often close to unity for a relativistic beam), and \(m_a\) is the axion mass. In Eq.\,\eqref{eq:omega_aNN}, the factor \(v/c \sim 1\) for relativistic beams leads to a significant enhancement in sensitivity compared to laboratory-sitting experiments where \(v/c \sim 10^{-3}\) for nonrelativistic DM\,\cite{Graham2021}. In Eq.\,\eqref{eq:omega_EDM}, for the canonical QCD axion making up 100\% of DM with $\rho_\text{DM} \approx 0.45\,\text{GeV/cm}^3$, the resulting oscillatory EDM can be as large as \(\sim 10^{-34}\,e\cdot\mathrm{cm}\)\, independent of the axion mass\,\cite{Graham:2013gfa}.

Physically, the gradient coupling in Eq.\,\eqref{eq:omega_aNN} produces a spin-precession angular frequency vector that oscillates at the axion frequency, alternating between directions parallel and anti-parallel to
\(\bm{v}\).\footnote{When we consider relativistic particles in a storage ring, we can neglect the relative velocity of the Earth and the local dark matter, which is on the order of $10^{-3}$. This means that the ``gradient'' interaction is dominated by the motion of the particles and not the gradient of the axion field in the laboratory frame. This is different from the other NMR experiments discussed in this paper.} The axion-induced EDM term likewise yields a precession vector that oscillates between the radially inward and outward directions \((\mathbf{E} + \bm{v}\times \mathbf{B})\). Hence, an initially in-plane spin is tipped and may oscillate 
above and below the ring plane in step with the axion field.

For hadronic beams, the polarization is measured with a polarimeter that
detects the left–right asymmetry in elastic scattering from a light nucleus, typically carbon. The differential cross section for a polarized beam can be written as\,\cite{Brantjes2012}
\begin{equation}
  \sigma(\theta)_{\text{pol}}
  \;=\;
  \sigma(\theta)_{\text{unp}}
  \bigl[\,1 + P\,A(\theta)\,\sin\psi\,\cos\phi \, \bigr],
\end{equation}
where \(\sigma(\theta)_{\text{unp}}\) is the cross section for unpolarized beam,
\(\theta\) is the laboratory polar scattering angle of the detected proton, \(P\) is the beam polarization, \(A(\theta)\) is the analyzing power (spin–orbit sensitivity of the reaction), and \((\psi,\phi)\) are the polar and azimuthal angles of the spin vector, with \(\phi\) measured from the normal to the scattering plane.

%\DB{We need to symmetrize the EDM interaction to the gradient interaction, how does $\sigma_{d}$ convert to gradient interaction}
The statistical uncertainty on the axion-induced EDM (or, equivalently, on the coupling of interest) scales as\,\cite{Anastassopoulos2016, Kim2021}
%\OK{Adjust the numerator according to the axion-EDM paper, Also look at Graham 2021 for the gradient coupling sensitivity.} \DB{Hss this comment been implemented?}
\begin{equation}
  \sigma_d
  \;=\;
  \frac{2 s\,\hbar}
       {P_0 \,A\, E^{*}\,
        \sqrt{\kappa\,N_c\,T_{\mathrm{exp}}\,\tau_p}},
  \label{eq:sigma_d}
\end{equation}
where \(s\) is the particle spin quantum number, \(P_0\) the initial polarization, \(A\) the analyzing power, \(E^{*}\) the equivalent electric field experienced in the ring, \(\kappa\) the polarimeter efficiency, \(N_c\) the number of stored particles per fill, \(T_{\mathrm{exp}}\) the total running time, and \(\tau_p\) the spin-coherence time.  Maximizing the product \(P_0 A \sqrt{\kappa N_c \tau_p}\) and achieving the longest feasible \(\tau_p\) are therefore central to reaching competitive sensitivity.

Storage ring experiments employ two main approaches to detect such precession: the \emph{frozen-spin configuration}\,\cite{Chang2019}, which is highly sensitive over a wide range of low frequencies, and the \emph{resonant method}\,\cite{Chang2019, Karanth2023, Kim2021}, which is sensitive to one frequency at a time.
%\YG{We may need references for each method} -> OK: Citation attached.
%If all possible frequencies must be scanned, the statistical sensitivity will be limited due to the limited time spent at each frequency.
The \emph{frozen-spin configuration} keeps the polarization approximately fixed relative to the momentum, so any small oscillatory effect can accumulate over time at sufficiently low perturbation frequencies, leading to a monotonic growth of out-of-plane polarization. 
% \YG{Reference for `frozen-spin configuration' needs} -> OK: Citation attached.
% \YG{We stopped here.}
% This method is well-suited for ultralight axions whose oscillation periods exceed the storage time. 
The statistical reach of a frozen-spin run is limited by the shorter of the axion coherence time \(\tau_a\) and the spin-coherence (depolarization) time of the stored beam \(\tau_p\): $T_{\text{int}} \;=\; \min(\tau_a, \tau_p).$ State-of-the-art and proposed rings can sustain polarization for $\tau_p \sim 10^{3\text{--}4}\,\mathrm{s}$\,\cite{Guidoboni2016}. Therefore, axions whose field decoheres on a shorter timescale lose sensitivity in proportion to the reduced integration time.  When \(\tau_a\) exceeds \(\tau_p\), the spin coherence sets the ultimate limit, and the frozen-spin method operates at full statistical power.

Alternatively, the \emph{resonant method} searches for the axion oscillation frequency near a narrow resonance, such as the anomalous spin-precession ($g-2$) frequency\,\cite{Chang2019,Karanth2023} or the frequency of an applied RF field\,\cite{Kim2021}. At resonance, even a feeble axion-induced precession is coherently driven, so the vertical spin component grows linearly for a time, recovering high sensitivity at higher axion masses.  
% Because the search bandwidth is only \(\Delta f \simeq f/ \rm{max}(Q_{a},Q_{s})\) and 
Because the carrier can be tuned only within the available $g-2$ or RF range, typically \(\mathcal{O}(\mathrm{kHz})\)–\(\mathcal{O}(\mathrm{MHz})\), scanning a broad mass range requires a long total integration time. Consequently, the technique is best suited for testing specific target frequencies and, if a signal is observed, for precisely determining the associated axion couplings once the mass (oscillation frequency) is known.

A practical scanning strategy was demonstrated by \citet{Karanth2023}: the beam momentum was adiabatically ramped with an RF cavity while the magnetic field of the ring was simultaneously ramped to keep the orbit closed, thus sweeping the $g-2$ frequency. This approach covered axion frequencies from 119.997\,kHz to 121.457\,kHz.

Because of the long interaction time and the possibility of parasitic operation, storage-ring searches can be sensitive to ultralight axions. They can probe masses from near the fuzzy DM limit (\(\sim 10^{-22}\,\mathrm{eV}\)) up to \(\mathcal{O}(10^{-8})\,\mathrm{eV}\), depending on the chosen resonance mode and detection frequency [often up to \(\mathcal{O}(\mathrm{MHz})\)]. 
% \DB{Please say, what this MHz is related to? Revolution period in the ring?} \OK{$g-2$ or RF frequency that drives the resonance in the resonant method.} 
A prime example is the proposed proton EDM experiment\,\cite{Omarov2022}, which primarily aims to improve the current sensitivity on CP violation in strong interactions (a.k.a. ``strong-CP'') by three orders of magnitude, reaching the sensitivity to new physics at the \(\sim 10^3\,\mathrm{TeV}\) scale. Importantly, it can also carry out a ``parasitic'' axion search over a wide mass range, without interfering with its primary physics goals\,\cite{Chang2019}.

Overall, storage-ring experiments offer a powerful and complementary avenue for probing axion couplings in the ultralight mass regime. The combination of relativistic boosts, long interaction times, and resonance-enhanced signal amplification renders these methods particularly competitive for discovering or constraining axion-like dark matter.

\section{Comparison of haloscopes and scanning strategies}
\label{sec:SpinHaloscopes:ScanningStep}

In this section, our aim is to describe cavity haloscopes and alkali-noble-gas spin haloscopes from a common perspective and to compare the respective scanning strategies.

\subsection{Bandwidth and resolution hierarchy $Q_{c}$ vs. $Q_{a}$}
%\jl{$\bullet$ Categorized by Q factor.}
Resonant experiments can be categorized into two groups based on their quality factor ($Q_c$): in the first category, the quality factor is lower than that for the UBDM (which is $\sim 10^6$ for the case of the standard halo model) and the opposite case of the quality factor of the experiment exceeding that of the UBDM. In the case of the cavity experiments, this roughly corresponds to the use ``normal'' as opposed to superconducting radio frequency (SRF) cavities.

%\jl{$\bullet$ Traditional cavity exp.}
Traditional cavity experiments belong to the category where $Q_c < Q_a$, with $Q_a \approx 10^6$ representing the $Q$-factor of the axion signal. 
% \DB{We have to decide how we call the DM and explain that in the beginning of the paper. For example, we can can decide to call everything ``UBDM'' or ``axion'' generically and stick with this throughout the m/s!}
These experimental signals exhibit a flat residual PSD after removing the filter spectrum, which originates from white thermal noise. Examples of such experiments include ADMX\,\cite{ADMX_2025_PRL,ADMX:2011hrx, ADMX:2018gho, ADMX:2020ote}, CAPP\,\cite{CAPP-2020-1,CAPP-2020-2,CAPP-2021,CAPP-2023,CAPP-4,CAPP-2024-0,CAPP-2024-1}, HAYSTAC\,\cite{Brubaker:2017rna}, and CAST-CAPP\,\cite{Adair:2022rtw}. For illustrative purposes, we use CAPP-12T as a case study\,\cite{CAPP-2024-0}. 
%\DB{On the next path, we need to check that there is no unnecessary repetition of the discussion in the cavity section.} 

%\jl{$\bullet$ Feature of traditional cavity.}
Considering their target frequency $f_0$ in the range of sub-GHz to 10\,GHz, the signal linewidth is $\Gamma_{a}/(2\pi) \sim \mathcal{O}(1)$\,kHz, and the cavity linewidth is about $\Gamma_c/(2\pi)  \approx f_0/Q_c $. The frequency resolution in the measured power spectrum is $\Delta \nu_{\rm res} = 100$\,Hz for CAPP-12T \cite{CAPP-2024-0}, which is smaller than the signal width. This leads to the relationship:
\begin{align}
\text{Traditional cavity:} \quad  2\pi \Delta \nu_{\rm res} \ll \Gamma_a \ll \Gamma_c\,.
\end{align}

%\jl{$\bullet$ Scan step of traditional cavity.}
In these experiments, the scan step is chosen to be as large as possible while maintaining high sensitivity. For instance, in CAPP-12T\,\cite{CAPP-2024-0}, a scan step of $130$\,kHz is used, corresponding to about one third of the caivty linewidth, $\Gamma_c/(2\pi)\approx400\,\mathrm{kHz}$.

%\jl{$\bullet$ NMR feature and scan step.}
For comagnetometer experiments conducted in the NMR mode\,\cite{Jiang:2021dby, Bloch:2021vnn, Xu:2023vfn}, we have a similar relationship:
\begin{align}
\text{NMR:} \quad  2\pi\Delta \nu_{\rm res} \sim \Gamma_a \ll \Gamma_{\rm NMR}\,,
\end{align}
where the NMR linewidth is $\Gamma_{\rm NMR}=2\pi \delta\nu_{\rm NMR}\sim (2\pi)\times 0.01-0.018\,\unit{\s^{-1}}$\, in the ChangE-NMR experiment\,\cite{Xu:2023vfn}. Since the target axion frequencies are approximately in the $\mathcal{O}(10)$\,Hz range, this results in a $\Gamma_a/(2\pi) \sim \SI{10}{\micro\Hz} $. Each individual measurement scan lasts about 30\,h on average, barely meeting the condition $\Delta \nu_{\rm res} \sim \Gamma_a/(2\pi)$.

%\jl{$\bullet$ NMR is similar to traditional cavity in terms of scanning.}
The relation $\Delta \nu_{\rm res}\ll\Gamma_a/(2\pi)$ implies that cavity experiments have multiple frequency bins within a signal width, whereas the noble gas alkali NMR signal discussed in this paper is typically only present in a few bins since $\Delta \nu_{\rm res}\approx\Gamma_a/(2\pi)$. %ultimately yielding better sensitivity due to superior resolution. 
Despite this difference, traditional cavity and NMR experiments align in all other aspects with the signal being well-contained within the Lorentzian amplification linewidth $\delta\nu_{\rm NMR}$. Consequently, it is expected that NMR experiments would utilize a scan step of  $\delta\nu_{\rm NMR}$, as in experiments of \citet{Jiang:2021dby, Bloch:2021vnn} which satisfies $\kappa \lesssim 1$, where $\kappa$ is a characteristic quantity for NMR experiments defined as Eq.\,\eqref{eq:kappa-definition} in Sec.\,\ref{sec:sen-NMR-haloscopes}. However, the ChangE-NMR search\,\cite{Xu:2023vfn} employs a significantly wider scan step of $0.25$\,Hz, exceeding $\delta\nu_{\rm NMR}$, because the background noise is predominantly magnetic noise dominating over the non-magnetic noise, $\kappa \gg 1$, see Sec.\,\ref{sec:sen-NMR-haloscopes} and Fig.\,\ref{fig:flat_peak}. Therefore, the sensitivity band is wider than the width of the NMR resonance by a factor of $\sqrt{\kappa}$, effectively increasing the scanning speed. A similar speed-up can be achieved in lumped-element haloscopes \cite{Chaudhuri:2021}, see Sec.\,\ref{sec:Lumped_circuits_narrowband}. 

Lastly, our focus shifts to the SRF cavity experiments\,\cite{danhoahn2023thesis,Cervantes:2022gtv, SHANHE:2023kxz}, characterized by a flat residual PSD. However, owing to the high cavity Q-factors ($Q_c \gg Q_a$), the following relationship holds:
\begin{align}
\text{SRF cavity:}\quad 2\pi\Delta \nu_{\rm res} \sim  \Gamma_c \ll \Gamma_a\,,
\end{align}
meaning it can capture only a fraction of the axion-signal power within the cavity bandwidth. This is in contrast to traditional cavity and NMR measurements, which can detect the entire signal power. 
At the same time, the noise power in each frequency bin also decreases due to the narrow linewidth of the resonator. 
In high-$Q$ systems such as SRF cavities or long $T_{2}$ NMR systems, the stochastic nature of the axion signal becomes particularly important~\cite{Foster2018Revealing_PhysRevD,Gramolin2022_SpectralPhysRevD}: the effective number of independent averages achievable per unit time is reduced, and the hypothesis-testing procedure must be modified accordingly [cf.\ Sec.\,\ref{sec:sen-NMR-haloscopes} and \citet{kim_junu_arne_under_preparation}]\footnote{For ultralow-mass axions, due to the long coherence time of axions, stochastic behavior must also be considered, as described in Sec.\,\ref{sec:SpinHaloscopes:ScanningSpeed}. }. 
At the same time, because SRF cavity haloscopes are expected to remain in the flat-noise regime (e.g., due to the circulator and near quantum-limited amplification), their scan optimization differs from the peaked-noise NMR case discussed above.
Therefore, the situation is dual to that of a low quality factor (Sec.\,\ref{sec:CavityHaloscopes:ScanningStrategy}). {Specifically, the scan step is set to a fraction of the axion width $\Gamma_a/(2\pi)$.

\subsection{Noise property: peaked regime vs. flat regime}
As mentioned in Sec.\,\ref{sec:sen-NMR-haloscopes}, the spectrum of noise in a spin haloscope can be categorized into a flat regime ($\kappa \sim 1$) and a peaked regime ($\kappa \gg 1$), where the non-amplifiable noise dominates in the former and the amplifiable noise dominates in the latter. In principle, similar considerations can also be applied to cavity searches; see also the LC circuit Sec.\,\ref{sec:Lumped_circuits} and \citet{Chaudhuri:2018rqn, Chaudhuri:2021}. However, the situation with cavities is somewhat different. 
The noise of a cavity haloscope can be decomposed, as mentioned in Sec.\,\ref{sec:CavityHaloscopes:Noise}, into thermal noise from the cavity, Johnson noise, and the added noise of the circuit outside the cavity, as seen in Eq.\,\eqref{eq:system noise partition}. The cavity noise temperature profile seen by antenna $T_{\mathrm{cav}}(\nu)$ can be represented as a product of the effective temperature $T_{\mathrm{eff}}(\nu)$ in Eq.\,\eqref{eq:cav_effective_temp} and a Lorentzian function in Eq.\,\eqref{eq:eta-frequency}:
\begin{equation}
    T_{\mathrm{cav}}(\nu) = \frac{4b}{(1+b)^{2}}T_{\mathrm{eff}}(\nu) L(\nu, \nu_{c}) = \sigma_{\mathrm{am}}(\nu) \eta^{2}(\nu)\,.
\end{equation}
Therefore, the effective temperature can be expressed in terms of the amplifiable noise $\sigma_{\mathrm{am}}(\nu)$ and the cavity quality factor, $Q_{\mathrm{cav}}$ as $T_{\mathrm{eff}}(\nu) \equiv [(1+b)^{2}/4b]\sigma_{\mathrm{am}}(\nu) Q_{\mathrm{cav}}^{2} $.
Here, the frequency dependence of the amplifiable noise $\sigma_{\rm{am}}(\nu)$ differs from the Lorentzian response of the resonator. This frequency dependence is governed by Bose-Einstein statistics of photons, see Eq.\,\eqref{eq:cav_effective_temp}. 
\begin{comment}
\DB{In equation (50), we have a function of $\nu$ on the right and $Q_{cav}$ is a number, so there cannot be an equality!}\YG{I recognize your point when I write this expression. I think $\sigma_{\mathrm{am}}$ is frequency dependent, due to some quantum mechanical properties of noise source, e.g. photon of bose-einstein statistics gives frequency dependent noise power. Jia, can the $\sigma_{\mathrm{am}}$ have frequency dependency, even the noise itself is Gaussian? Done}
\end{comment}
For a cavity haloscope, as shown in Fig.\,\ref{fig:cavity_circulator_diagram}, a circulator is introduced between the cavity and the first-stage low-noise amplifier to reduce impedance mismatch-induced loss between the cavity and the amplifier.
Then, the total system noise in Eq.\,\eqref{eq:system noise partition} has contributions from the thermal noise of the cavity and the 50$\,\Omega$ load of the circulator. When the 50$\,\Omega$-load temperature and the cavity temperature are equal, we define the total noise contribution as $T_{\rm{out}}(\nu)$, with contributions from the cavity, $T_{\mathrm{cav}}(\nu)$, and the circulator, $T_{\mathrm{circ}}(\nu)$. This can be expressed as:
\begin{equation}
\begin{split}
T_{\rm{out}}(\nu) &=T_{\mathrm{cav}}(\nu) +T_{\rm{circ}}(\nu)=T_{\mathrm{cav}}(\nu) + \frac{(1-b)^2+4(Q_{l}(\nu-\nu_{c})/\nu_{c})^{2}}{(1+b)^2+4(Q_{l}(\nu-\nu_{c})/\nu_{c})^{2}}T_{\mathrm{eff}}(\nu)\\
&=\frac{4b}{(1+b)^2}L(\nu,\nu_{c})T_{\mathrm{eff}}(\nu) + \frac{(1-b)^2+4(Q_{l}(\nu-\nu_{c})/\nu_{c})^{2}}{(1+b)^2+4(Q_{l}(\nu-\nu_{c})/\nu_{c})^{2}}T_{\mathrm{eff}}(\nu) = T_{\mathrm{eff}}(\nu)\,,
\end{split}
\label{eq: noise split}
\end{equation}
where $Q_{l}=Q_{\rm{cav}}/(1+b)$. In this case, the cavity-related Lorentzian term cancels, yielding $T_{\mathrm{out}}(\nu)=T_{\mathrm{eff}}(\nu)$.
%The Lorentzian term associated with the cavity cancels.

%However, due to the circulator, the first stage amplifier sees flat noise at $T_{\mathrm{eff}}$, and the frequency dependence of the resonator is suppressed by the additional thermal noise reflected from the circulator
 Moreover, as discussed in Sec.\,\ref{sec:CavityHaloscopes:Noise}, due to vacuum fluctuations, the minimum achievable thermal noise temperature $T_{\mathrm{eff}}(\nu)$ in Eq.\,\eqref{eq:cav_effective_temp} is limited to the noise level corresponding to ``half-a-photon noise power'' $h\nu/(2k_{B})$\,\cite{PhysRevD.88.035020}. Similarly, the minimum achievable noise for the amplifier is constrained to the noise level of half-a-photon by the Heisenberg uncertainty principle. Note that most haloscopes use phase-insensitive detection, effectively detecting both quadratures at the same time and thus enabling operation at the half-a-photon noise level\,\cite{PhysRevD.88.035020}. 
In state-of-the-art cavity-haloscope experiments operating at this limit, we have $ T_{\mathrm{eff}} \sim T_{\mathrm{add}} $. 
Since the thermal noise from the cavity plus the load contribution are close to the added noise temperature (from the receiver chain), this equality implies that the function $ \kappa$ defined in Eq.\,\eqref{eq:kappa-definition} is close to unity. Consequently, the amplifiable noise and non-amplifiable noise are of the same order for the cavity, whereas in the NMR spin haloscope case, as in the work of \citet{Xu:2023vfn}, the amplifiable noise is larger than the non-amplifiable noise. As a result, there is an enhancement in the scan step in the NMR case, but not in the cavity case (see discussion in Sec.\,\ref{sec:CavityHaloscopes:casestudy}). This leads to differences in the scanning-rate optimization between gigahertz-range cavity haloscopes (Eq.\,\eqref{eq: multi bin scanning rate}) and low-frequency ($\lesssim$10\,Hz) spin haloscopes (Eq.\,\eqref{eq:spin Scanning speed}). 

A fundamental difference arises from the distinction between the flat and peaked regimes, see Sec.\,\ref{sec:sen-NMR-haloscopes}.
In a typical gigahertz-range cavity haloscope, thanks to the use of a circulator and a near-quantum-limited low-noise amplifier, the system operates in the flat regime ($\kappa \sim 1$). Therefore, unlike in the NMR-haloscope approaches, the sensitive bandwidth is not enhanced by $\kappa$, but is determined by either the cavity or axion bandwidth. 
\subsection{Other differences}
Another key distinction in scanning speed, as in Eq.\,\eqref{eq: multi bin scanning rate}, \eqref{eq:lumped_scan_rate}, \eqref{eq:spin Scanning speed} arises from how the SNR scales with the measurement time $\Delta T$, which depends on whether $\Delta T$ is shorter or longer than the axion coherence time $\tau_a$.
In the incoherent regime $\Delta T \gg \tau_a$, signals from different coherence intervals add incoherently, and the power SNR grows as $\sqrt{\Delta T}$, whereas in the coherent regime $\Delta T \lesssim \tau_a$, the signal can accumulate coherently over the measurement interval, leading to an approximately linear dependence on $\Delta T$ in the PSD-based formulation (see Sec.\,\ref{sec:CavityHaloscopes:ScanningStrategy},\ref{sec:SpinHaloscopes:ScanningSpeed}).
The corresponding SNR scalings (with a weak axion signal compared to the noise) can be expressed as:
\begin{equation}
\begin{cases}
&\displaystyle\mathrm{SNR}_{\mathrm{cav}} \propto \frac{g_{a\gamma\gamma}^{2} Q_{c}}{k_{B} T_{\mathrm{sys}}} \sqrt{\frac{\Delta T}{\Delta \nu_{a}}}\,, \\\\
&\displaystyle\mathrm{SNR}_{\mathrm{NMR}} \propto \frac{g_\mathrm{aNN}^{2}\Delta T}{\Sigma_{am}}\,.
\end{cases}
\label{eq: SNR compare cav NMR} 
\end{equation}
As a result, the scan-rate expressions differ because the integration time required to reach a fixed target SNR scales differently in the two regimes.
For cavity haloscopes, where the power SNR grows as $\sqrt{\Delta T}$, achieving a fixed SNR implies $\Delta T \propto \mathrm{SNR}^2$, which leads to a scan-rate of the form of Eq.\,\eqref{eq: multi bin scanning rate} when written as $\Delta\nu/\Delta T$.
In contrast, for NMR haloscopes in the regime where the SNR scales approximately linearly with the $\Delta T$, resulting in the scan-rate form given by Eq.\,\eqref{eq:spin Scanning speed}.

In summary, the difference in scanning-speed equations between cavity and NMR haloscopes originates from the distinct time dependence of signal and noise, which, in turn, is determined by the axion-mass–dependent coherence time and constraints on the achievable measurement time.

\section{Conclusion and outlook}

In this paper, we discuss the principles, general features, sensitivity, and scanning-rate optimization for various types of experiments searching for ultralight bosonic dark matter. The main features of these experiments are summarized in Table~\ref{tab:summary}. 
\begin{table*}[t]
\caption{Comparison of haloscopes. The figures of merit for spin haloscopes are written in terms of different quantities (number of spins, coherence time, etc.) compared to the other cases and are not presented here.}
\label{tab:summary}
\small
\renewcommand{\arraystretch}{1.2}
\setlength{\tabcolsep}{3pt}

\begin{ruledtabular}
\begin{tabular}{ccccc}
%\begin{tabular}{|l|c|c|c|c|}
 & \cbox{0.18\textwidth}{Cavity\\haloscopes}
 & \cbox{0.18\textwidth}{Lumped\\circuits}
 & \cbox{0.18\textwidth}{Atomic and NMR\\haloscopes}
 & \cbox{0.18\textwidth}{Storage\\rings} \\
\hline
Interaction
 & \cbox{0.18\textwidth}{Axion-photon}
 & \cbox{0.18\textwidth}{Axion-photon}
 & \multicolumn{2}{c}{\cbox{0.36\textwidth}{Axion-nucleon and\\axion-gluon}} \\
\hline
Working principle
 & \lbox{0.18\textwidth}{Conversion of axions into photons in a resonant cavity under a strong magnetic field}
 & \lbox{0.18\textwidth}{Same as for cavity haloscope but photon wavelength is larger than the detector}
 & \multicolumn{2}{c}{\lbox{0.36\textwidth}{Axion field gradient or axion-induced EDM causing spin precession}} \\
\hline
Tuning method \\for resonance searches
 & \cbox{0.18\textwidth}{Adjusting cavity geometry}
 & \cbox{0.18\textwidth}{Circuit resonance}
 & \multicolumn{2}{c}{\cbox{0.36\textwidth}{Adjusting resonance frequency via bias magnetic field}} \\
\hline
Capability of\\broadband search
 & \cbox{0.18\textwidth}{No}
 & \cbox{0.18\textwidth}{Yes}
 & \multicolumn{2}{c}{\cbox{0.36\textwidth}{Yes}} \\
\hline
Parameters to\\optimize
 & \lbox{0.18\textwidth}{Cavity parameters (Q-factor, volume, form factor), antenna coupling and system noise temperature}
 & \lbox{0.18\textwidth}{Q-factor, magnetic field, volume, temperature, amplifier noise}
 & \multicolumn{2}{c}{\lbox{0.36\textwidth}{Sample composition, relaxation time $T_2$, spin polarization, and system noise}} \\
\hline
Figure of merit
 & \cbox{0.18\textwidth}{$B_{0}^{4}V^2C^{2}Q_{c}$}
 & \cbox{0.18\textwidth}{$B_{0}^{4}V^{10/3}c_{PU}^{4}Q_r$}
 & \multicolumn{2}{c}{\cbox{0.36\textwidth}{--}} \\
\hline
Mass range
 & \lbox{0.18\textwidth}{0.1--\SI{100}{\micro\eV} (MHz--GHz)}
 & \lbox{0.18\textwidth}{$\sim 10^{-11}$\,eV -- \SI{1}{\micro\eV} (kHz--GHz)}
 & \lbox{0.18\textwidth}{$\sim 10^{-22}$\,eV -- \SI{1}{\micro\eV} (\SI{100}{\nano\Hz}--GHz)}
 & \lbox{0.18\textwidth}{$\sim 10^{-22}$\,eV -- \SI{10}{\nano\eV} (100\,nHz--MHz)} \\
\hline
Examples
 & \lbox{0.18\textwidth}{ADMX~\cite{ADMX_2025_PRL}, HAYSTAC~\cite{HAYSTAC-2021}, CAPP~\cite{CAPP-2024-1}, QUAX~\cite{Quax_2025}, ORGAN~\cite{ORGAN_2024}}
 & \lbox{0.18\textwidth}{SHAFT~\cite{Gramolin2021NaturePhysics}, ABRACADABRA~\cite{Ouellet:2019prl}, DMRadio~\cite{Rapidis:2023}, ADMX SLIC~\cite{ADMX:2020prl}, WISPLC~\cite{ZZhang:2022prd}}
 & \lbox{0.18\textwidth}{NASDUCK~\cite{Bloch2023NASDUCK_SERF}, ChangE-NMR~\cite{xu2023_ChangE}, CASPEr~\cite{CASPEr-LF-2025}, QUAX~\cite{QUAX:2020adt}}
 & \lbox{0.18\textwidth}{COSY~\cite{Karanth2023}, sr-EDM~\cite{Chang2019}} \\
\end{tabular}
\end{ruledtabular}
\end{table*}
In the course of the discussion, we have attempted to unify the language used to describe the noise characteristics of various detectors.

Typically, the searches employ resonant detectors, so covering a finite mass range of the UBDM particles requires scanning the resonance frequency. Central questions are how fast such scanning can be done and what the optimum compromise is between sensitivity and scanning speed? The key aspect of our considerations is understanding the sources of noise. In all cases, it turns out to be useful to decompose the noise into amplifiable and non-amplifiable components.

In resonant cavity-based dark matter searches, the amplifiable noise (thermal noise) is comparable to the non-amplifiable noise (added noise from the amplifier chain). In addition, the circulator placed between the cavity and the amplifier flattens the thermal noise response from the cavity. Therefore, a cavity haloscope is operated in the ``flat'' regime described in the NMR haloscope section (\ref{sec:sen-NMR-haloscopes}) under ideal conditions. However, the axion signals are determined by the properties of the cavity and the kinetic energy distribution of the dark matter axion. Accordingly, tuning is performed in frequency steps on the order of the narrower of the cavity and the axion bandwidths. Since the axion mass is unknown, the cavity haloscope experiments are designed to maximize the scanning rate by optimizing the cavity under given experimental constraints, such as the limited cavity volume imposed by the superconducting magnet and the frequency range of the amplifier. Typically, the tuning step size is set to less than or equal to half the cavity or axion bandwidth. In addition, the scanning-rate equation remains valid for a high-Q resonator cavity, since the noise is reduced by the narrow cavity bandwidth while the signal power remains constant and is ultimately limited by the quality factor of the axion. The stochastic nature of the axion field must be taken into account in the analysis, especially if the number of averages is reduced due to the higher scanning speed enabled by high-Q cavities.

In the case of resonant spin searches, when amplifiable noise dominates, it is possible to make scan steps significantly larger than the resonance width, speeding up the search. 
% The same conclusion was reported for LC circuits\,\cite{Chaudhuri:2021}. 
A similar conclusion was reported for resonant lumped-element circuits in the case of thermal noise shaped by the resonator exceeding the white noise level set by the amplifier input\,\cite{Chaudhuri:2021}. 

The field of axion searches is very active, with new techniques being proposed and explored. For example, employing Josephson junction-based quantum detectors has been proposed\,\cite{HAYSTAC-2021, Colorado_Squeezing_1, Colorado_Squeezing_2}. These leverage vacuum squeezing to measure a single quadrature, surpassing the standard quantum limit (also referred to as the single-photon noise limit), which is a fundamental limitation of traditional double quadrature measurements. In addition to squeezing experiments, photon-counting experiments, such as those using Rydberg atoms, have been pursued at Kyoto University\,\cite{CARRACKII_1999_NPB}, and are currently being developed at Yale\,\cite{Yale_rydberg_2024}. 

Recently, a new detection scheme using electro-optics technology has been proposed to read out the axion-induced electric field from the cavity\,\cite{davoudiasl2025EO_Heterodyne} that can potentially reach near the standard quantum limit. Seeding the cavity and measuring the fluctuations of variance could be helpful \cite{Variance:Omarov}, however, this may not necessarily be advantageous in all cases \cite{Aybas:2021cdk}.

As we still do not know what the nature of DM is, one might wonder if all the considerable effort to detect axions may ultimately be in vain. In our opinion, this will not be the case, as the techniques with unprecedented sensitivity may and do find applications in other areas of science and technology. For example, it has been realized recently that many of the developments made in the context of axion searches are also immediately applicable to searches for gravitational waves \cite{Aggrawal:reviewGW}.
More broadly, developments in quantum- and near-quantum-limited microwave reception (e.g., parametric amplification and squeezing) and in ultrastable, low-loss resonant structures are of broad utility in precision microwave measurements, including radio astronomy \cite{bryerton2013ultra} and deep-space communication receivers \cite{bautista2001cryogenic}. Related advances in ultra-sensitive magnetometry and comagnetometry, central to spin-haloscope experiments, likewise translate to applied sensing contexts \cite{budker2007optical,bud13OpticalMagnetometry,kornack2005nuclear,Kitching:25}.

\section*{Acknowledgments}

The authors acknowledge helpful discussions with Junu Jeong. 
%This work has been supported by the Cluster of Excellence “Precision Physics, Fundamental Interactions, and Structure of Matter” (PRISMA+ EXC 2118/2) funded by the German Research Foundation (DFG) within the German Excellence Strategy (Project ID 390831469).
This research was supported in part by the DFG Project ID 390831469: EXC 2118 (PRISMA++ Cluster of Excellence), by the COST Action within the project COSMIC WISPers (Grant No. CA21106), by ERC grant ERC-2024-SYG 101167211, and the Munich Institute for Astro-, Particle and BioPhysics (MIAPbP), which is funded by the Deutsche Forschungsgemeinschaft (DFG, German Research Foundation) under Germany´s Excellence Strategy – EXC-2094 – 390783311.
The work of D. F. J. K. was supported by the U.S. National Science Foundation under grant PHYS-2510625. The work of Y. Kim was supported by the Alexander von Humboldt Foundation. The work of J.L. is supported by the National Science Foundation of China under Grant No. 12235001 and No. 12475103. The work of D. A. was supported by the Scientific and Technological Research Council of Türkiye under grant number 122C341.
O. K. is supported by the U.S. DOE under Contract No. DE-SC0021616.
The work of A.~O.~S. is supported by the U.S. National Science Foundation CAREER grant PHY-2145162, the U.S. Department of Energy grant DE-SC0025942, and the Gordon and Betty Moore Foundation, grant DOI 10.37807/gbmf12248.

\bibliographystyle{apsrev4-1}
\bibliography{references}

\end{document}